\documentclass[twocolumn,a4paper]{aastex63}

\usepackage{xcolor}
\usepackage{graphicx}
\usepackage{xspace}

\usepackage{amsmath} % for multiline 
\usepackage{amstext}
\usepackage{amssymb} %for less sim
\usepackage{mathtools} % for overset
\usepackage{paralist} % for modified counter environments, e.g. compactenum

\let\tablenum\relax
\usepackage{siunitx}
\usepackage[utf8]{inputenc} % for umlaute
\usepackage[T5,T1]{fontenc}

\usepackage{natbib}
\usepackage{hyperref}

\graphicspath{{./}{figures/}}

\newcommand{\anaAC}{point-source correlation analysis\xspace} % 
\newcommand{\anaGE}{UHECR stacking analysis\xspace}
\newcommand{\anaPA}{2pt-correlation analysis\xspace}
\newcommand{\txs}{TXS~0506+056\xspace}
\newcommand{\ngc}{NGC~1068\xspace}

\DeclareSIUnit{\EeV}{EeV}
\DeclareSIUnit{\PeV}{PeV}
\DeclareSIUnit{\TeV}{TeV}
\DeclareSIUnit{\Mpc}{Mpc}
\DeclareSIUnit{\Gpc}{Gpc}
\DeclareSIUnit{\erg}{erg}
\DeclareSIUnit{\year}{yr}

\newcommand{\derive}[3][]{\ensuremath{\frac{{\rm d}^{#1}#2}{{\rm d}#3^{#1}}}}
\newcommand{\drv}[1]{\ensuremath{{\rm d}#1}}
\newcommand{\dg}{\ensuremath{{}^\circ}}

\newcommand{\CR}{\ensuremath{{\rm CR}}}
\newcommand{\fcorr}{\ensuremath{f_{\rm corr}}}

\newcommand{\ecuth}{\ensuremath{E_{\rm cut}=\SI{100}{\EeV}}}

\published{2022 August 3 in \textit{The Astrophysical Journal}}

\shorttitle{Search for Correlations of Neutrinos and UHECRs}
\shortauthors{ANTARES, IceCube, Pierre Auger, and Telescope Array Collaborations}

\begin{document}

\title{Search for Spatial Correlations of Neutrinos with Ultra-High-Energy Cosmic Rays}

\begin{abstract}
For several decades, the origin of ultra-high-energy cosmic rays (UHECRs) has been an unsolved question of high-energy astrophysics. 
One approach for solving this puzzle is to correlate UHECRs with high-energy neutrinos, since neutrinos are a direct probe of hadronic interactions of cosmic rays and are not deflected by magnetic fields. 
In this paper, we present three different approaches for correlating the arrival directions of neutrinos with the arrival directions of UHECRs.
The neutrino data is provided by the IceCube Neutrino Observatory and ANTARES, while the UHECR data with energies above $\sim$\SI{50}{\EeV} is provided by the Pierre Auger Observatory and the Telescope Array.
All experiments provide increased statistics and improved reconstructions with respect to our previous results reported in 2015. 
The first analysis uses a high-statistics neutrino sample optimized for point-source searches to search for excesses of neutrinos clustering in the vicinity of UHECR directions. 
The second analysis searches for an excess of UHECRs in the direction of the highest-energy neutrinos. 
The third analysis searches for an excess of pairs of UHECRs and highest-energy neutrinos on different angular scales.
None of the analyses has found a significant excess, and previously reported over-fluctuations are reduced in significance.
Based on these results, we further constrain the neutrino flux spatially correlated with UHECRs.
\end{abstract}

\collaboration{10000}{The ANTARES collaboration}

\author{A.~Albert}
\affiliation{Universit\'e de Strasbourg, CNRS,  IPHC UMR 7178, F-67000 Strasbourg, France}
\affiliation{Universit\'e de Haute Alsace, F-68100 Mulhouse, France}

\author{S.~Alves}
\affiliation{IFIC - Instituto de F\'isica Corpuscular (CSIC - Universitat de Val\`encia) c/ Catedr\'atico Jos\'e Beltr\'an, 2 E-46980 Paterna, Valencia, Spain}

\author{M.~Andr\'e}
\affiliation{Technical University of Catalonia, Laboratory of Applied Bioacoustics, Rambla Exposici\'o, 08800 Vilanova i la Geltr\'u, Barcelona, Spain}

\author{M.~Anghinolfi}
\affiliation{INFN - Sezione di Genova, Via Dodecaneso 33, 16146 Genova, Italy}

\author{M.~Ardid}
\affiliation{Institut d'Investigaci\'o per a la Gesti\'o Integrada de les Zones Costaneres (IGIC) - Universitat Polit\`ecnica de Val\`encia. C/  Paranimf 1, 46730 Gandia, Spain}

\author{S.~Ardid}
\affiliation{Institut d'Investigaci\'o per a la Gesti\'o Integrada de les Zones Costaneres (IGIC) - Universitat Polit\`ecnica de Val\`encia. C/  Paranimf 1, 46730 Gandia, Spain}

\author{J.-J.~Aubert}
\affiliation{Aix Marseille Univ, CNRS/IN2P3, CPPM, Marseille, France}

\author{J.~Aublin}
\affiliation{Universit\'e de Paris, CNRS, Astroparticule et Cosmologie, F-75013 Paris, France}

\author{B.~Baret}
\affiliation{Universit\'e de Paris, CNRS, Astroparticule et Cosmologie, F-75013 Paris, France}

\author{S.~Basa}
\affiliation{Aix Marseille Univ, CNRS, CNES, LAM, Marseille, France}

% ...

\collaboration{10000}{The IceCube collaboration}

\author{R. Abbasi}
\affiliation{ Department of Physics, Loyola University Chicago, Chicago, IL 60660, USA}

\author{M. Ackermann}
\affiliation{ DESY, D-15738 Zeuthen, Germany}

\author{J. Adams}
\affiliation{ Dept. of Physics and Astronomy, University of Canterbury, Private Bag 4800, Christchurch, New Zealand}

\author{J. A. Aguilar}
\affiliation{ Universit{\'e} Libre de Bruxelles, Science Faculty CP230, B-1050 Brussels, Belgium}

\author{M. Ahlers}
\affiliation{ Niels Bohr Institute, University of Copenhagen, DK-2100 Copenhagen, Denmark}

\author{M. Ahrens}
\affiliation{ Oskar Klein Centre and Dept. of Physics, Stockholm University, SE-10691 Stockholm, Sweden}

\author{J.M. Alameddine}
\affiliation{ Dept. of Physics, TU Dortmund University, D-44221 Dortmund, Germany}

\author{C. Alispach}
\affiliation{ D{\'e}partement de physique nucl{\'e}aire et corpusculaire, Universit{\'e} de Gen{\`e}ve, CH-1211 Gen{\`e}ve, Switzerland}

\author{A. A. Alves Jr.}
\affiliation{ Karlsruhe Institute of Technology, Institute for Astroparticle Physics, D-76021 Karlsruhe, Germany}

\author{N. M. Amin}
\affiliation{ Bartol Research Institute and Dept. of Physics and Astronomy, University of Delaware, Newark, DE 19716, USA}

% ...

\collaboration{10000}{The Pierre Auger collaboration}

\author{P.~Abreu}
\affiliation{ Laborat\'orio de Instrumenta\c{c}\~ao e F\'\i{}sica Experimental de Part\'\i{}culas -- LIP and Instituto Superior T\'ecnico -- IST, Universidade de Lisboa -- UL, Lisboa, Portugal}

\author{M.~Aglietta}
\affiliation{ Osservatorio Astrofisico di Torino (INAF), Torino, Italy}
\affiliation{ INFN, Sezione di Torino, Torino, Italy}

\author{J.M.~Albury}
\affiliation{ University of Adelaide, Adelaide, S.A., Australia}

\author{I.~Allekotte}
\affiliation{ Centro At\'omico Bariloche and Instituto Balseiro (CNEA-UNCuyo-CONICET), San Carlos de Bariloche, Argentina}

\author{K.~Almeida Cheminant}
\affiliation{ Institute of Nuclear Physics PAN, Krakow, Poland}

\author{A.~Almela}
\affiliation{ Instituto de Tecnolog\'\i{}as en Detecci\'on y Astropart\'\i{}culas (CNEA, CONICET, UNSAM), Buenos Aires, Argentina}
\affiliation{ Universidad Tecnol\'ogica Nacional -- Facultad Regional Buenos Aires, Buenos Aires, Argentina}

\author{J.~Alvarez-Mu\~niz}
\affiliation{ Instituto Galego de F\'\i{}sica de Altas Enerx\'\i{}as (IGFAE), Universidade de Santiago de Compostela, Santiago de Compostela, Spain}

\author{R.~Alves Batista}
\affiliation{ IMAPP, Radboud University Nijmegen, Nijmegen, The Netherlands}

\author{G.A.~Anastasi}
\affiliation{ Osservatorio Astrofisico di Torino (INAF), Torino, Italy}
\affiliation{ INFN, Sezione di Torino, Torino, Italy}

\author{L.~Anchordoqui}
\affiliation{ Department of Physics and Astronomy, Lehman College, City University of New York, Bronx, NY, USA}

% ...

\collaboration{10000}{The Telescope Array collaboration}

\author{T.~Abu-Zayyad}
\affiliation{ Department of Physics, Loyola University Chicago, Chicago, Illinois, USA}
\affiliation{ High Energy Astrophysics Institute and Department of Physics and Astronomy, University of Utah, Salt Lake City, Utah, USA}

\author{M.~Allen}
\affiliation{ High Energy Astrophysics Institute and Department of Physics and Astronomy, University of Utah, Salt Lake City, Utah, USA}

\author{Y.~Arai}
\affiliation{ Graduate School of Science, Osaka City University, Osaka, Osaka, Japan}

\author{R.~Arimura}
\affiliation{ Graduate School of Science, Osaka City University, Osaka, Osaka, Japan}

\author{E.~Barcikowski}
\affiliation{ High Energy Astrophysics Institute and Department of Physics and Astronomy, University of Utah, Salt Lake City, Utah, USA}

\author{J.W.~Belz}
\affiliation{ High Energy Astrophysics Institute and Department of Physics and Astronomy, University of Utah, Salt Lake City, Utah, USA}

\author{D.R.~Bergman}
\affiliation{ High Energy Astrophysics Institute and Department of Physics and Astronomy, University of Utah, Salt Lake City, Utah, USA}

\author{S.A.~Blake}
\affiliation{ High Energy Astrophysics Institute and Department of Physics and Astronomy, University of Utah, Salt Lake City, Utah, USA}

\author{I.~Buckland}
\affiliation{ High Energy Astrophysics Institute and Department of Physics and Astronomy, University of Utah, Salt Lake City, Utah, USA}

\author{R.~Cady}
\affiliation{ High Energy Astrophysics Institute and Department of Physics and Astronomy, University of Utah, Salt Lake City, Utah, USA}

% ...


\correspondingauthor{The ANTARES, IceCube, Pierre Auger and Telescope Array Collaborations}
\email{analysis@icecube.wisc.edu}

\section{Introduction}

The Earth is continuously bombarded by high-energy cosmic rays, most of which are charged atomic nuclei \citep{Zyla:2020zbs}.
It is generally believed that cosmic rays with energies above \SI{1}{\EeV} (\SI{e18}{eV}), known as ultra-high-energy cosmic rays (UHECRs), mostly originate from extra-galactic sources in the nearby universe.
Based on the estimated magnitude of galactic magnetic fields~\citep{Nagano:2000ve}, cosmic rays below this energy are believed to diffuse within their host galaxy, whereas cosmic rays above this energy escape from the galaxy. 
These assumptions are confirmed by the observation of large-scale anisotropies in the arrival directions of UHECRs above \SI{8}{EeV} with the excess flux directed from outside of our Galaxy~\citep{Aab:2017tyv}.
The two largest observatories for UHECRs are the Pierre Auger Observatory (Auger)~\citep{ThePierreAuger:2015rma} in Argentina in the Southern hemisphere and Telescope Array (TA) in the United States in the Northern hemisphere~\citep{AbuZayyad:2012kk}. 
They have performed detailed analyses of the arrival directions of detected UHECRs
and have revealed interesting features, such as possible correlations with the directions
of known starburst and active galaxies in the nearby universe observed by Auger \citep{Aab:2018chp,Abreu:2021eL},
and intermediate-scale directional clustering observed by TA \citep{Abbasi:2014lda,Tkachev:20211D}.
However, due to the small number of events detected at the highest energies,
these indications of anisotropic arrival directions have not yet been confirmed on the $5\sigma$ level.
Thus, neither Auger nor TA report an unambiguous identification of the UHECR sources to date. 
One of the main reasons is the deflection of cosmic rays by magnetic fields during propagation from their sources to Earth, which alters their directional information with respect to the source positions.
The deflection of UHECRs increases with increasing charge, which is not well determined at the highest energies above \SI{50}{\EeV} considered in this work.
This uncertainty additionally complicates the identification of UHECR sources.
Nevertheless, UHECRs at the highest energies are deflected the least due to their extremely high magnetic rigidity \citep{Alves_Batista_2019}, which makes them suitable for directional correlation searches.

We search for the sources of the highest-energy cosmic rays with a multi-messenger approach using high-energy neutrinos and UHECRs with energies above $\sim$\SI{50}{\EeV}.
High-energy neutrinos are direct tracers of hadronic interactions of cosmic rays, and are thus expected to be produced at the acceleration site or during propagation (see e.g.~\citealp{Murase:2014tsa, meszaros2018astrophysical, Anchordoqui:2007tn}).
Furthermore, the electrically-neutral neutrinos carry the directional information of their origin,
since they are not deflected by magnetic fields.
A combined analysis of UHECRs and high-energy neutrinos is further motivated by the observation of a diffuse flux of high-energy neutrinos of astrophysical origin above the background of atmospheric neutrinos.
The flux follows a hard power law that extends to energies of multiple \si{PeV} and possibly beyond \citep{Aartsen:2013jdh,Aartsen:2016xlq,Stettner:2019tok,Aartsen:2020aqd,Abbasi:2020jmh}. 
The measured flux is closely below the Waxman-Bahcall bound \citep{Waxman:1998yy,Bahcall:1999yr}, which is a theoretical upper bound on the astrophysical neutrino flux derived from the observed UHECR flux, suggesting a connection between UHECRs and high-energy neutrinos \citep{Murase:2013rfa,Ahlers:2018fkn}. 
We note that the Seyfert and Starburst Galaxy \ngc appears both as a neutrino source candidate in~\cite{aartsen10yrIntegratedPS2020} and as a UHECR source candidate in~\cite{Aab:2018chp}.
Recently, the potential of high-energy neutrinos in a multi-messenger approach has been demonstrated together with observations of high-energy photons \citep{IceCube:2018dnn,IceCube:2018cha}, finding compelling evidence for neutrino emission from the blazar \txs.
At the same time, it can be concluded that blazars seem insufficient to saturate the total observed neutrino flux \citep{Murase:2016gly,Aartsen:2016lir} based on the non-observation of steady neutrino fluxes from known blazars in the universe.
This result motivates different approaches to identify hadronic sources in the local universe;
a promising method is to correlate the sources traced with UHECRs with high-energy neutrinos, and vice versa.

In this paper, we report results from three conceptually different approaches. 
All are based on correlating the arrival directions of high-energy neutrinos detected by the IceCube Neutrino Observatory (IC) \citep{Aartsen:2016nxy}
and the ANTARES experiment (ANT) \citep{Collaboration:2011nsa}
with the arrival directions of UHECRs with energies above $\sim$\SI{50}{\EeV} measured by Auger and TA:
\begin{itemize}
    \item Analysis~1 (described in \autoref{sec:anaAC-method})
uses the measured UHECR directions as well as magnetic deflection estimates for identifying regions in which we search for point-like neutrino sources.
    \item Analysis~2 (described in \autoref{sec:anaGE-method})
uses the arrival directions of neutrinos with a high probability of astrophysical origin to search for a correlated clustering of UHECR arrival directions, while also accounting for the magnetic deflection.
    \item Analysis~3 (described in \autoref{sec:anaPA-method})
is based on a largely model-independent two-point correlation analysis of the arrival directions of UHECR and neutrinos with a high probability of astrophysical origin.
\end{itemize}
With these three approaches, we follow up on previous searches performed by the Pierre Auger, Telescope Array and IceCube collaborations \citep{Aartsen:2015dml}, which showed an interesting correlation at a significance level close to three standard deviations.

Since 2015, the analyses have been improved with substantially enlarged data sets, extending the Auger data from \SIrange{10}{13}{\year} (see \autoref{sec:pao}) and TA data from \SIrange{6}{9}{\year} (see \autoref{sec:ta}).
The different data sets from IceCube are enlarged from \SIrange{4}{10.5}{\year}, from \SIrange{4}{7.5}{\year}, and from \SIrange{2}{8}{\year}, depending on the specific detection channel as described in \autoref{sec:ic}.
Newly included are two data sets of \SI{9}{\year} and \SI{11}{\year} from ANTARES, as further specified in \autoref{sec:ant}.
Due to the updated data sets by Auger and TA, their respective shift of the energy scale is updated from a sole shift of TA energies by $-13\%$ to a shift of $+14\%$ for Auger and $-14\%$ TA energies above the energy threshold of $\sim$\SI{50}{\EeV} \citep{biteauCoveringCelestialSphere2019}. 
Furthermore, we include an improved magnetic deflection model that distinguishes between the Northern and Southern hemisphere for analysis 2.
We report the results from the three improved correlation searches, 
which update the preliminary reported results in \citet{Schumacher:2019qdx,Aublin:2019irc,barbanoICRC2019}.
In addition, we report upper limits on the correlated fluxes of UHECRs and neutrinos based on benchmark models for the magnetic deflections.

\section{Observatories and data sets}

All data sets used in this paper are used in previous work by the four respective collaborations.
This section focuses on the main aspects relevant for our analyses.

\subsection{The IceCube Neutrino Observatory}
\label{sec:ic}

The IceCube Neutrino Observatory \citep{Aartsen:2016nxy} is an ice-Cherenkov detector sensitive to neutrinos with energies $\ge \SI{5}{GeV}$.
It is located at the geographic
South Pole, about \SIrange{1.45}{2.45}{km} deep in the ice. 
Its main component consists of a volume of about
\SI{1}{km^3} glacial ice instrumented with 5160 photo-multipliers which are connected to the surface by 86 cable strings.

Two classes of neutrino-induced events can be phenomenologically distinguished:
elongated, track-like events are produced by muons that originate mostly from 
charged-current $\nu_\mu$ interactions;
and the spherical, cascade-like events that originate from charged-current $\nu_e$ and $\nu_\tau$ interactions with hadronic and electromagnetic decays, as well as neutral-current interactions of all flavors.
Typically, track-like events enable a better angular resolution than cascade-like events due to their different topologies, but provide a poorer energy resolution \citep{icEnergyReco2014,wandkowsky2018latest}.
One method of suppressing the dominant background of down-going  muons produced by cosmic-ray interactions in the atmosphere is by selecting events with the interaction vertex within the detector \citep{Aartsen:2014gkd,Kopper:2017zzm,wandkowsky2018latest}.
Alternatively, through-going tracks with either horizontal or up-going directions are selected, such that the atmospheric muons are blocked out by Earth
\citep{Aartsen:2016xlq,haack2018measurement}.
In the case of down-going tracks, a high energy threshold together with elaborate selection procedures are necessary to filter out atmospheric muons \citep{IceCube:2018cha,Aartsen:2016lmt,Aartsen:2016oji}.
In all cases, the remaining event rate is dominated by atmospheric neutrinos.
The selection of astrophysical neutrinos can be achieved on a statistical basis 
by selecting very energetic events or, in case of the very down-going region, by vetoing events where an atmospheric shower is observed in IceTop, IceCube's surface detector for cosmic rays~\citep{IceTop:2013}. 

\begin{figure*}[htp]
\epsscale{1.1}
\plotone{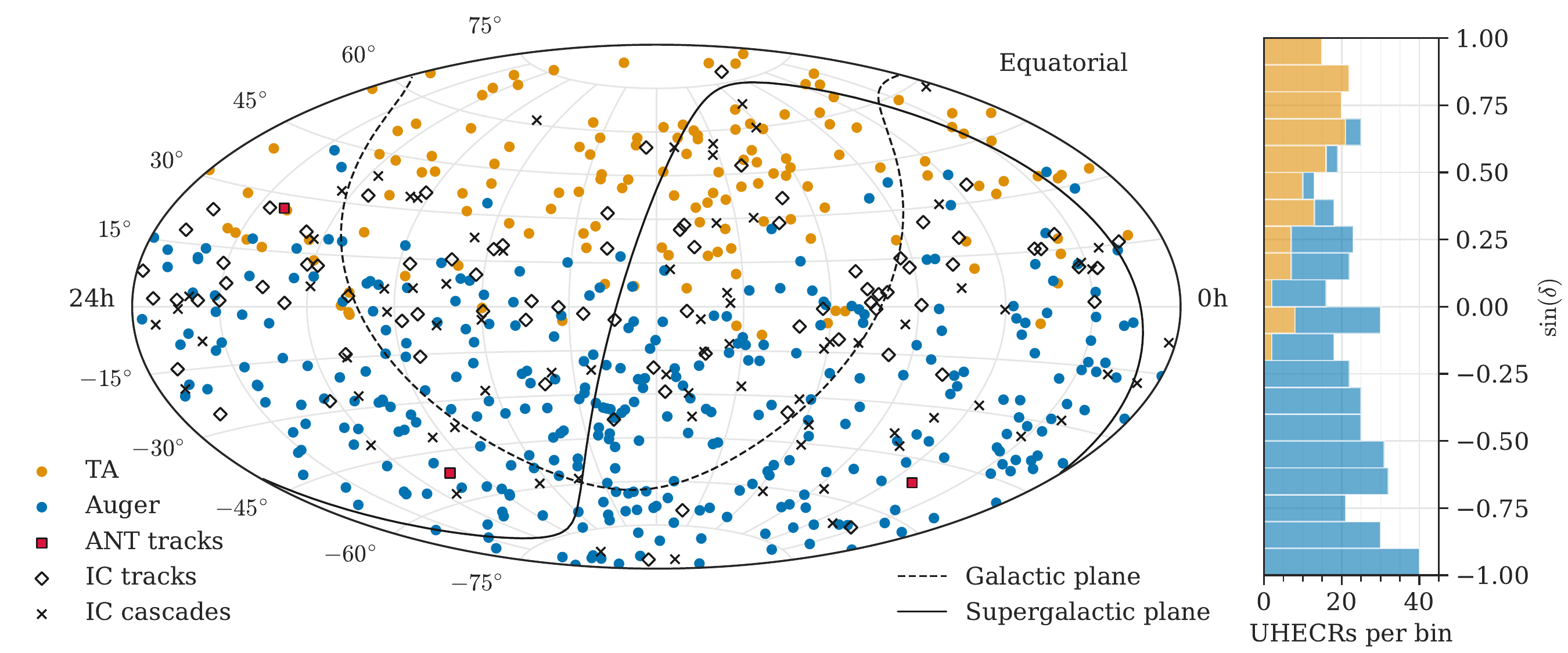}
\caption{
Left: Skymap of the arrival directions of UHECR events and high-energy neutrinos.
The high-energy neutrino track-like events from IceCube  consists of the HESE,  HENU and EHE data sets, while the cascade-like events are only of the IceCube-HESE data sets.
From ANTARES, only high-energy tracks are selected for the analyses.
Right: A histogram of the declination of UHECR events, separated into Auger and TA contribution.}
\label{fig:HE-nu_UHECR_skymap}
\end{figure*}

For the three analyses, data from multiple detection channels
are used, which are:
(i) a data set of through-going tracks from the full sky optimized for point-source searches (PS),
(ii) a data set of high-energy starting events (HESE) of both topologies from the full sky,
(iii) a data set of high-energy neutrinos (HENU) selected from through-going tracks with horizontal and up-going directions, 
and (iv) a data set of tracks from a selection of extremely high-energy events (EHE).
The PS data set is used for analysis 1, while the HESE, HENU and EHE data sets are used for analyses 2 and 3.
For analyses 2 and 3, track-like events from the  HESE, HENU, and EHE data sets are combined,
while multiple instances of identical events are removed.
This results in a data set of 81 track-like events.
In analyses 2 and 3, the 76 cascade-like events from the HESE data set are analyzed separately due to their larger directional uncertainty.
The sky distribution of selected events is shown in \autoref{fig:HE-nu_UHECR_skymap} and an overview of the nomenclature is presented in \autoref{tab:nu_overview}.

The PS data set consists of a combination of data collected from 7 years of operation between 2008 and 2015 that was used for point-source searches \citep{Aartsen:2016oji} and data from 3.5 years of operation between 2015 and 2018
that was selected for the real-time gamma-ray follow-up (GFU) program of IceCube \citep{IceCube:2018cha,Aartsen:2016lmt}.
The combined data set consists of about \num{1.4} million track-like events above $\sim\SI{100}{GeV}$.
It is dominated by atmospheric neutrinos in the Northern hemisphere and by atmospheric muons in the Southern hemisphere. 
The median of the angular resolution ($\Psi$) is better than \SI{0.5}{\degree} above energies of a few TeV.
\autoref{fig:median_psf} shows the median of the angular resolution for the different detector configurations of IceCube, and for data from the ANTARES detector (see \autoref{sec:ant}).
Note that analysis 1 is not based on the median of the angular resolution, but on the estimator for the angular resolution on an event-by-event basis ($\sigma$).

The HESE data set contains 76 cascade-like events and 26 track-like events that have been collected between 2010 and 2017, as presented in \citet{wandkowsky2018latest}. 
It consists of neutrinos of all flavors that interacted inside the detection volume, called starting events, with deposited energies ranging between about \SI{20}{TeV} and \SI{2}{PeV}.
Integrated above \SI{60}{TeV}, which corresponds to 60 events in total, the percentage of events of astrophysical origin, i.e. the astrophysical purity, is larger than \SI{75}{\percent}~\citep{wandkowsky2018latest}, while the percentage is lower below \SI{60}{TeV}.
The angular resolution is about \SI{1}{\degree} for track-like events and \SI{15}{\degree}
for cascade-like events above \SI{100}{TeV}.
The resolution of the deposited energy for tracks and cascades is around \SI{10}{\percent}~\citep{icEnergyReco2014} without accounting for systematic uncertainties, but the cascades have a better correlation with the primary neutrino energy since they deposit most of their energy inside the detector, while tracks do not.

The HENU data set consists of the 35 highest energy track-like events with a reconstructed declination $\geq -5\dg$, which have been collected between 2009 and 2016 \citep{haack2018measurement} for the measurement of the diffuse muon-neutrino flux \citep{Aartsen:2016xlq}.
From the original data set starting at $\sim\SI{100}{GeV}$, only events of
high probability of non-atmospheric origin have been selected by applying an energy threshold of $\ge \SI{200}{TeV}$ on the reconstructed muon energy.
This corresponds to an astrophysical purity of more than \SI{50}{\percent}. 
The astrophysical purity here is defined as the flux ratio of the astrophysical to the sum of atmospheric and astrophysical differential fluxes and thus depends on the assumed astrophysical flux.
For the estimation of the astrophysical purity, the best fit astrophysical flux from \citet{Aartsen:2016xlq}
$ {\drv \phi / \drv E } =
\SI{1.01e-18}{GeV^{-1} cm^{-2} s^{-1} sr^{-1}}
(E/\SI{100}{TeV})^{-2.19}$
is used as well as the best fit atmospheric flux from the same reference.

The EHE data set consists of 20 events that have been collected between 2008 and 2017 \citep{Aartsen:2018vtx,Aartsen:2016lmt}.
The selection is targeting high-energy track-like events
of good angular resolution $\le \SI{1}{\degree}$.
The selection has been optimized to be sensitive to events in the energy range of \SIrange{0.5}{10.0}{PeV}.
The integrated astrophysical purity depends on the assumption on the spectrum of astrophysical events at the highest energies, but can be estimated as approximately \SI{60}{\percent} purity (see Table~2 in~\cite{Aartsen:2016lmt}).

\subsection{The ANTARES Neutrino Telescope}
\label{sec:ant}

\begin{figure*}[!ht]
\epsscale{0.8}
\centering
\plotone{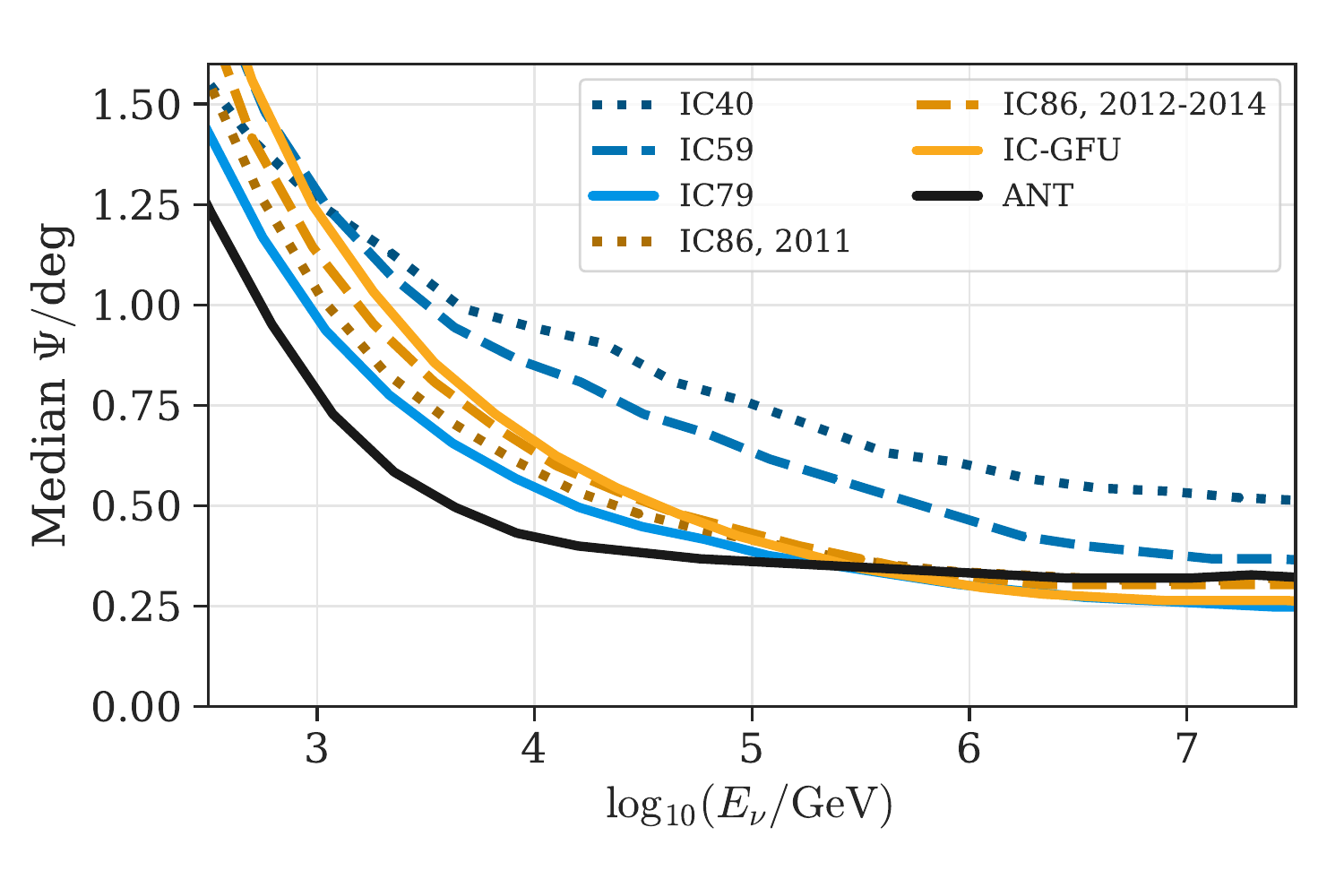}
\caption{Median of the angular resolution, $\Psi$, which is the angle between the true neutrino direction and reconstructed muon direction for IceCube and ANTARES point-source data as a function of the true neutrino energy. 
This is calculated based on simulation data sets.
The data from partial detector configurations of IceCube are denoted by the number of operating strings, (IC40, IC59, IC79; \citealp{Aartsen:2014cva}) and are shown in blue.
The data of the full detector configuration with 86 strings including the GFU data set \citep{Aartsen:2016lmt, Aartsen:2016oji, IceCube:2018cha} are identified by the years of data taking and are shown in orange.
The angular resolution of ANTARES data for the 11-year data set \citep{Albert:2018kjg,Aublin:2019zzn} is shown in black.}
\label{fig:median_psf}
\end{figure*}

\begin{table*}[ht]
\centering
\begin{tabular}{lccc}
\hline
Detector &  Analysis 1 &  Analysis 2 &  Analysis 3 \\
\hline
ANTARES & PS & HENU & HENU\\
IceCube & PS & HESE + HENU + EHE & HESE + HENU + EHE\\
\hline
\multicolumn{4}{c}{ } \\
\hline
Data set & \multicolumn{2}{c}{Description} & Topology \\
\hline
PS & \multicolumn{2}{c}{Optimized for point-source searches, $\nu_\mu$ candidates.} & Tracks\\
HESE & \multicolumn{2}{c}{High-energy starting events, all flavors.} & Tracks and cascades \\
HENU & \multicolumn{2}{c}{High-energy selection of $\nu_\mu$ candidates.} & Tracks\\
EHE & \multicolumn{2}{c}{Extremely-high energy $\nu_\mu$ candidates.} & Tracks\\
\hline
\end{tabular}
\caption{Overview over different neutrino data sets used in the different analyses.}
\label{tab:nu_overview}
\end{table*}

The ANTARES telescope \citep{Collaboration:2011nsa} is a deep-sea Cherenkov neutrino detector located 40 km offshore from Toulon, France, in the Mediterranean Sea.
The detector is composed of 12 vertical strings anchored at the sea floor at a depth of \SI{2475}{m}.
The strings are spaced at distances of about \SI{70}{m} from each other, instrumenting a total volume of $\sim$\SI{0.01}{km^3}.
Each string is equipped with 25 storeys of 3 optical modules \citep{Amram:2001mi}, vertically spaced by \SI{14.5}{m}, for a total of 885 optical modules. Each optical module houses a 10-inch photomultiplier tube facing 45$^\circ$ downward to optimize the detection of light from upward going charged particles. The detector was completed in 2008.

The ANTARES and IceCube detection principles are very similar (see \autoref{sec:ic}).
Particles above the Cherenkov threshold induce coherent radiation emitted in a cone with a characteristic angle of 42$^\circ$ in water. 
The position, time, and collected charge of the signals induced in the photomultiplier tubes by detected photons are used to reconstruct the direction and energy of events induced by neutrino interactions and atmospheric muons. Trigger conditions based on combinations of local coincidences are applied to identify signals due to physics events over the environmental light background due to $^{40}$K decays and bioluminescence \citep{Aguilar:2006pd}. 
ANTARES is thus also capable of detecting neutrino charged- and neutral-current interactions of all flavors.

For analysis 1, we use all track-like events from the 11-year data set
used for point-source searches (PS) recorded between 2007 and 2017 \citep{Albert:2018kjg,Aublin:2019zzn}.
The high-energy events (HENU) for analyses 2 and 3 are selected from an earlier data set of track-like and cascade-like events collected between 2007 and 2015 \citep{Albert:2017ohr}. 
In order to ensure a high probability of astrophysical origin for analyses 2 and 3, we require an astrophysical purity $\ge \SI{40}{\%} $ based on the same definition 
as used for the HENU data set of IceCube.
This selection results in a total of three track-like events and no cascade-like events, of which the arrival directions are shown in \autoref{fig:HE-nu_UHECR_skymap}.
An overview of the nomenclature is presented in \autoref{tab:nu_overview}.
The track-like events are combined with the respective IceCube data sets.
All events have a good angular resolution 
that is below \SI{0.4}{\degree} above \SI{10}{TeV} \citep{Albert:2017ohr}.
Overall, the angular resolution of ANTARES tracks is better than for IceCube tracks for energies below \SI{100}{TeV}, and comparable around \SI{100}{TeV} and above.
The median angular resolution, $\Psi$, of the 11-year data set compared to the IceCube PS data set is shown in \autoref{fig:median_psf}.

Despite the smaller detection volume, the inclusion of ANTARES data significantly improves the all-sky coverage of the neutrino data set used in analysis 1.
This is shown in \autoref{sec:anaAC-method} and \autoref{fig:neutrino_acceptance} through the relative contribution to the expected number of signal events for the individual PS data sets of ANTARES and IceCube.
Particularly in the Southern hemisphere, where the background from atmospheric muons
results in a higher energy threshold in IceCube,
the ANTARES data contributes substantially to the signal acceptance for soft source spectra.

\subsection{The Pierre Auger Observatory}
\label{sec:pao}

The Pierre Auger Observatory is located in Argentina (\SI{32}{\degree S}, \SI{69}{\degree W}) at a mean altitude of \SI{1.4}{\kilo\meter} above sea level \citep{ThePierreAuger:2015rma}.
The observatory has a hybrid design combining an array of particle detectors at ground (surface detector, SD) and an atmospheric fluorescence detector (FD)
for detecting the air showers caused by UHECRs interacting with the atmosphere.
The SD array is composed of \num{1660} water-Cherenkov detectors spread over an area of \SI{3000}{km^2}. 
The SD area is overlooked by the FD, which consists of 27 wide-angle optical telescopes located at four sites.
The reconstruction of SD events is described in detail in \citet{ThePierreAuger:2020SDrec}. 
The energy estimate is based on the signal at \SI{1000}{m} from the reconstructed intersection of the shower axis with the ground, which is extracted with a fit of the lateral distribution of signals in the individual detectors. 
This value is then corrected to take into account the different absorption suffered by showers coming at different angles. 
Due to the hybrid design of the observatory, this energy estimator is calibrated via the correlation with the near-calorimetric energy measured by the FD with the events observed by both SD and FD. Through this calibration, the energy estimate for SD is done
without relying on Monte Carlo. 
At the energies considered in this work, the systematic uncertainty on the energy scale is \SI{14}{\percent} \citep{Dawson:2020bkp}, 
the statistical uncertainty on the energy due to the number of triggered SD stations and uncertainties of their response is smaller than \SI{12}{\percent} \citep{Abreu:2011pj, ThePierreAuger:2020SDrec}, and 
the angular uncertainty is less than \SI{0.9}{\degree}
\citep{Bonifazi:2009ma}.
 
The UHECR data set used here consists of 324 events observed with the SD between January 2004 and April 2017 \citep{Aab:2018chp} with reconstructed energies $\ge\SI{52}{EeV}$ and zenith angles $\le \SI{80}{\degree}$. 
The data statistics is enlarged with respect to the previously used data set \citep{PierreAuger:2014yba} by 93 events and includes an updated energy calibration based on a larger number of hybrid events and an improved absorption correction.
The angular acceptance translates into a field of view ranging from \SIrange{-90}{+45}{\degree} in declination.
All energies are scaled up by a constant factor of \SI{+14}{\percent} to match both Auger and TA energies on a common energy scale (see \autoref{sec:ta}), following the recommendation of the Auger-TA joint working group \citep{biteauCoveringCelestialSphere2019}.
This scaling factor was determined by cross-calibrating the measured UHECR fluxes in the declination band around the celestial equator covered by both observatories.
It is chosen such that the fluxes in the common declination band match at \SI{40}{\EeV} of the Auger data set and at \SI{53}{\EeV} of the TA data set, respectively.
The celestial distribution of the UHECR arrival directions is shown in \autoref{fig:HE-nu_UHECR_skymap}. The directional exposure as function of declination can be found in \autoref{fig:uhecr-exposure}, which amounts to an integrated geometric exposure of \SI{91300}{km^2\,yr\,sr}.

\subsection{The Telescope Array}
\label{sec:ta}

\begin{figure}[!ht]
\epsscale{1.15}
\plotone{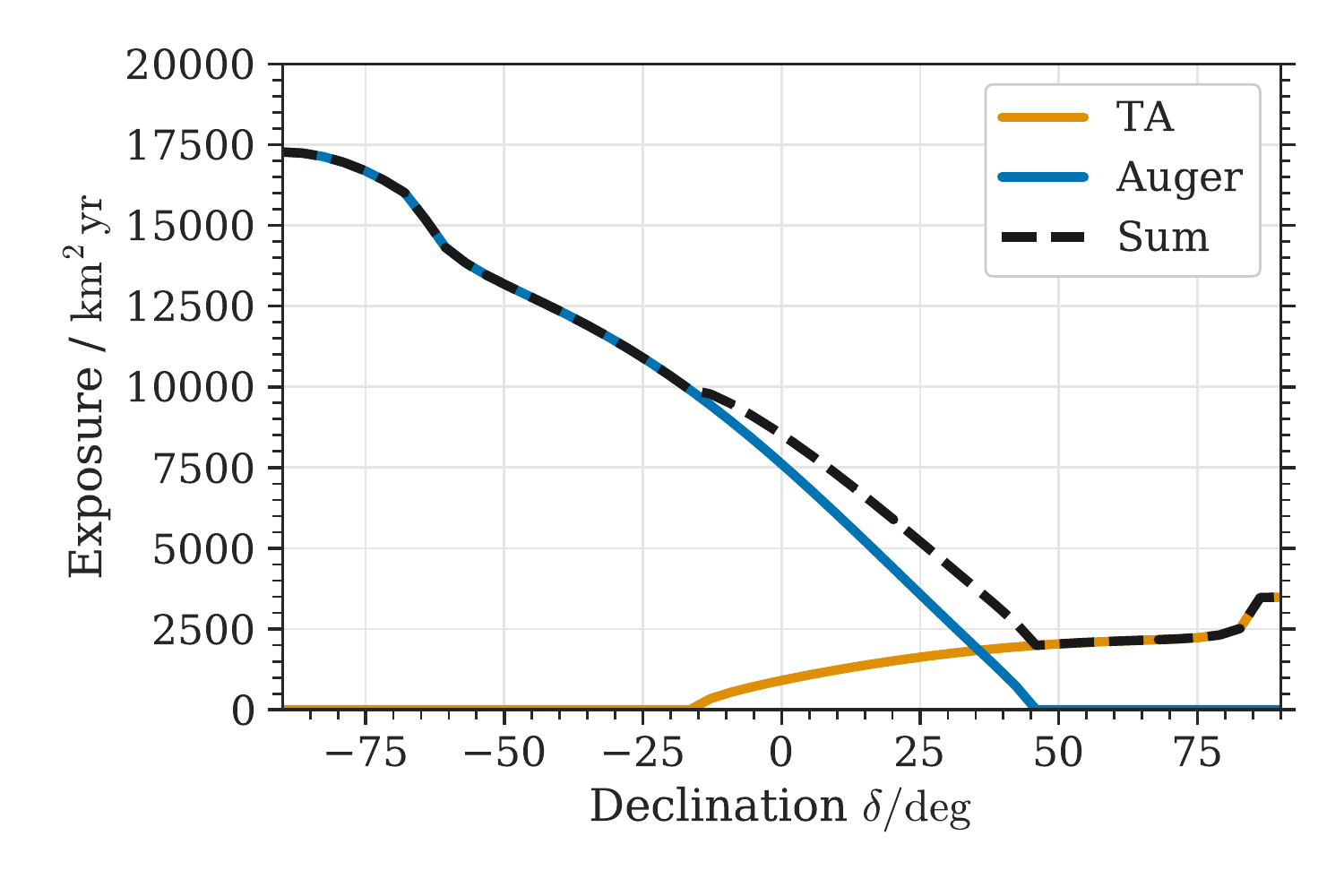}
\caption{Directional exposures of TA, Auger, and their sum.
The underlying geometric exposure functions are derived from~\cite{SOMMERS2001271}.
See also Fig.~1 of \cite{biteauCoveringCelestialSphere2019}.}
\label{fig:uhecr-exposure}
\end{figure}

The Telescope Array (TA) is located in Millard County, Utah, USA (\SI{39.3}{\degree N}, \SI{112.9}{\degree W}) at an altitude of about \SI{1,4}{\kilo\meter} \citep{Kawai:2008zza}. 
It consists of a surface detector (SD) array, composed of \num{507} plastic scintillation detectors of \SI{3}{m^2} each.
The SD stations are located on a square grid with \SI{1.2}{\kilo\meter} separation, which extends over an area of \SI{700}{km^2} \citep{AbuZayyad:2012kk}. 
The TA SD array observes UHECRs with a duty cycle near \SI{100}{\percent}.
With its wide field of view, the SD array covers a range from \SIrange{-15}{+90}{\degree} in declination.
In addition to the SD, there are three fluorescence telescope stations, instrumented with \numrange{12}{14} telescopes each \citep{Tokuno:2012mi}.
The telescope stations observe the sky above the SD array and measure the longitudinal development of the air showers as they traverse the atmosphere.

The data set for this analysis follows the selection in~\citet{Abbasi:2014lda}, which has been updated to 9 years of data in~\citet{abbasi:2018ApJ867L27A}.
The data set is identical to the data set used for anisotropy analyses presented in~\citet{Troitsky:20171D}.
It consists of 143 events observed with the SD between May 2008 and May 2017 with reconstructed energies $\ge\SI{57}{EeV}$ and zenith angles $\le \SI{55}{\degree}$. 
At these energies, the angular uncertainty is about \SI{1.5}{\degree}. 
The statistical uncertainty on the reconstructed energy is about \SIrange{15}{20}{\percent},
with an additional systematic uncertainty on the energy scale of \SI{21}{\percent} \citep{abbasi:2018ApJ867L27A}.
All energies are scaled down by \SI{-14}{\percent} to match both Auger and TA energies on a common energy scale (see \autoref{sec:pao}), following the recommendation of the Auger--TA joint working group \citep{biteauCoveringCelestialSphere2019}.

The celestial distribution of selected events is shown in \autoref{fig:HE-nu_UHECR_skymap}.
The right-hand side of the figure is a histogram of the sine of declination of all UHECR events, showing that TA data substantially contributes to the full-sky exposure of the combined UHECR data set.
\autoref{fig:uhecr-exposure} shows the directional exposure as a function of declination, with an integrated total exposure of \SI{11600}{km^2\,yr\,sr} \citep{biteauCoveringCelestialSphere2019}.

\section{Magnetic deflections}
\label{sec:mag}

\begin{figure*}[!t]
\epsscale{1.1}
\plottwo{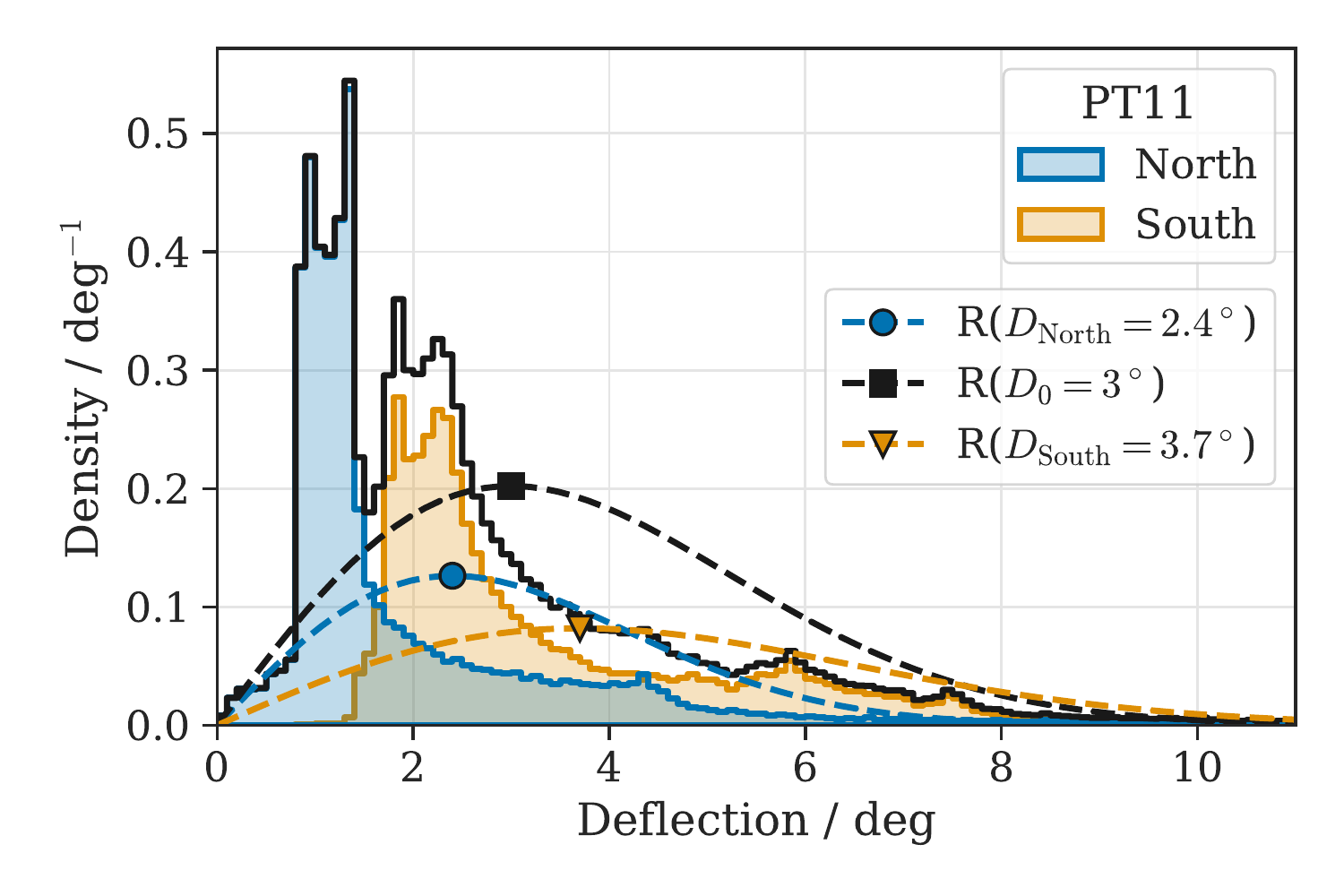}{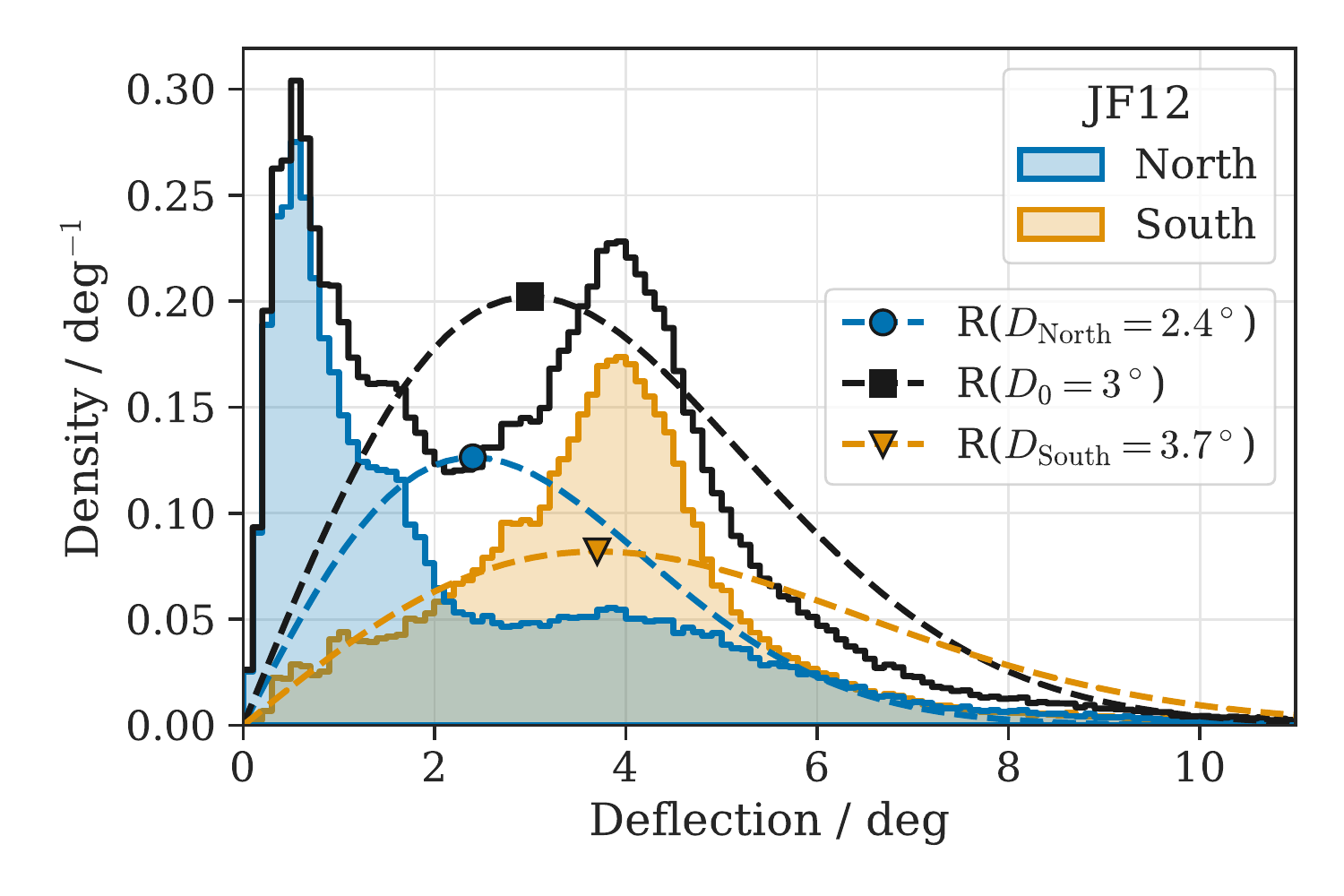}
\caption{Deflection simulation for protons of $100$\,EeV and two different galactic magnetic models: PT2011 \citep{PTgmf2011} and
JF2012 \citep{JFgmf2012}. 
Shaded areas in blue and orange show the histogram split into Galactic Northern and Southern hemispheres, with their sum shown as black line.
Additionally, Rayleigh PDFs are shown in the same colors;
a Rayleigh function with mode $\sigma$ is the 1D projection of a symmetric 2D-Gaussian function with width $\sigma$, with the projection being $(x,y)\rightarrow\sqrt{x^2+y^2}$.
The position of the markers indicate the mode of the Rayleigh PDF and thus the value of the default deflection, $D$.
Note that the integrals of the split PDFs are normalized to 0.5, while the integrals of the full PDFs are normalized to 1.
}
\label{fig:mag_deflection}
\end{figure*}

Despite their extremely high rigidity, UHECRs are deflected in galactic and in extragalactic magnetic fields (GMF/EGMF) by a non-negligible amount.
Neither the strength and correlation length of the extragalactic magnetic fields \citep{Durrer:2013pga,Kronberg:1993vk} nor the distance of the UHECR sources are known well.
Measurements of the Faraday rotation of extragalactic sources indicate that the extragalactic magnetic fields are weaker than \SI{1}{nG} 
\citep{Pshirkov:2015tua}.
Assuming a correlation length of $\sim$\SI{1}{\Mpc}, this results in a deflection less than \SI{2}{\degree} for protons of \SI{100}{\EeV}, even at source distances of \SI{50}{Mpc}.
In line with the previously reported results \citep{Aartsen:2015dml}, the deflection outside of our galaxy is assumed to be generally weaker than within our galaxy, and it is not modeled explicitly but benchmarked within the uncertainty of the deflection by GMF and the uncertainty of the rigidity due to the unknown composition. 

The measurement and modeling of the GMF is a complex task and subject of on-going discussion\footnote{See \href{https://icrc2021-venue.desy.de}{https://icrc2021-venue.desy.de} 
for a recent review by Tess Jaffe on the GMF at ICRC2021: \\ "Constraining Magnetic Fields at Galactic Scales".}.
%https://icrc2021-venue.desy.de/channel/video/Review-Constraining-Magnetic-Fields-at-Galactic-Scales/a68f4bf56265fc21c3d4abbfabc89315/62
Among different proposed models,
we use the JF2012 \citep{JFgmf2012} model and the PT2011 \citep{PTgmf2011} model to estimate the deflection of the UHECRs in the GMF.
Both models consist of a disc and a halo component,
while the JF2012 model has an additional x-shaped field component perpendicular to the disc.
A propagation simulation has been conducted with a Monte-Carlo approach to estimate the deflection of protons with an energy of \SI{100}{\EeV} that are distributed isotropically outside of the GMF.
The resulting deflections are shown in \autoref{fig:mag_deflection}, split into the 
Galactic Northern and Southern hemisphere based on the arrival direction of the proton.
The estimated deflection shows a complex structure
which differs considerably for the two models.
However, the mean deflection is about \SI{3}{\degree} in both cases.
Due to the heavy tails of the distributions, the mean is consistently larger than the median, thus we choose the mean as a conservative deflection estimate.
The split into the two hemispheres shows a considerably smaller deflection in the Galactic North than in the Galactic South due to North-South asymmetries present in both models of the GMF.
For this work we have chosen a robust benchmark modeling of the effect of the GMF, thus avoiding biases of detailed model uncertainties.
The deflection process is assumed to be random, resulting in a symmetric 2D Gaussian distribution of UHECR arrival directions around a given source direction.
In reverse, the source position is assumed to be located within the 2D Gaussian distribution around the arrival direction of a UHECR event.
The standard deviation of the Gaussian deflection, $\sigma_{\rm MD}$,
depends on a scaling factor, $C$, and inversely on the energy of the UHECR event, $E_{\rm CR}$
\begin{equation}
\label{eq:deflection}
\sigma_{\rm MD}
= \frac{D_0 \cdot C}{E_{\rm CR}/\SI{100}{EeV}}.
\end{equation}
The default deflection, $D_0$, for $C=1$ and $E_{\rm CR}=\SI{100}{EeV}$ is derived from the mean of the deflection values obtained with the simulation, which are shown in \autoref{fig:mag_deflection}. 
Note that the deflection is usually a 2D quantity with $(x,y)$ coordinates, which in \autoref{fig:mag_deflection} is projected to the absolute value, i.e.$\sqrt{x^2+y^2}$.
Furthermore, larger and thus more conservative values of the deflection will be tested to include an uncertainty of the GMF model.
The scaling factor $C$ thus accounts for uncertainties of both GMF model and UHECR charge, which is not known on an event-by-event basis at highest energies.

Analysis 1 uses a default deflection of $D_0=\SI{3}{\degree}$ over the whole sky,
while analysis 2 uses $D_{\rm North/South}=(\SI{2.4}{\degree}, \SI{3.7}{\degree})$ for UHECRs with arrival directions from the Galactic Northern and Southern hemisphere, respectively.
The different choices have been made based on the respectively better sensitivity for the two analyses.
Analysis 1 and 2 implement the deflection into their methods as described in the respective sections~\ref{sec:anaAC-method} and~\ref{sec:anaGE-method}.
Analysis 3 employs a model-independent approach with respect to the UHECR deflection.

\section{Analysis methods}

\subsection{Unbinned neutrino point-source search with UHECR information}
\label{sec:anaAC-method}

The goal of analysis 1 is to find point-like neutrino sources that are spatially correlated with UHECR arrival directions within a region derived from their magnetic deflection estimate.
The search for neutrino sources utilizes
the unbinned maximum-likelihood analysis commonly used in IceCube~\citep{Aartsen:2016oji, aartsen8yrDiffusePS2019,
aartsen10yrIntegratedPS2020}.
This enables an easy combination of the high-statistics, full-sky neutrino data sets, which are the PS data sets of IceCube and ANTARES described in \autoref{sec:ic} and \autoref{sec:ant}, respectively.
In addition to this standard method,
the magnetic deflection regions, defined by the UHECR arrival directions, energy, and scaling factor, 
are used for constraining the possible source regions.

\begin{figure*}
\epsscale{1.1}
\plotone{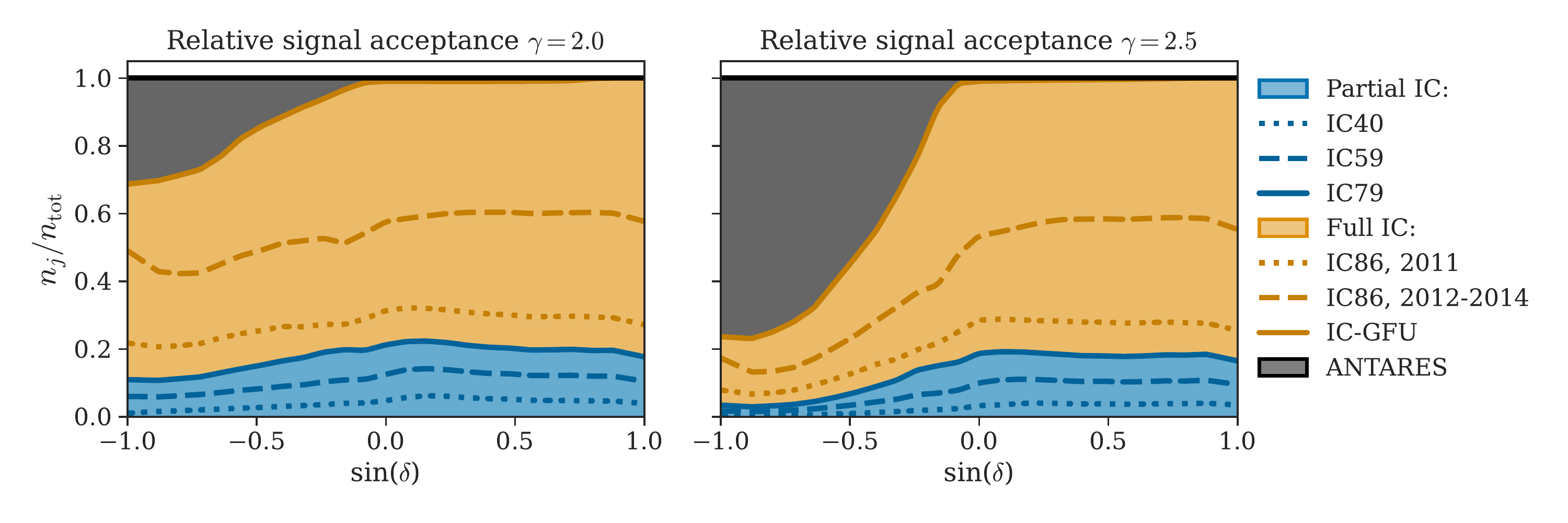}
\caption{The relative signal contribution, $n_j/n_{\rm tot}$, for analysis 1 of the different configurations of IceCube and the ANTARES point-source data set as a function of the declination $\delta $. 
The left plot shows a hard underlying
power-law signal spectrum (\autoref{eq:nu-flux}) with index $\gamma =2.0$, while the right plot shows a softer spectrum with $\gamma =2.5$.
The relative signal contributions from partial detector configurations of IceCube are shown in blue,
while the contributions of the full detector configuration of IceCube with 86 strings are shown in orange.
The ANTARES contribution is shown in gray.
The lines as listed in the rightmost legend indicate the contributions from each data set individually.}
\label{fig:neutrino_acceptance}
\end{figure*}

The unbinned neutrino likelihood in the source direction $\vec{\Omega} = (\alpha, \delta)$ in right ascension and declination consists of the sum of a signal PDF, $S$, and a background PDF, $B$
\begin{align}
\mathcal{L} \left(n_s, \gamma \right) 
& = \prod_{j=1}^{7} \prod_{i=1}^{N_j} \left[ f_j(\sin \delta, \gamma) \frac{n_s}{N_j} \cdot S(\vec{\Omega}_i, E_i, \sigma_i | \gamma, \vec{\Omega}) \right. \nonumber \\ 
& + \left. \left(1- f_j(\sin \delta, \gamma) \frac{n_s}{N_j} \right) \cdot B( \delta_i, E_i) \right] . \label{eq:anaC-nu-llh}
\end{align}
The likelihood product with index $i$ runs over all neutrino events within each data set with index $j$ (see \autoref{sec:ic} and~\autoref{sec:ant}).
The likelihood product with index $j$ runs over all data sets to yield the final likelihood function.
The likelihood combination of the seven data sets by ANTARES and IceCube is thus handled in the same way as described in~\cite{Aartsen:2016oji}.
The signal and background PDFs, $S$ and $B$, are evaluated for four observables per neutrino event, weighted with the number of signal events over total number of events per data set, $n_s/N_j$.
The observables are the reconstructed right ascension and declination, summarized as $\vec{\Omega}_i=(\alpha_i,\,\delta_i)$, reconstructed energy, $E_i$, and angular error estimator, $\sigma_i$.
The signal PDF consists of two terms:
the first term is a declination-dependent reconstructed energy distribution, where the underlying neutrino flux is modeled as a power law
\begin{equation}
\derive{\Phi^\nu}{E} = \frac{\drv N^\nu}{\drv E \, \drv A \, \drv t \, \drv \Omega}
= \Phi^\nu_0 \cdot \left(\frac{E}{\SI{1}{\GeV}} \right)^{-\gamma}
\label{eq:nu-flux}.
\end{equation}
Here, $\Phi^\nu_0$ is the flux normalization, \SI{1}{\GeV} is the corresponding pivot energy, and $\gamma$ is the spectral index of the power law.
The second term of the signal PDF is a spatial term modeled as a Gaussian with the width, $\sigma_i$, given by the angular error estimator of each neutrino candidate on an event-by-event basis.
The background PDF
is determined as function of reconstructed energy, $E_i$, and declination, $\delta_i$, from randomized experimental data.
A full description of the signal and background PDFs, $S$ and $B$, can be found in \citet{Aartsen:2016oji}.
The proportionality factor, $f_j(\sin \delta, \gamma)$, is the relative signal acceptance per neutrino data set calculated from the expected number of signal events via
\begin{align}
\label{eq:prop_factor}
f_j(\sin \delta, \gamma) &= \frac{n_j}{n_{\rm tot}} \nonumber \\
&=\frac{T^j \int (\drv \Phi^\nu/ \drv E) A_{\rm eff}^j(E, \sin \delta) ~\drv E}{\sum_{j=1}^{7} T^j \int (\drv \Phi^\nu/ \drv E) A_{\rm eff}^j(E, \sin \delta) ~\drv E}.
\end{align}
This factor is used to correctly weight the signal contribution of each data set $j$ for a given source declination and spectral index.
Per data set $j$, the livetime is denoted with $T^j$ and the effective area is denoted with $A_{\rm eff}^j$.
\autoref{fig:neutrino_acceptance} shows the proportionality factor of each data set as a function of declination for two spectral indices, $\gamma=2.0$ and $2.5$.
Note that all data sets are evaluated with the same formulation of the likelihood function as used by IceCube, instead of combining different formulations as described in~\cite{illuminatiANTARESIceCube2020}.

The best-fit signal parameters, $\hat{n}_s$ and $\hat{\gamma}$, at a given source position, $\vec{\Omega}$, are obtained with the maximum-likelihood method.
The number of events and the proportionality factor are related to the neutrino flux using the respective livetime and effective area of each data set via
\begin{equation}
\label{eq:n-phi}
n_s = \sum_{j=1}^{7} T^j \int \derive{\Phi^\nu}{E} A_{\rm eff}^j(E, \sin \delta) ~\drv E.
\end{equation}

The corresponding significance of a source at position $\vec{\Omega}$ is evaluated using a likelihood ratio test,
which yields the test statistic (TS)
\begin{equation}
\mathrm{TS}_\nu(\vec{\Omega}) = 
2 \left.\log \left(\frac{ \mathcal{L}\left(\hat{n}_s, \hat{\gamma}\right)}{\mathcal{L}(n_s=0)} \right) \right|_{\vec{\Omega}}.
\label{eq:nu-ts}
\end{equation}
Here, the null hypothesis is defined via $n_s=0$, representing the case of no neutrino sources in the vicinity of UHECR events.
Instead of searching for a single neutrino source, the whole sky is searched
on a HEALpix~\citep{gorskiHEALPix2005} grid with a resolution of approximately \SI{0.2}{\degree}.
This results in a skymap of the TS values at each grid center, $\vec{\Omega}_{\rm grid}$.

The second step is to combine the neutrino likelihood with the information provided by UHECR events and their deflection estimate.
The deflection estimate for one UHECR with index $k$ and arrival direction $\vec{\Omega}_k^{\rm CR}$, which possibly originated in the direction of $\vec{\Omega}_{\rm grid}$, is defined as a 2D Gaussian.
Its width is the quadratic sum of the magnetic deflection estimate, $\sigma_{\rm MD}$ of \autoref{eq:deflection}, and the UHECR angular reconstruction error, $\sigma_{\CR}^2$,
which is \SI{0.9}{\degree} for Auger events and \SI{1.5}{\degree} for TA events
\begin{equation}
\label{eq:spatial_prior}
\mathcal{P}_k \propto \exp \left(-\frac{(\vec{\Omega}_{\rm grid} - \vec{\Omega}_k^{\rm CR} )^2}{2(\sigma_{\rm MD}^2 + \sigma_{\CR}^2)} \right).
\end{equation}
The term $\mathcal{P}_k$ acts as a spatial constraint, which is multiplied to the neutrino likelihood function defined in \autoref{eq:anaC-nu-llh} via $\mathcal{L} \rightarrow \mathcal{L} \cdot \mathcal{P}_k$.
Effectively, the constraint is added as a logarithmic term to the neutrino TS defined for each grid point in \autoref{eq:nu-ts} via $\mathrm{TS} \rightarrow \mathrm{TS} + 2\cdot \log (\mathcal{P}_k )$.
Finally, the maximum of the combined UHECR-neutrino TS is found at the best-fit grid point, $\hat{\vec{\Omega}}_{\rm grid}$, via
\begin{align}
\label{eq:single_uhecr_ts}
\mathrm{TS}^{\rm comb}_k(\hat{\vec{\Omega}}_{\rm grid})
&= \max \left( \mathrm{TS}_\nu(\vec{\Omega}_{\rm grid} )
- \frac{( \vec{\Omega}_{\rm grid} - \vec{\Omega}_k^{\rm CR} )^2}{\sigma_{\rm MD}^2 + \sigma_{\CR}^2} \right)
\end{align}
The maximum marks the neutrino-source candidate which is the most-likely counterpart to the respective UHECR event used to calculate the spatial constraint.
The normalization in \autoref{eq:spatial_prior} adds only a constant term to the TS, which can be omitted when calculating significances and $p$-values based on pseudo-experiments (see below).
As a third, final step, this procedure is repeated for all UHECRs and all obtained TS values are added up yielding the final test statistic.
This search strategy was first developed in the context of this \anaAC and first outlined in sec.~4 of~\cite{Schumacher:2019qdx}. 
It has already been applied also to
other neutrino correlation searches with spatially extended source localization, namely the correlation with ANITA events \citep{Anita:2020} and with gravitational wave events \citep{GWIcecubeLIGO2020}.
Note that the analysis described here is improved with respect to the previous search \citep[sec.~5]{Aartsen:2015dml}, as it models the displacement of a point-like neutrino source with respect to the UHECR arrival direction based on the assumed magnetic deflection.
Here, the position and flux of a point-like source are fit in the vicinity of the UHECR direction,
while in the previous search, a spatially-extended neutrino emission around the UHECR direction was fit.

Six different signal hypotheses are tested,
which are characterized by a lower cut on the UHECR energy, $E_{\rm CR} > E_{\rm cut}$ with $E_{\rm cut} \in \{70,\, 85,\, 100\}\,\mathrm{EeV}$, and the scaling factor of the deflection estimate, $C \in \{1,2\}$.
The lower energy cut is a threshold for selecting only the highest-energy UHECRs with the lowest deflection for this analysis.
No analogous energy cut is applied to the neutrino data.
The scaling factor $C$ is a model parameter for scaling the baseline magnetic deflection, and it is not derived from the UHECR data.
The baseline magnetic deflection is $D_0=\SI{3}{\degree}$ for all UHECRs, which is then converted into the spatial constraint of an individual UHECR event using \autoref{eq:deflection} and~\autoref{eq:spatial_prior}.
The corresponding sensitivity and $3\sigma$-discovery potential are evaluated based on the normalization of the neutrino flux per source, $\Phi^\nu_0$ (see \autoref{eq:nu-flux}).
The $3\sigma$ threshold is chosen, since the background expectation needs to be calculated based on pseudo-experiments, as described in the next paragraph. 
A $5\sigma$-discovery potential is computationally too expensive to be calculated for the various hypotheses.
Sensitivity is defined as the expected median upper limit at 90\% confidence level on $\Phi^\nu_0$ in case of a null measurement,
while the discovery potential is defined as the median of $\Phi^\nu_0$ required for a rejection of the background hypothesis with $3\sigma$ significance.

Both sensitivity and discovery potential are calculated based on data challenges, for which signal-like and background-like pseudo-experiments (PEs) are generated.
Since the calculation is based on PEs, the constant term in \autoref{eq:spatial_prior} and~\autoref{eq:single_uhecr_ts}, i.e.~the normalization of the constraint term, can be omitted.
For all signal and background PEs, the UHECR arrival directions are kept fixed.
The background PEs are obtained by randomizing the experimental neutrino data in right ascension.
The signal PEs reflect the six different signal hypotheses;
one signal PE is based on experimental neutrino data randomized in right ascension, to which Monte-Carlo neutrino events, representing a neutrino source, are added.
The location of this neutrino source is chosen randomly within the spatial deflection constraint of one UHECR event.
This way, we mimic a neutrino source that is displaced with respect to the UHECR direction due to the UHECR deflection.
In the baseline model of the signal PEs, all UHECR events have such an artificial neutrino source in their vicinity.
For one PE, all neutrino sources have an $E^{-2}$ power-law spectrum and the same flux normalization.
Note that the UHECR energy cut and scaling factor used to generate the spatial constraints for the likelihood function are the same as for determining the neutrino source location.

\begin{figure*}
\epsscale{1.1}
\plotone{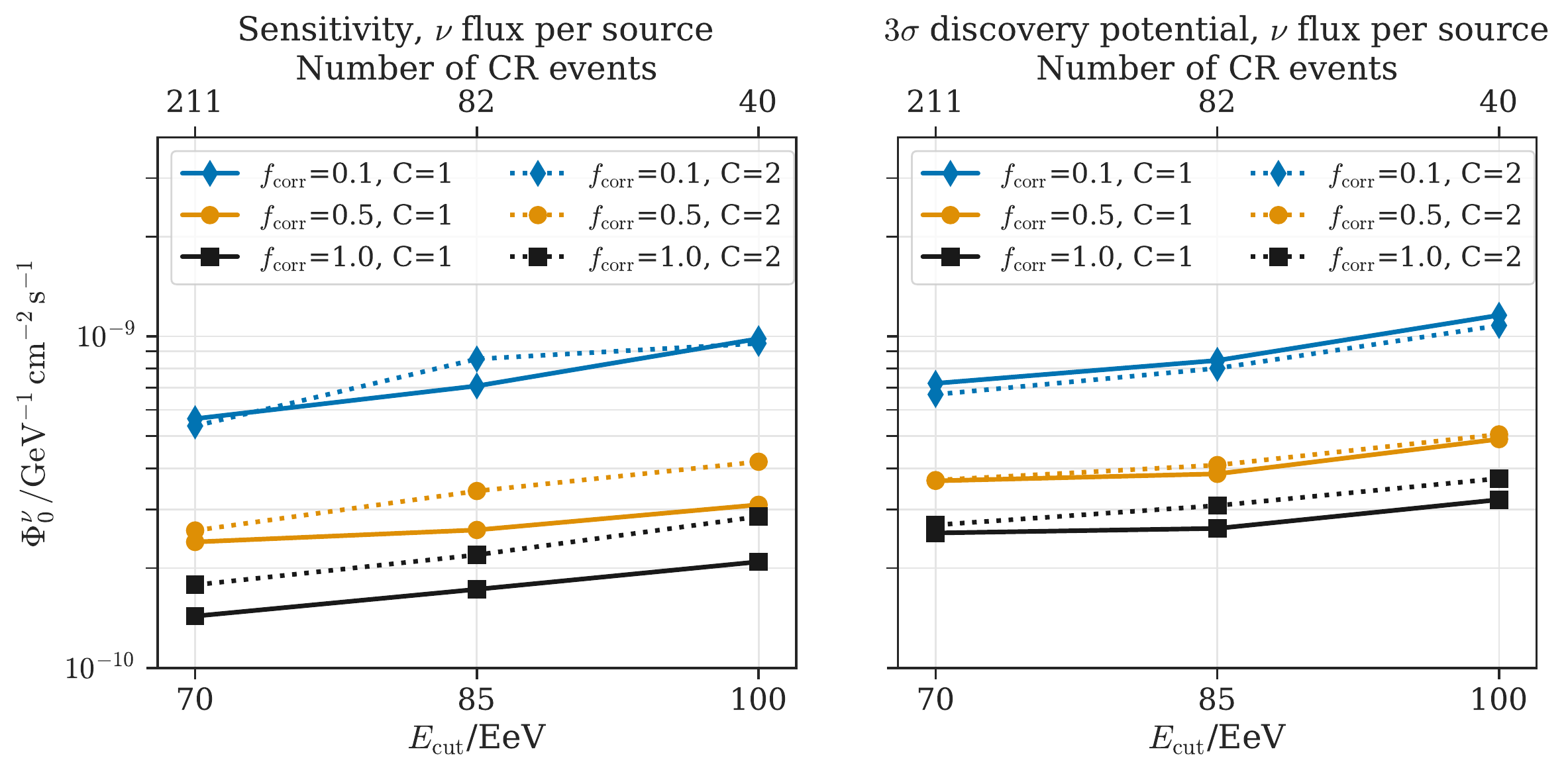}
\caption{Sensitivity (left) and $3\sigma$-discovery potential (right) of the neutrino flux per source as function of the UHECR energy cut in EeV, for the three correlation fractions and for the two scaling factors we used.
}
\label{fig:anaC_sens}
\end{figure*}

We generate additional signal PEs by varying the fraction of UHECR events, $\fcorr$, 
with a neutrino source in their vicinity.
This mimics signal models where not all UHECRs have a corresponding neutrino source in the vicinity of their arrival direction.
Thus, the number of neutrino sources is determined by the rounded-up product of the correlation fraction with the numbers of UHECRs, $N_\nu^\mathrm{src} =\left \lceil{\fcorr N_{\rm CR}}\right \rceil$,
where $N_{\rm CR}$ is the number of UHECRs passing the energy cut.
We choose the correlation fraction from three discrete values, $\fcorr \in \{1,\,0.5,\,0.1\}$, 
where $\fcorr=1$ represents the baseline model.
Note that the correlation fraction is only used to choose the number of neutrino sources in the signal PEs,
while in a real experimental data, it is not known a priori which UHECRs have a correlated neutrino source and which do not.
For $\fcorr<1$, only a number of $N_\nu^\mathrm{src} < N_{\rm CR}$ of randomly selected UHECRs will have a correlated neutrino source in the signal PEs.
Independent of the correlation fraction, the TS values of all UHECR constraints are added up, since in real experimental data it would not be known which UHECRs have a correlated neutrino source.
The sensitivity and $3\sigma$-discovery potential in terms of the flux normalization per neutrino source for all signal models and hypotheses are reported in \autoref{tab:anaC-sens-lim}, which are also shown in \autoref{fig:anaC_sens}.
We see that a smaller correlation fraction increases the flux normalization per neutrino source for both sensitivity and $3\sigma$-discovery potential.
This is expected since a smaller correlation fraction corresponds to fewer neutrino sources,
which in turn need to have a higher flux normalization to be detected by the analysis.

It is noticeable that for a small correlation fraction of $\fcorr=0.1$, the signal model with a larger scaling factor of $C=2$ yields a better $3\sigma$-discovery potential than the smaller scaling factor of $C=1$.
This is unexpected, as a smaller UHECR deflection and thus a smaller spatial constraint should yield more accuracy in detecting the corresponding neutrino sources.
However, the amount of sky covered by the constraints does not grow uniformly with the scaling factor and numbers of UHECRs, since the UHECR spatial constraints can overlap.
In the case of a small correlation fraction, a high neutrino flux per source is required to reach the $3\sigma$-discovery potential due to the small number of sources present\footnote{There are 22, 9 and 4 sources for $\fcorr=0.1$ and the energy cuts of $E_{\rm cut} \in \{70,\, 85,\, 100\}\,\mathrm{EeV}$, respectively.}.
A closer study showed that in the case of few, but strong neutrino sources, it is beneficial to search a larger area of the sky for sources so that fewer of them are missed.
This effect is thus caused by an interplay between deflection size, number of UHECRs and relatively high neutrino flux per source in this particular case.

\subsection{Unbinned likelihood-stacking analysis of UHECRs and high-energy neutrinos}
\label{sec:anaGE-method}
The second analysis is based on using the neutrinos with a high probability of astrophysical origin as markers for the location of the sources of UHECRs and neutrinos.
The UHECR events are stacked using an unbinned likelihood analysis with the arrival directions of the high-energy neutrinos as the source positions.
The signal hypothesis is defined by the number of UHECR events, which are clustered around the neutrino arrival directions, as well as the size of the magnetic deflection.
The background hypothesis is defined by an isotropic flux of UHECRs.
This approach is thus complementary to the \anaAC, where the UHECR events are used as source markers and an isotropic flux of neutrinos defines the background hypothesis.
The logarithm of the likelihood function is defined as
\begin{align} \label{eq:anaB-llh}
& \log \,\mathcal{L}(n_s) = \nonumber \\
& \sum_{i=1}^{N_{\text{Auger}}} \log {\bigg( }\frac{n_s}{N_{\text{CR}}}S^i_{\text{Auger}} + \frac{N_{\text{CR}}-n_s}{N_{\text{CR}}}B^i_{\text{Auger}}{\bigg) } \nonumber \\
& +  \sum_{i=1}^{N_{\text{TA}}} \log {\bigg( }\frac{n_s}{N_{\text{CR}}}S^i_{\text{TA}} + \frac{N_{\text{CR}}-n_s}{N_{\text{CR}}}B^i_{\text{TA}}{\bigg) },
\end{align}
where the free parameter is the total number of UHECR signal events, $n_s$.
The sums run over all UHECR events in the respective data set, i.e.~$N_{\text{Auger}}$ and $N_{\text{TA}}$,
where the total number of UHECR events is $N_{\rm CR} = N_{\text{Auger}} + N_{\text{TA}}$.
In contrast to the likelihood function of analysis 1, the signal PDFs,
$S^i_{\text{Auger}}$ and $S^i_{\text{TA}}$, and the background PDFs, $B^i_{\text{Auger}}$ and $B^i_{\text{TA}}$, are stacked for all high-energy neutrinos such that $n_s$ is a global parameter.
The background PDFs per UHECR detector, $B^i_{\rm det}$, are the normalized expectations for an isotropic UHECR flux as function of declination, which correspond to the normalized detector exposures
(see \autoref{fig:uhecr-exposure}).
The signal PDF per UHECR detector is defined as
\begin{align}
& S^i_{\rm det} (\vec{\Omega_i}, E_i) = R_{\rm det} (\delta_i) \cdot
\sum_{j=1}^{N_\nu} S_j(\vec{\Omega_i}, \sigma(E_i)) \nonumber \\
& {\rm with}\ \sigma^2 = \sigma_{\rm MD}^2 + \sigma_{\CR}^2,
\label{eq:anaGE_signal}
\end{align}
as a function of the UHECR arrival direction, $\vec{\Omega_i}$, and energy, $E_i$.
The term $R_{\rm det}$ is the relative exposure of each detector as a function of the declination per UHECR event, $\delta_i$.
\autoref{fig:uhecr-exposure} shows the absolute directional exposures, which are each normalized to 1 over the whole sky to obtain $R_{\rm det}$.
The sum runs over all neutrino events, $N_\nu$, where the terms of $S_j(\vec{\Omega_i}, \sigma(E_i))$ are the spatial likelihood maps obtained from the neutrino directional reconstruction smeared with the UHECR uncertainty with width $\sigma(E_i)$ (see \autoref{eq:spatial_prior}).
Thus, these terms are PDFs representing the total uncertainty on the common source position of UHECR event $i$ and neutrino event $j$, evaluated at the UHECR arrival direction, $\vec{\Omega_i}$.
Similar to analysis 1, three different deflections corresponding to scaling factors of $C\in \{1, 2, 3\}$ are
represented in the signal terms of the likelihood function.
Again, $C$ is a model parameter that scales the magnitude of the deflection and it is not derived from UHECR data.
Depending on whether the UHECR arrival direction is in the Northern or Southern Galactic Hemisphere (see \autoref{fig:mag_deflection}), the baseline magnetic deflection is thus
\begin{align}
    & D =  C \cdot \left(D_{\rm North},D_{\rm South} \right)
    = C \cdot \left(\SI{2.4}{\degree}, \SI{3.7}{\degree}\right)\nonumber \\
    & \text{with}\ C\in \{1, 2, 3\}.
\end{align}
The final value of the deflection, $\sigma_{\rm MD}$, is then calculated based on the UHECR energy using \autoref{eq:deflection}.

The best-fit of the number of signal events, $\hat{n}_s$, is found with a maximum likelihood estimation.
The resulting test statistic is defined as the likelihood ratio of the maximum likelihood over the background likelihood with $n_s=0$
\begin{equation}
    {\rm TS} = 2\log \frac{\mathcal{L}(\hat{n}_s)}{\mathcal{L}(n_s=0)}.
\end{equation}
The significance is then determined based on the assumption that the background expectation is distributed according to a $\chi^2$ function with one degree of freedom, which has been verified using background-like PEs.
The test statistic is calculated separately for all track-like and all cascade-like neutrino events, and separately for the three different deflections.
The analysis approach is essentially the same as published in \citet{Aartsen:2015dml},
except for the updated magnetic deflection values, which are split into the Galactic Northern and Southern Hemisphere as described in \autoref{sec:mag}.

The sensitivity and the $3\sigma$-discovery potential in terms of $n_s$ are obtained with data challenges based on PEs.
Here, the neutrino arrival directions are kept fixed per PE, while the UHECR arrival directions and energies are generated resembling the signal and background hypothesis.
For the background PEs, all UHECR arrival directions are derived from an isotropic flux and thus according to the exposure of the respective UHECR experiments.
The energies of the UHECR events are sampled from a power-law proportional to $E^{-4.2}$ for the Auger events and to $E^{-4.5}$ for the TA events, same as in \citet{Aartsen:2015dml}.
For the signal PEs, a number of $n_s$ UHECR arrival directions are distributed randomly based on their respective spatial signal terms in \autoref{eq:anaGE_signal}, $S_j(\vec{\Omega_i}, \sigma(E_i))$.
Note that all UHECRs have the same scaling factor, corresponding to the respective signal hypothesis as implemented in the likelihood function.
The sensitivity and $3\sigma$-discovery potential for the three different benchmark values of $C$ are presented in \autoref{tab:anaB-res}, separately for the track-like and cascade-like neutrino events.
We find that the sensitivity and $3\sigma$-discovery potential using neutrino tracks requires much fewer UHECRs than when using neutrino cascades, which is expected due to the large differences in angular reconstruction uncertainty of the two event topologies.

\subsection{Two-point angular correlation of UHECRs and high-energy neutrinos}
\label{sec:anaPA-method}
The \anaPA relies on counting pairs of UHECRs and the high-energy neutrinos, where the angular separation between their arrival directions is within a maximum angular separation, $\alpha$.
The observed number of pairs within this radius, $n_{\rm obs}(\alpha)$, is compared to the mean number of pairs expected from pure background, i.e.~uncorrelated arrival directions, $\langle n_{\rm bckg}(\alpha)\rangle$.
The relative excess of pairs is defined as
\begin{equation}
    \frac{n_{\rm obs}(\alpha)}{\langle n_{\rm bckg}(\alpha)\rangle} - 1.
\label{eq:pair-excess}
\end{equation}
As the separation angle is not known a-priori, angles between \SI{1}{\degree} and \SI{30}{\degree} in steps of \SI{1}{\degree} are tested.
The angle with the largest deviation from the background expectation is chosen after the scan.

The significance of the experimental result is calculated with respect to background PEs.
Similar to analysis 2, the background PEs are generated with a fixed set of neutrino arrival directions, and
uncorrelated UHECR arrival directions are generated according to the exposure of the respective UHECR experiments.
The energies are sampled from the same spectra as described in \autoref{sec:anaGE-method}.
As a cross-check, additional background PEs are generated with the fixed set of UHECR arrival directions, and neutrino arrival directions are randomized in right ascension.
The different types of background PEs approximate an isotropic flux of UHECR and high-energy neutrinos, respectively.

Note that this analysis does not include an assumption of the magnetic deflection of UHECRs, which makes it a robust and model-independent approach.
The analysis approach is the same as published in \citet{Aartsen:2015dml}.

\section{Results}
\subsection{Unbinned neutrino point-source search with UHECR information}
\label{sec:ana1-res}
The test statistic of experimental data is obtained with the neutrino \anaAC as described in \autoref{sec:anaAC-method},
assuming the six different combinations of signal parameters, i.e.~ combinations of scaling factor, $C$, and lower energy cut, $E_{\rm cut}$.
Each of the six experimental TS values is evaluated with respect to the corresponding distribution of TS obtained from background-like PEs.
The resulting $p$-values for all six tests are listed in \autoref{tab:anaC-sens-lim}.
The smallest $p$-value ($9.7\%$) is found for the scaling factor of $C=2$ and an
UHECR energy cut of $E_{\rm cut} = \SI{85}{EeV}$.
The $p$-value increases to $24\%$ when correcting for the trials due to the six correlated tests.
This correction is based on the combined distribution of all minimum $p$-values of the background PEs.
All $p$-values are fully compatible with
the background hypothesis of no resolved neutrino sources in spatial correlation with the UHECRs considering the assumed signal models.
Based on the experimental TS value, $90\%$ C.L.~upper limits on the normalization of the flux of neutrino sources correlated with UHECR arrival directions are reported in the last three rows of \autoref{tab:anaC-sens-lim}.
In the case of the single $p$-value larger than 50\% found for $C=1$ and $\ecuth)$,
the limits are set equal to the sensitivity in order to not over-estimate the limits based on an under-fluctuation in data.
In addition to the baseline correlation fraction of 100\%, the limits are calculated for two smaller correlation fractions of 50\% and 10\%.
In these cases, the limits are relaxed by about \SIrange{40}{60}{\percent} for $\fcorr=0.5$ compared to $\fcorr=1$, and by about a factor of 3 to almost 5 for $\fcorr=0.1$ compared to $\fcorr=1$.

\begin{table*}[htbp]
\begin{center}
\begin{tabular}{lcccccc}
\hline
\textbf{Analysis parameters} &  \multicolumn{3}{c}{} & \multicolumn{3}{c}{}  \\
$D_0 \cdot C$ & $3\dg$ & $3\dg$ & $3\dg$ &$6\dg$ &$6\dg$ &$6\dg$ \\
%$D_0 \cdot C$ & \multicolumn{3}{c}{$3\dg$} & \multicolumn{3}{c}{$6\dg$} \\
$E_{\rm cut}$ & \SI{70}{\EeV} & \SI{85}{\EeV}  & \SI{100}{\EeV} & \SI{70}{\EeV}  & \SI{85}{\EeV}  & \SI{100}{\EeV} \\
\hline
{\textbf{Number of neutrino sources}, $\mathbf{N_\nu^\mathrm{src}}$} &  \multicolumn{3}{c}{} & \multicolumn{3}{c}{}  \\
$\fcorr=0.1$ & 22  &  9  & 4   & 22  &  9  & 4  \\
$\fcorr=0.5$ & 106 &  41 & 20  & 106 &  41 & 20 \\
$\fcorr=1$   & 211 &  82 & 40  & 211 &  82 & 40 \\
\hline 
\textbf{Sensitivity} &  \multicolumn{3}{c}{} & \multicolumn{3}{c}{}  \\
$f_{\rm corr}$=0.1 &  5.6 & 7.1 & 9.8 &  5.4 & 8.5 & 9.5 \\
$f_{\rm corr}$=0.5 &  2.4 & 2.6 & 3.1 &  2.6 & 3.4 & 4.2 \\
$f_{\rm corr}$=1   &  1.4 & 1.7 & 2.1 &  1.8 & 2.2 & 2.9 \\
\hline
\textbf{$\mathbf{3\sigma}$ disc.~potential} &  \multicolumn{3}{c}{} & \multicolumn{3}{c}{}  \\
$f_{\rm corr}$=0.1 &   7.2 & 8.5 & 11.6 &   6.7 & 8.0 & 10.8 \\
$f_{\rm corr}$=0.5 &   3.7 & 3.8 &  4.9 &   3.7 & 4.1 &  5.1 \\
$f_{\rm corr}$=1   &   2.6 & 2.6 &  3.2 &   2.7 & 3.1 &  3.7 \\
\hline
\textbf{90\% C.L.~upper limit} &   \multicolumn{3}{c}{} & \multicolumn{3}{c}{}  \\
$f_{\rm corr}$=0.1 &   6.4 & 9.2 & 9.8 &   6.7 & 10.9 & 10.6 \\
$f_{\rm corr}$=0.5 &   2.8 & 3.6 & 3.1 &   3.3 &  4.7 &  4.5 \\
$f_{\rm corr}$=1   &   1.7 & 2.3 & 2.1 &   2.3 &  3.2 &  3.1 \\
\hline
\textbf{Pre-trial $\mathbf{p}$-value} &  0.33 &  0.23 &  $>$0.5 &  0.19 &  0.097 &  0.43 \\
\hline
\end{tabular}
\caption{Sensitivity, discovery potential and 90\% C.L.~upper limit for the different analysis parameters, which are the UHECR energy cut, $E_{\rm cut}$, magnetic deflection, $D_0 \cdot C$,
and correlation fraction, $f_{\rm corr}$, for the \anaAC.
The values are given as normalization of the neutrino flux per source in the unit of
$10^{-10}\,\mathrm{GeV^{-1}\, cm^{-2} \, s^{-1}}$.
The neutrino flux per source is modeled with a power-law of the form $\drv{\Phi}/ \drv{E} = \Phi_0 \cdot \left( E/\SI{1}{\GeV}\right)^{-2}$.
The last row states all six experimentally obtained pre-trial $p$-values with respect to the null hypothesis of an isotropic neutrino flux.
}
\label{tab:anaC-sens-lim}
\end{center}
\end{table*}

\subsection{Unbinned likelihood-stacking analysis of UHECRs and high-energy neutrinos}

The \anaGE is performed for track-like and cascade-like high-energy neutrinos separately for three different scaling factors of $C\in \{1, 2, 3\}$.
This results in six $p$-values with respect to the background hypothesis of an isotropic flux of UHECRs, of which none is significant.
The smallest $p$-value is 0.26, which is found for cascades and the largest scaling factor of $C=3$, corresponding to the benchmark deflection in the Northern and Southern Galactic hemisphere of (7.2$^{\circ}$, 11.1$^{\circ}$).
Based on the observed TS values, 90\% C.L.~limits on the number of UHECR events correlated to the high-energy neutrinos are calculated using the PEs of the corresponding signal hypothesis for each of the six tests.
Since all $p$-values using the track-like neutrinos are larger than 0.5, the limits are set equal to the sensitivity in order to not over-estimate the limits based on an under-fluctuation in data.
All $p$-values and the corresponding limits are reported in \autoref{tab:anaB-res} together with the sensitivity and discovery potential.

\begin{table}[htbp]
\centering
\begin{tabular}{lccc}
\hline
\textbf{Analysis parameters} & {~} & {~} & \\
$D_{N/S} \cdot C$ & (2.4$^{\circ}$, 3.7$^{\circ}$) & $\times 2$ & $\times 3$ \\
% $D_{N/S} \cdot C$ & (2.4$^{\circ}$, 3.7$^{\circ}$) & (4.8$^{\circ}$, 7.4$^{\circ}$) & (7.2$^{\circ}$, 11.1$^{\circ}$) \\
\hline
\textbf{Sensitivity} & {~} & {~} & \\
tracks & 9.8 & 9.8 & 10.3 \\
cascades & 57 & 61 & 64 \\
\hline
\textbf{$\mathbf{3\sigma}$ disc.~potential} & {~} & {~} & \\
tracks & 20 & 21 & 21 \\
cascades & 113 & 129 & 135 \\
\hline
\textbf{90\% C.L.~UL} & {~} & {~} & \\
tracks & 9.8 & 9.8 & 10.3 \\
cascades & 55 & 76 & 101 \\
\hline
\textbf{$\mathbf{p}$-values} & {~} & {~} & \\
tracks & $>$0.5 & $>$0.5 & $>$0.5 \\
cascades & $>$0.5 & 0.38 & 0.26 \\ % 0.51
\hline
\end{tabular}
\caption{Sensitivity, $3\sigma$ discovery potential, and 90\% C.L. upper limits in terms of number of UHECRs ($n_s$, c.f.~\autoref{eq:anaB-llh}) as well as the pre-trial $p$-values for the \anaGE with the samples of high-energy tracks and cascades
assuming an isotropic flux of UHECRs.}
\label{tab:anaB-res}
\end{table}

\subsection{Two-point angular correlation of UHECRs and high-energy neutrinos}

\begin{table}[htbp]
\centering
\begin{tabular}{cccc} % lrrr
\hline
Event type & bg. hypothesis & separation & {\it p}-value \\
\hline
tracks & isotropic neutrinos & 14$^{\circ}$ & 0.23 \\
cascades & isotropic neutrinos & 16$^{\circ}$ & 0.15 \\
\hline
tracks & isotropic UHECRs & 10$^{\circ}$  & $>$0.5\\ %& 0.84
cascades & isotropic UHECRs & 16$^{\circ}$ & 0.18 \\
\hline
\end{tabular}
\caption{Post-trial {\it p}-value and best-fit angular separation for the \anaPA obtained with the neutrino data sets of high-energy tracks and cascades, as stated in the first column. 
The $p$-values are corrected for choosing the largest deviation out of all maximum angular separations.
The respective background hypotheses are stated in the second column.}
\label{tab:anaA-res}
\end{table}

\begin{figure*}%[htbp]
\epsscale{1.1}
\begin{center}
\includegraphics[width=\textwidth]{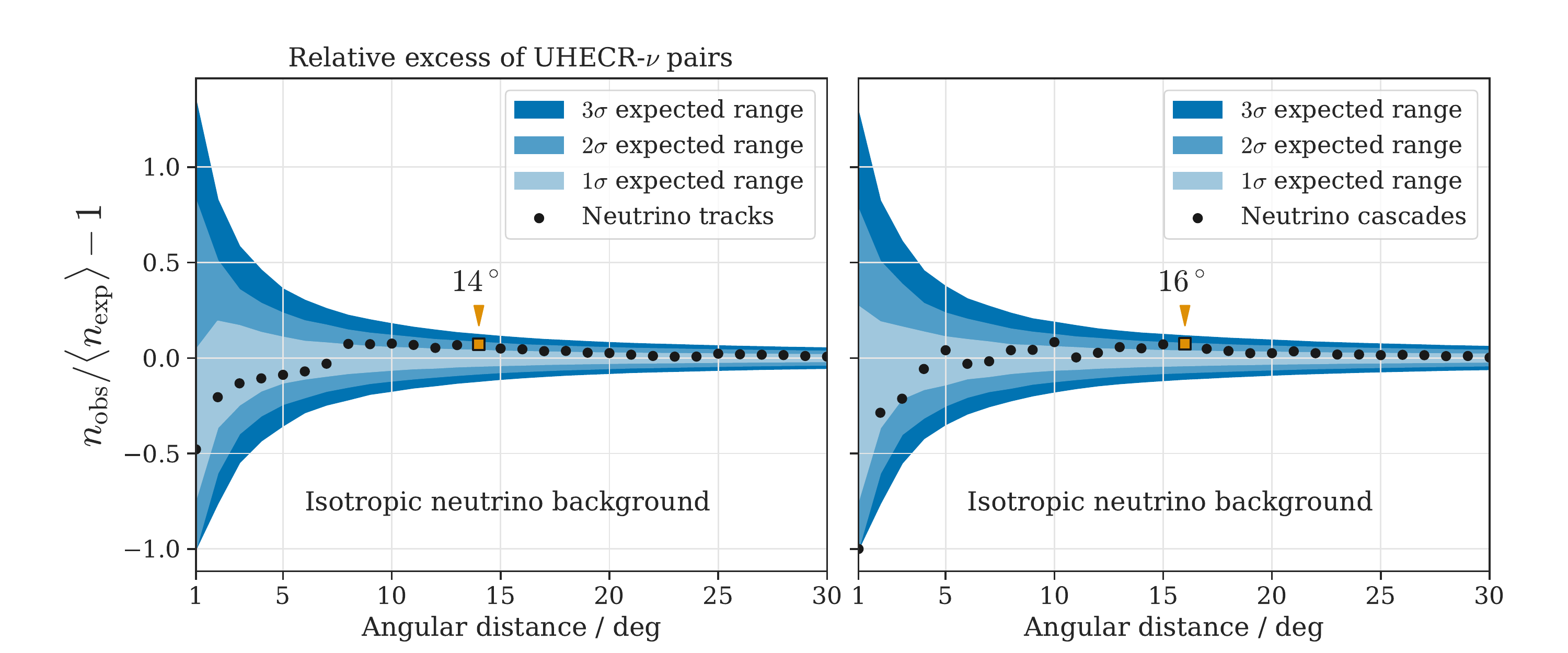}
\caption{Relative excess of pairs, $n_{\rm obs}/\langle n_{\rm exp}\rangle -1$, as a function of the maximum angular
separation of the neutrino and UHECR pairs.
The experimental result is shown as black dots, while the blue color bands show the regions containing the 1, 2 and 3$\sigma$ fluctuations from background PEs based on an isotropic distribution of high-energy neutrinos.
The results for the track-like neutrinos are shown on the left, and for the cascade-like neutrino on the right.} 
\label{fig:anaA-result}
\end{center}
\end{figure*}

The \anaPA is applied to the sets of neutrino tracks and cascades separately, as well as for the background hypothesis of an isotropic UHECRs flux and an isotropic high-energy neutrino flux.
The significance of the result with respect to the background hypothesis is corrected for the scan over the separation angles.
This results in four $p$-values and four respective best-fit angular separation, as listed in \autoref{tab:anaA-res}.
The largest deviation from an isotropic flux of high-energy neutrinos ($p$-value=15\%) is found using cascades and a maximum angular separation of \SI{16}{\degree}.
None of the $p$-values show a significant result, thus the results are all compatible with their respective background hypothesis.
The relative excesses of pairs with respect to an isotropic distribution of neutrinos as a function of the separation angle are shown in \autoref{fig:anaA-result}, separately for neutrino tracks and cascades.

\section{Conclusions}
Three complementary analyses have been performed on the UHECR data sets provided by the Pierre Auger
and the Telescope Array Collaborations combined with the high-energy and full-sky neutrino data sets provided by
the IceCube and ANTARES Collaborations.
For both the UHECR and neutrino data sets, the combination of data from the two respective observatories provides a field of view over the entire sky.
None of the analyses found a result incompatible with the assumed background hypotheses of either an isotropic neutrino flux or an isotropic UHECR flux.
This is an important update on the results reported in \citet{Aartsen:2015dml}, where $p$-values close to the $3\sigma$ level were reported when applying analyses 2 and 3 to cascade-like neutrino events.

Based on the results, 90\% C.L.~upper limits are calculated for analysis 1 on the flux of neutrinos from point-like sources
correlated with UHECRs, and for analysis 2 on the number of UHECR events correlated with high-energy neutrinos.
Analysis 1 reports upper limits on the correlated neutrino flux per source given different sets of parameters that make assumptions about the lower cut on the UHECR energy and the fraction of UHECRs with a correlated neutrino source, as well as about a scaling factor of the magnetic deflection~(see \autoref{tab:anaC-sens-lim}).
Analysis 2 reports upper limits on the number of correlated UHECR events with either track-like or cascade-like neutrinos depending on the scaling factor of the magnetic deflection
(see \autoref{tab:anaB-res}). 
This is the first time that upper limits from a direct correlation search of UHECRs and neutrinos are reported.

These limits are calculated based on the assumed signal and background hypotheses and are thus based on the parameters of the respective signal models.
In the following, we discuss the limitations of these signal models in the light of underlying assumptions made for the neutrino-UHECR correlation.
There are several plausible explanations why we have not observed a significant correlation of UHECRs and neutrinos.

Neutrino sources are presumably distributed over the whole universe, and the emitted neutrinos are able to reach Earth without deflection or absorption. 
For example, the first neutrino source identified with compelling evidence, \txs~\citep{IceCube:2018dnn,IceCube:2018cha}, 
is located at a redshift of $z=0.34$ \citep{paiano:2018txs}, corresponding to a comoving distance of about \SI{1.3}{Gpc}. 
This is beyond the assumed horizon for UHECR sources in the local universe of up to about \SI{200}{Mpc}.
Therefore, the fraction of correlated sources within the UHECR horizon is small compared to the total number of neutrino sources.
This was quantified by~\cite{palladino:2020} for muon neutrinos above \SI{200}{\TeV} based on the non-observation of neutrino multiplets above this energy.
They concluded that the high-energy flux measured with IceCube must come from a large number of sources such that no multiplets are observed.
Necessarily, most of the neutrino sources lie beyond the UHECR horizon, which is used in~\cite{palladino:2020} to explain the lack of UHECR-neutrino correlations found in~\cite{Aartsen:2015dml} and~\cite{Schumacher:2019qdx}, and the same applies to the current study.
In these studies, however, the energy threshold for neutrinos lies around \SI{20}{\TeV} for the combined high-energy data sets and as low as \SI{100}{\GeV} for the PS data set.
The non-observation of UHECR-neutrino correlations thus extends to even lower energies than considered by~\cite{palladino:2020}.

Even if there were local sources of both UHECRs and neutrinos, they could be transient phenomena, such that the UHECRs emitted over a short period of time arrive much later due to their deflection,
whereas the neutrinos travel on a straight path.
It has been estimated by \cite{Davoudifar:2011nxv} that the deflection in the EGMF causes a typical time delay on the order of $10^5$ years considering a source distance of \SI{50}{Mpc} and an EGMF strength of \SI{2}{nG}.
A delay of a couple of decades as expected from propagation in the GMF is already sufficient to de-correlate the observed UHECRs and neutrinos, as the livetime of the neutrino and UHECR data sets ranges between 7 and 13 years.

Another source of uncertainty is the mass composition and thus the charge of UHECRs at the highest energies.
The measurements of the UHECR composition above $\sim$\SI{50}{\EeV} are not yet conclusive.
However, several composition constraints at the highest energies~\citep{aab_2014_xmax1,aab_2014_xmax2}
and the lack of significant, magnetically-induced structures in the UHECR arrival directions~\citep{Aab_2020_magn} suggest that less than 10\% of UHECRs above $\sim$\SI{50}{\EeV} might be light elements.
Only the light UHECRs are expected to have their source in the vicinity of their arrival direction as derived from the benchmark deflection model.
The sources of heavier UHECRs, e.g.~iron nuclei, are spatially almost uncorrelated to the UHECR arrival directions due to the 26-times larger magnetic deflection compared to protons.
We see in analysis 1 that the limits on the neutrino flux are significantly relaxed when assuming a correlation fraction of 50\% or 10\%. 
This correlation fraction, defined as the number of UHECRs with a neutrino source in their vicinity, is a simplified model of the UHECR composition.

The modeling of the magnetic deflection as a 2D Gaussian as assumed for analysis 1 and 2 is a simplification with respect to the PT11 and JF12 GMF models \citep{PTgmf2011,JFgmf2012}, which themselves are subject to large uncertainties.
Especially in the region of the Galactic Plane, we expect deflections that are larger than the assumed mean value of around \SI{3}{\degree}.
In addition, coherent deflection of UHECRs in the GMF or the deflection of UHECRs in the IGMF are not explicitly accounted for. 
Overall, a coherent shift of UHECRs depending on their arrival direction or an overall significantly larger deflection in the GMF and IGMF can dilute the spatial correlation of UHECRs and neutrinos.
Nevertheless, a scaling of the overall deflection is covered partially with the scaling factor applied in analyses 1 and 2, since the overall strength of the deflection and the UHECR charge are largely degenerate parameters.

From a theoretical perspective, the connection of UHECRs and high-energy neutrinos is plausible based on their similar energy budget \citep{Waxman:1998yy,Bahcall:1999yr,Murase:2013rfa,Ahlers:2018fkn}.
However, this connection could not be proven with the current data and analysis approaches.
It is to note that neutrinos with energies in the TeV--PeV range originate most likely from cosmic rays with energies below $\sim$\SI{100}{PeV}~\citep{Murase:2016gly}, which is below the energy threshold of $>\SI{50}{EeV}$ of the UHECR data sets used here.
A direct connection to UHECRs can only be proven with ultra-high-energy neutrinos in the EeV range that have not been discovered yet.
The non-observation of a UHECR-neutrino correlation rather points to the possibility that efficient UHECR sources are less efficient neutrino sources and vice versa:
Sources with efficient UHECR acceleration and emission require an optically thin proton and radiation environment,
while sources with dense proton and radiation targets are efficient in neutrino production, but not in UHECR emission~\citep{murase_agn_2014,rodrigues_agn_2021}. 
This argument, however, can be relaxed when considering models
where cosmic rays below $\sim$\SI{100}{PeV} are confined in a calorimetric environment and subsequently produce TeV--PeV neutrinos,
while a fraction of the cosmic rays are accelerated to the highest energies and escape the source before interacting~\citep{Murase:2016gly,Ahlers:2018fkn}.
Although no indication of such a scenario has been found in this analysis,
the first indication of such a connection could be the observation of the Seyfert Galaxy \ngc as a potential neutrino source candidate \citep{aartsen10yrIntegratedPS2020,Inoue_2020} and UHECR source candidate \citep{Aab:2018chp}.
As such sources are numerous also within the UHECR horizon, dedicated future searches
correlating UHECR, photon, and neutrino observations might be able to set constraints on specific source candidates, instead of relying on UHECR and neutrino data alone.
\ngc as a potential neutrino source candidate \citep{aartsen10yrIntegratedPS2020,Inoue_2020} and UHECR source candidate \citep{Aab:2018chp}
could serve as blueprint for a catalog of potential common sources to be constrained with dedicated analyses.

In summary, the three analyses presented here reflect complementary approaches for tackling
the question of a common origin of UHECRs and high-energy neutrinos.
Despite substantially enlarged data sets and improved methods, the initially reported results in \citet{Aartsen:2015dml} could not be strengthened.
For future analyses, we expect substantial gains in sensitivity if the charge of the UHECRs could be estimated on an event-by-event basis, as it is expected for future measurements by Auger Prime \citep{augerPrime2016}.

\section*{Acknowledgements}
%% ANTARES
The authors of the ANTARES collaboration acknowledge the financial support of the funding agencies:
% France:
Centre National de la Recherche Scientifique (CNRS), Commissariat \`a
l'\'ener\-gie atomique et aux \'energies alternatives (CEA),
Commission Europ\'eenne (FEDER fund and Marie Curie Program),
Institut Universitaire de France (IUF), LabEx UnivEarthS (ANR-10-LABX-0023 and ANR-18-IDEX-0001),
R\'egion \^Ile-de-France (DIM-ACAV), R\'egion
Alsace (contrat CPER), R\'egion Provence-Alpes-C\^ote d'Azur,
D\'e\-par\-tement du Var and Ville de La
Seyne-sur-Mer, France;
% Germany: 
Bundesministerium f\"ur Bildung und Forschung
(BMBF), Germany; 
% Italy
Istituto Nazionale di Fisica Nucleare (INFN), Italy;
% Netherlands
Nederlandse organisatie voor Wetenschappelijk Onderzoek (NWO), the Netherlands;
% Russia
Council of the President of the Russian Federation for young
scientists and leading scientific schools supporting grants, Russia;
% Romania
Executive Unit for Financing Higher Education, Research, Development and Innovation (UEFISCDI), Romania;
% Spain
Ministerio de Ciencia, Innovaci\'{o}n, Investigaci\'{o}n y
Universidades (MCIU): Programa Estatal de Generaci\'{o}n de
Conocimiento (refs. PGC2018-096663-B-C41, -A-C42, -B-C43, -B-C44)
(MCIU/FEDER), Generalitat Valenciana: Prometeo (PROMETEO/2020/019),
Grisol\'{i}a (refs. GRISOLIA/2018/119, /2021/192) and GenT
(refs. CIDEGENT/2018/034, /2019/043, /2020/049, /2021/023) programs, Junta de
Andaluc\'{i}a (ref. A-FQM-053-UGR18), La Caixa Foundation (ref. LCF/BQ/IN17/11620019), EU: MSC program (ref. 101025085), Spain;
% Marocco
Ministry of Higher Education, Scientific Research and Innovation, Morocco, and the Arab Fund for Economic and Social Development, Kuwait.
% A.O.B.:
We also acknowledge the technical support of Ifremer, AIM and Foselev Marine
for the sea operation and the CC-IN2P3 for the computing facilities.
The ANTARES collaboration acknowledges the significant contributions to this manuscript from Julien Aublin.
\newline
\newline
%% IceCube
The IceCube collaboration acknowledges the significant contributions to this manuscript from Cyril Alispach, Anastasia Barbano and Lisa Schumacher.
The authors of the IceCube Collaboration acknowledge the support from the following agencies: 
USA {\textendash} U.S. National Science Foundation-Office of Polar Programs,
U.S. National Science Foundation-Physics Division,
U.S. National Science Foundation-EPSCoR,
Wisconsin Alumni Research Foundation,
Center for High Throughput Computing (CHTC) at the University of Wisconsin{\textendash}Madison,
Open Science Grid (OSG),
Extreme Science and Engineering Discovery Environment (XSEDE),
Frontera computing project at the Texas Advanced Computing Center,
U.S. Department of Energy-National Energy Research Scientific Computing Center,
Particle astrophysics research computing center at the University of Maryland,
Institute for Cyber-Enabled Research at Michigan State University,
and Astroparticle physics computational facility at Marquette University;
Belgium {\textendash} Funds for Scientific Research (FRS-FNRS and FWO),
FWO Odysseus and Big Science programmes,
and Belgian Federal Science Policy Office (Belspo);
Germany {\textendash} Bundesministerium f{\"u}r Bildung und Forschung (BMBF),
Deutsche Forschungsgemeinschaft (DFG),
Helmholtz Alliance for Astroparticle Physics (HAP),
Initiative and Networking Fund of the Helmholtz Association,
Deutsches Elektronen Synchrotron (DESY),
and High Performance Computing cluster of the RWTH Aachen;
Sweden {\textendash} Swedish Research Council,
Swedish Polar Research Secretariat,
Swedish National Infrastructure for Computing (SNIC),
and Knut and Alice Wallenberg Foundation;
Australia {\textendash} Australian Research Council;
Canada {\textendash} Natural Sciences and Engineering Research Council of Canada,
Calcul Qu{\'e}bec, Compute Ontario, Canada Foundation for Innovation, WestGrid, and Compute Canada;
Denmark {\textendash} Villum Fonden and Carlsberg Foundation;
New Zealand {\textendash} Marsden Fund;
Japan {\textendash} Japan Society for Promotion of Science (JSPS)
and Institute for Global Prominent Research (IGPR) of Chiba University;
Korea {\textendash} National Research Foundation of Korea (NRF);
Switzerland {\textendash} Swiss National Science Foundation (SNSF);
United Kingdom {\textendash} Department of Physics, University of Oxford.
L.S. and C.H.W. acknowledge financial support from the "Deutsche Forschungsgemeinschaft".
\newline
\newline
%% Auger
The successful installation, commissioning, and operation of the Pierre
Auger Observatory would not have been possible without the strong
commitment and effort from the technical and administrative staff in
Malarg\"ue. We are very grateful to the following agencies and
organizations for financial support:

Argentina -- Comisi\'on Nacional de Energ\'\i{}a At\'omica; Agencia Nacional de
Promoci\'on Cient\'\i{}fica y Tecnol\'ogica (ANPCyT); Consejo Nacional de
Investigaciones Cient\'\i{}ficas y T\'ecnicas (CONICET); Gobierno de la
Provincia de Mendoza; Municipalidad de Malarg\"ue; NDM Holdings and Valle
Las Le\~nas; in gratitude for their continuing cooperation over land
access; Australia -- the Australian Research Council; Belgium -- Fonds
de la Recherche Scientifique (FNRS); Research Foundation Flanders (FWO);
Brazil -- Conselho Nacional de Desenvolvimento Cient\'\i{}fico e Tecnol\'ogico
(CNPq); Financiadora de Estudos e Projetos (FINEP); Funda\c{c}\~ao de Amparo \`a
Pesquisa do Estado de Rio de Janeiro (FAPERJ); S\~ao Paulo Research
Foundation (FAPESP) Grants No.~2019/10151-2, No.~2010/07359-6 and
No.~1999/05404-3; Minist\'erio da Ci\^encia, Tecnologia, Inova\c{c}\~oes e
Comunica\c{c}\~oes (MCTIC); Czech Republic -- Grant No.~MSMT CR LTT18004,
LM2015038, LM2018102, CZ.02.1.01/0.0/0.0/16{\textunderscore}013/0001402,
CZ.02.1.01/0.0/0.0/18{\textunderscore}046/0016010 and \\
CZ.02.1.01/0.0/0.0/17{\textunderscore}049/0008422; France -- Centre de Calcul
IN2P3/CNRS; Centre National de la Recherche Scientifique (CNRS); Conseil
R\'egional Ile-de-France; D\'epartement Physique Nucl\'eaire et Corpusculaire
(PNC-IN2P3/CNRS); D\'epartement Sciences de l'Univers (SDU-INSU/CNRS);
Institut Lagrange de Paris (ILP) Grant No.~LABEX ANR-10-LABX-63 within
the Investissements d'Avenir Programme Grant No.~ANR-11-IDEX-0004-02;
Germany -- Bundesministerium f\"ur Bildung und Forschung (BMBF); Deutsche
Forschungsgemeinschaft (DFG); Finanzministerium Baden-W\"urttemberg;
Helmholtz Alliance for Astroparticle Physics (HAP);
Helmholtz-Gemeinschaft Deutscher Forschungszentren (HGF); Ministerium
f\"ur Innovation, Wissenschaft und Forschung des Landes
Nordrhein-Westfalen; Ministerium f\"ur Wissenschaft, Forschung und Kunst
des Landes Baden-W\"urttemberg; Italy -- Istituto Nazionale di Fisica
Nucleare (INFN); Istituto Nazionale di Astrofisica (INAF); Ministero
dell'Istruzione, dell'Universit\'a e della Ricerca (MIUR); CETEMPS Center
of Excellence; Ministero degli Affari Esteri (MAE); M\'exico -- Consejo
Nacional de Ciencia y Tecnolog\'\i{}a (CONACYT) No.~167733; Universidad
Nacional Aut\'onoma de M\'exico (UNAM); PAPIIT DGAPA-UNAM; The Netherlands
-- Ministry of Education, Culture and Science; Netherlands Organisation
for Scientific Research (NWO); Dutch national e-infrastructure with the
support of SURF Cooperative; Poland -- Ministry of Education and
Science, grant No.~DIR/WK/2018/11; National Science Centre, Grants
No.~2016/22/M/ST9/00198, 2016/23/B/ST9/01635, and 2020/39/B/ST9/01398;
Portugal -- Portuguese national funds and FEDER funds within Programa
Operacional Factores de Competitividade through Funda\c{c}\~ao para a Ci\^encia
e a Tecnologia (COMPETE); Romania -- Ministry of Research, Innovation
and Digitization, CNCS/CCCDI -- UEFISCDI, projects PN19150201/16N/2019,
PN1906010, TE128 and PED289, within PNCDI III; Slovenia -- Slovenian
Research Agency, grants P1-0031, P1-0385, I0-0033, N1-0111; Spain --
Ministerio de Econom\'\i{}a, Industria y Competitividad (FPA2017-85114-P and
PID2019-104676GB-C32), Xunta de Galicia (ED431C 2017/07), Junta de
Andaluc\'\i{}a (SOMM17/6104/UGR, P18-FR-4314) Feder Funds, RENATA Red
Nacional Tem\'atica de Astropart\'\i{}culas (FPA2015-68783-REDT) and Mar\'\i{}a de
Maeztu Unit of Excellence (MDM-2016-0692); USA -- Department of Energy,
Contracts No.~DE-AC02-07CH11359, No.~DE-FR02-04ER41300,
No.~DE-FG02-99ER41107 and No.~DE-SC0011689; National Science Foundation,
Grant No.~0450696; The Grainger Foundation; Marie Curie-IRSES/EPLANET;
European Particle Physics Latin American Network; and UNESCO.
\newline
\newline
%% TA

The Telescope Array experiment is supported by the Japan Society for
the Promotion of Science(JSPS) through
Grants-in-Aid
for Priority Area
%"Highest Energy Cosmic Rays"
431,
for Specially Promoted Research
%``Extreme Phenomena in the Universe Explored by Highest Energy Cosmic Rays''
%Grant Number
JP21000002,
%Grant-in-Aid
for Scientific  Research (S)
%"Quest for the unified picture of the explosion mechanism of supernovae and the central engine of gamma-ray bursts"
%Grant Number
JP19104006,
%Grant-in-Aid
for Specially Promoted Research
%"Extended Telescope Array Experiment - Nearby Extreme Universe Elucidated by Highest-energy Cosmic Rays"
%Grant Number
JP15H05693,
%Grant-in-Aid
for Scientific  Research (S)
%"Study of the ultra high energy cosmic ray source evolution by detailed measurement of cosmic rays in the wide energy range"
%Grant Number
JP15H05741, for Science Research (A) JP18H03705,
%Grant-in-Aid
for Young Scientists (A)
%"hoge hoge"
%Grant Number
JPH26707011,
and for Fostering Joint International Research (B)
%"Search for Ultra-High Energy Cosmic Ray origin using the extended Telescope Array experiment"
%Grant Number
JP19KK0074,
by the joint research program of the Institute for Cosmic Ray Research (ICRR), The University of Tokyo;
by the Pioneering Program of RIKEN for the Evolution of Matter in the Universe (r-EMU);
by the U.S. National Science
Foundation awards PHY-1404495, PHY-1404502, PHY-1607727, PHY-1712517, PHY-1806797, PHY-2012934, and PHY-2112904;
by the National Research Foundation of Korea
% \linebreak
(2017K1A4A3015188, 2020R1A2C1008230, \& 2020R1A2C2102800) ;
%\linebreak
by the Ministry of Science and Higher Education of the Russian Federation under the contract 075-15-2020-778, IISN project No. 4.4501.18, and Belgian Science Policy under IUAP VII/37 (ULB). This work was partially supported by the grants ofThe joint research program of the Institute for Space-Earth Environmental Research, Nagoya University and Inter-University Research Program of the Institute for Cosmic Ray Research of University of Tokyo. The foundations of Dr. Ezekiel R. and Edna Wattis Dumke, Willard L. Eccles, and George S. and Dolores Dor\'e Eccles all helped with generous donations. The State of Utah supported the project through its Economic Development Board, and the University of Utah through the Office of the Vice President for Research. The experimental site became available through the cooperation of the Utah School and Institutional Trust Lands Administration (SITLA), U.S. Bureau of Land Management (BLM), and the U.S. Air Force. We appreciate the assistance of the State of Utah and Fillmore offices of the BLM in crafting the Plan of Development for the site.  Patrick A.~Shea assisted the collaboration with valuable advice and supported the collaboration’s efforts. The people and the officials of Millard County, Utah have been a source of steadfast and warm support for our work which we greatly appreciate. We are indebted to the Millard County Road Department for their efforts to maintain and clear the roads which get us to our sites. We gratefully acknowledge the contribution from the technical staffs of our home institutions. An allocation of computer time from the Center for High Performance Computing at the University of Utah is gratefully acknowledged.

\facility{ANTARES, IceCube Neutrino Observatory,  Pierre Auger Observatory, Telescope Array}

\clearpage

\bibliographystyle{aasjournal}
\bibliography{uhecr-nu-paper}

\begin{thebibliography}{}
\expandafter\ifx\csname natexlab\endcsname\relax\def\natexlab#1{#1}\fi
\providecommand{\url}[1]{\href{#1}{#1}}
\providecommand{\dodoi}[1]{doi:~\href{http://doi.org/#1}{\nolinkurl{#1}}}
\providecommand{\doeprint}[1]{\href{http://ascl.net/#1}{\nolinkurl{http://ascl.net/#1}}}
\providecommand{\doarXiv}[1]{\href{https://arxiv.org/abs/#1}{\nolinkurl{https://arxiv.org/abs/#1}}}

\bibitem[{Aab {et~al.}(2014{\natexlab{a}})Aab, Abreu, Aglietta, Ahn,
  Al~Samarai, Albuquerque, Allekotte, Allen, Allison, Almela, Alvarez~Castillo,
  Alvarez-Mu\~niz, Alves~Batista, Ambrosio, Aminaei, Anchordoqui, Andringa,
  Aramo, Aranda, Arqueros, Asorey, Assis, Aublin, Ave, Avenier, Avila, Awal,
  Badescu, Barber, B\"auml, Baus, Beatty, Becker, Bellido, Berat, Bertaina,
  Bertou, Biermann, Billoir, Blaess, Blanco, Bleve, Bl\"umer,
  Boh\'a\ifmmode~\check{c}\else \v{c}\fi{}ov\'a, Boncioli, Bonifazi, Bonino,
  Borodai, Brack, Brancus, Bridgeman, Brogueira, Brown, Buchholz, Bueno,
  Buitink, Buscemi, Caballero-Mora, Caccianiga, Caccianiga, Candusso, Caramete,
  Caruso, Castellina, Cataldi, Cazon, Cester, Chavez, Chiavassa, Chinellato,
  Chudoba, Cilmo, Clay, Cocciolo, Colalillo, Coleman, Collica, Coluccia,
  Concei\ifmmode \mbox{\c{c}}\else~\c{c}\fi{}\ ao, Contreras, Cooper, Cordier,
  Coutu, Covault, Cronin, Curutiu, Dallier, Daniel, Dasso, Daumiller, Dawson,
  de~Almeida, De~Domenico, de~Jong, de~Mello~Neto, De~Mitri, de~Oliveira,
  de~Souza, del Peral, Deligny, Dembinski, Dhital, Di~Giulio, Di~Matteo, Diaz,
  D\'{\i}az~Castro, Diogo, Dobrigkeit, Docters, D'Olivo, Dorofeev,
  Dorosti~Hasankiadeh, Dova, Ebr, Engel, Erdmann, Erfani, Escobar, Espadanal,
  Etchegoyen, Facal San~Luis, Falcke, Fang, Farrar, Fauth, Fazzini, Ferguson,
  Fernandes, Fick, Figueira, Filevich, Filip\ifmmode \check{c}\else
  \v{c}\fi{}i\ifmmode~\check{c}\else \v{c}\fi{}, Fox, Fratu, Fr\"ohlich, Fuchs,
  Fujii, Gaior, Garc\'{\i}a, Garcia~Roca, Garcia-Gamez, Garcia-Pinto, Garilli,
  Gascon~Bravo, Gate, Gemmeke, Ghia, Giaccari, Giammarchi, Giller, Glaser,
  Glass, G\'omez~Berisso, G\'omez~Vitale, Gon\ifmmode~\mbox{\c{c}}\else
  \c{c}\fi{}alves, Gonzalez, Gonz\'alez, Gookin, Gordon, Gorgi, Gorham,
  Gouffon, Grebe, Griffith, Grillo, Grubb, Guarino, Guedes, Hampel, Hansen,
  Harari, Harrison, Hartmann, Harton, Haungs, Hebbeker, Heck, Heimann, Herve,
  Hill, Hojvat, Hollon, Holt, Homola, H\"orandel, Horvath, Hrabovsk\'y, Huber,
  Huege, Insolia, Isar, Jandt, Jansen, Jarne, Josebachuili, K\"a\"ap\"a,
  Kambeitz, Kampert, Kasper, Katkov, K\'egl, Keilhauer, Keivani, Kemp,
  Kieckhafer, Klages, Kleifges, Kleinfeller, Krause, Krohm, Kr\"omer,
  Kruppke-Hansen, Kuempel, Kunka, LaHurd, Latronico, Lauer, Lauscher,
  Lautridou, Le~Coz, Le\~ao, Lebrun, Lebrun, Leigui~de Oliveira,
  Letessier-Selvon, Lhenry-Yvon, Link, L\'opez, Lopez~Ag\"uera, Louedec,
  Lozano~Bahilo, Lu, Lucero, Ludwig, Malacari, Maldera, Mallamaci, Maller,
  Mandat, Mantsch, Mariazzi, Marin, Mari\ifmmode~\mbox{\c{s}}\else \c{s}\fi{},
  Marsella, Martello, Martin, Martinez, Mart\'{\i}nez~Bravo, Martraire,
  Mas\'{\i}as~Meza, Mathes, Mathys, Matthews, Matthews, Matthiae, Maurel,
  Maurizio, Mayotte, Mazur, Medina, Medina-Tanco, Meissner, Melissas, Melo,
  Menshikov, Messina, Meyhandan, Mi\ifmmode \acute{c}\else
  \'{c}\fi{}anovi\ifmmode~\acute{c}\else \'{c}\fi{}, Micheletti, Middendorf,
  Minaya, Miramonti, Mitrica, Molina-Bueno, Mollerach, Monasor,
  Monnier~Ragaigne, Montanet, Morello, Mostaf\'a, Moura, Muller, M\"uller,
  M\"uller, M\"unchmeyer, Mussa, Navarra, Navas, Necesal, Nellen, Nelles,
  Neuser, Nguyen, Niechciol, Niemietz, Niggemann, Nitz, Nosek, Novotny,
  No\ifmmode~\check{z}\else \v{z}\fi{}ka, Ochilo, Olinto, Oliveira, Pacheco,
  Pakk Selmi-Dei, Palatka, Pallotta, Palmieri, Papenbreer, Parente, Parra,
  Paul, Pech, P\ifmmode~\mbox{\c{e}}\else \c{e}\fi{}kala, Pelayo, Pepe,
  Perrone, Petermann, Peters, Petrera, Petrov, Phuntsok, Piegaia, Pierog,
  Pieroni, Pimenta, Pirronello, Platino, Plum, Porcelli, Porowski, Prado,
  Privitera, Prouza, Purrello, Quel, Querchfeld, Quinn, Rautenberg, Ravel,
  Ravignani, Revenu, Ridky, Riggi, Risse, Ristori, Rizi, Rodrigues~de Carvalho,
  Rodriguez~Cabo, Rodriguez~Fernandez, Rodriguez~Rojo,
  Rodr\'{\i}guez-Fr\'{\i}as, Rogozin, Ros, Rosado, Rossler, Roth, Roulet,
  Rovero, Saffi, Saftoiu, Salamida, Salazar, Saleh, Salesa~Greus, Salina,
  S\'anchez, Sanchez-Lucas, Santo, Santos, Santos, Sarazin, Sarkar, Sarmento,
  Sato, Scharf, Scherini, Schieler, Schiffer, Schmidt, Scholten, Schoorlemmer,
  Schov\'anek, Schulz, Schulz, Schumacher, Sciutto, Segreto, Settimo, Shadkam,
  Shellard, Sidelnik, Sigl, Sima, \ifmmode~\acute{S}\else
  \'{S}\fi{}mia\l{}kowski, \ifmmode~\check{S}\else \v{S}\fi{}m\'{\i}da, Snow,
  Sommers, Sorokin, Squartini, Srivastava, Stani\ifmmode~\check{c}\else
  \v{c}\fi{}, Stapleton, Stasielak, Stephan, Stutz, Suarez, Suomij\"arvi,
  Supanitsky, Sutherland, Swain, Szadkowski, Szuba, Taborda, Tapia, Tartare,
  Tepe, Theodoro, Timmermans, Todero~Peixoto, Toma, Tomankova, Tom\'e,
  Tonachini, Torralba~Elipe, Torres~Machado, Travnicek, Trovato, Tueros,
  Ulrich, Unger, Urban, Vald\'es~Galicia, Vali\~no, Valore, van Aar, van
  Bodegom, van~den Berg, van Velzen, van Vliet, Varela, Vargas~C\'ardenas,
  Varner, V\'azquez, V\'azquez, Veberi\ifmmode~\check{c}\else \v{c}\fi{},
  Verzi, Vicha, Videla, Villase\~nor, Vlcek, Vorobiov, Wahlberg, Wainberg,
  Walz, Watson, Weber, Weidenhaupt, Weindl, Werner, Widom, Wiencke,
  Wilczy\ifmmode~\acute{n}\else \'{n}\fi{}ska, Wilczy\ifmmode~\acute{n}\else
  \'{n}\fi{}ski, Will, Williams, Winchen, Wittkowski, Wundheiler, Wykes,
  Yamamoto, Yapici, Yuan, Yushkov, Zamorano, Zas, Zavrtanik, Zavrtanik, Zaw,
  Zepeda, Zhou, Zhu, Zimbres~Silva, Ziolkowski, \& Zuccarello}]{aab_2014_xmax1}
Aab, A., Abreu, P., Aglietta, M., {et~al.} 2014{\natexlab{a}}, Phys. Rev. D,
  90, 122005, \dodoi{10.1103/PhysRevD.90.122005}

\bibitem[{Aab {et~al.}(2014{\natexlab{b}})Aab, Abreu, Aglietta, Ahn,
  Al~Samarai, Albuquerque, Allekotte, Allen, Allison, Almela, Alvarez~Castillo,
  Alvarez-Mu\~niz, Alves~Batista, Ambrosio, Aminaei, Anchordoqui, Andringa,
  Aramo, Aranda, Arqueros, Asorey, Assis, Aublin, Ave, Avenier, Avila, Awal,
  Badescu, Barber, B\"auml, Baus, Beatty, Becker, Bellido, Berat, Bertania,
  Bertou, Biermann, Billoir, Blaess, Blanco, Bleve, Bl\"umer,
  Boh\'a\ifmmode~\check{c}\else \v{c}\fi{}ov\'a, Boncioli, Bonifazi, Bonino,
  Borodai, Brack, Brancus, Bridgeman, Brogueira, Brown, Buchholz, Bueno,
  Buitink, Buscemi, Caballero-Mora, Caccianiga, Caccianiga, Candusso, Caramete,
  Caruso, Castellina, Cataldi, Cazon, Cester, Chavez, Chiavassa, Chinellato,
  Chudoba, Cilmo, Clay, Cocciolo, Colalillo, Coleman, Collica, Coluccia,
  Concei\ifmmode \mbox{\c{c}}\else~\c{c}\fi{}\ ao, Contreras, Cooper, Cordier,
  Coutu, Covault, Cronin, Curutiu, Dallier, Daniel, Dasso, Daumiller, Dawson,
  de~Almeida, De~Domenico, de~Jong, de~Mello~Neto, De~Mitri, de~Oliveira,
  de~Souza, del Peral, Deligny, Dembinski, Dhital, Di~Giulio, Di~Matteo, Diaz,
  D\'{\i}az~Castro, Diogo, Dobrigkeit, Docters, D'Olivo, Dorofeev,
  Dorosti~Hasankiadeh, Dova, Ebr, Engel, Erdmann, Erfani, Escobar, Espadanal,
  Etchegoyen, Facal San~Luis, Falcke, Fang, Farrar, Fauth, Fazzini, Ferguson,
  Fernandes, Fick, Figueira, Filevich, Filip\ifmmode \check{c}\else
  \v{c}\fi{}i\ifmmode~\check{c}\else \v{c}\fi{}, Fox, Fratu, Fr\"ohlich, Fuchs,
  Fuji, Gaior, Garc\'{\i}a, Garcia~Roca, Garcia-Gamez, Garcia-Pinto, Garilli,
  Gascon~Bravo, Gate, Gemmeke, Ghia, Giaccari, Giammarchi, Giller, Glaser,
  Glass, G\'omez~Berisso, G\'omez~Vitale, Gon\ifmmode~\mbox{\c{c}}\else
  \c{c}\fi{}alves, Gonzalez, Gonz\'alez, Gookin, Gordon, Gorgi, Gorham,
  Gouffon, Grebe, Griffith, Grillo, Grubb, Guarino, Guedes, Hampel, Hansen,
  Harari, Harrison, Hartmann, Harton, Haungs, Hebbeker, Heck, Heimann, Herve,
  Hill, Hojvat, Hollon, Holt, Homola, H\"orandel, Horvath, Hrabovsk\'y, Huber,
  Huege, Insolia, Isar, Jandt, Jansen, Jarne, Josebachuili, K\"a\"ap\"a,
  Kambeitz, Kampert, Kasper, Katkov, K\'egl, Keilhauer, Keivani, Kemp,
  Kieckhafer, Klages, Kleifges, Kleinfeller, Krause, Krohm, Kr\"omer,
  Kruppke-Hansen, Kuempel, Kunka, LaHurd, Latronico, Lauer, Lauscher,
  Lautridou, Le~Coz, Le\~ao, Lebrun, Lebrun, Leigui~de Oliveira,
  Letessier-Selvon, Lhenry-Yvon, Link, L\'opez, Lopez~Ag\"uera, Louedec,
  Lozano~Bahilo, Lu, Lucero, Ludwig, Malacari, Maldera, Mallamaci, Maller,
  Mandat, Mantsch, Mariazzi, Marin, Mari\ifmmode~\mbox{\c{s}}\else \c{s}\fi{},
  Marsella, Martello, Martin, Martinez, Mart\'{\i}nez~Bravo, Martraire,
  Mas\'{\i}as~Meza, Mathes, Mathys, Matthews, Matthews, Matthiae, Maurel,
  Maurizio, Mayotte, Mazur, Medina, Medina-Tanco, Meissner, Melissas, Melo,
  Menshikov, Messina, Meyhandan, Mi\ifmmode \acute{c}\else
  \'{c}\fi{}anovi\ifmmode~\acute{c}\else \'{c}\fi{}, Micheletti, Middendorf,
  Minaya, Miramonti, Mitrica, Molina-Bueno, Mollerach, Monasor,
  Monnier~Ragaigne, Montanet, Morello, Mostaf\'a, Moura, Muller, M\"uller,
  M\"uller, M\"unchmeyer, Mussa, Navarra, Navas, Necesal, Nellen, Nelles,
  Neuser, Nguyen, Niechciol, Niemietz, Niggemann, Nitz, Nosek, Novotny,
  No\ifmmode~\check{z}\else \v{z}\fi{}ka, Ochilo, Olinto, Oliveira, Pacheco,
  Pakk Selmi-Dei, Palatka, Pallotta, Palmieri, Papenbreer, Parente, Parra,
  Paul, Pech, P\ifmmode~\mbox{\c{e}}\else \c{e}\fi{}kala, Pelayo, Pepe,
  Perrone, Petermann, Peters, Petrera, Petrov, Phuntsok, Piegaia, Pierog,
  Pieroni, Pimenta, Pirronello, Platino, Plum, Porcelli, Porowski, Prado,
  Privitera, Prouza, Purrello, Quel, Querchfeld, Quinn, Rautenberg, Ravel,
  Ravignani, Revenu, Ridky, Riggi, Risse, Ristori, Rizi, Rodrigues~de Carvalho,
  Rodriguez~Cabo, Rodriguez~Fernandez, Rodriguez~Rojo,
  Rodr\'{\i}guez-Fr\'{\i}as, Rogozin, Ros, Rosado, Rossler, Roth, Roulet,
  Rovero, Saffi, Saftoiu, Salamida, Salazar, Saleh, Salesa~Greus, Salina,
  S\'anchez, Sanchez-Lucas, Santo, Santos, Santos, Sarazin, Sarkar, Sarmento,
  Sato, Scharf, Scherini, Schieler, Schiffer, Schmidt, Scholten, Schoorlemmer,
  Schov\'anek, Schulz, Schulz, Schumacher, Sciutto, Segreto, Settimo, Shadkam,
  Shellard, Sidelnik, Sigl, Sima, \ifmmode~\acute{S}\else
  \'{S}\fi{}mia\l{}kowski, \ifmmode~\check{S}\else \v{S}\fi{}m\'{\i}da, Snow,
  Sommers, Sorokin, Squartini, Srivastava, Stani\ifmmode~\check{c}\else
  \v{c}\fi{}, Stapleton, Stasielak, Stephan, Stutz, Suarez, Suomij\"arvi,
  Supanitsky, Sutherland, Swain, Szadkowski, Szuba, Taborda, Tapia, Tartare,
  Tepe, Theodoro, Timmermans, Todero~Peixoto, Toma, Tomankova, Tom\'e,
  Tonachini, Torralba~Elipe, Torres~Machado, Travnicek, Trovato, Tueros,
  Ulrich, Unger, Urban, Vald\'es~Galicia, Vali\~no, Valore, van Aar, van
  Bodegom, van~den Berg, van Velzen, van Vliet, Varela, Vargas~C\'ardenas,
  Varner, V\'azquez, V\'azquez, Veberi\ifmmode~\check{c}\else \v{c}\fi{},
  Verzi, Vicha, Videla, Villase\~nor, Vlcek, Vorobiov, Wahlberg, Wainberg,
  Walz, Watson, Weber, Weidenhaupt, Weindl, Werner, Widom, Wiencke,
  Wilczy\ifmmode~\acute{n}\else \'{n}\fi{}ska, Wilczy\ifmmode~\acute{n}\else
  \'{n}\fi{}ski, Will, Williams, Winchen, Wittkowski, Wundheiler, Wykes,
  Yamamoto, Yapici, Yuan, Yushkov, Zamorano, Zas, Zavrtanik, Zavrtanik, Zaw,
  Zepeda, Zhou, Zhu, Zimbres~Silva, Ziolkowski, \& Zuccarello}]{aab_2014_xmax2}
---. 2014{\natexlab{b}}, Phys. Rev. D, 90, 122006,
  \dodoi{10.1103/PhysRevD.90.122006}

\bibitem[{{Aab} {et~al.}(2015){Aab}, {Abreu}, {Aglietta}, {Ahn}, {Samarai},
  {Albuquerque}, {Allekotte}, {Allen}, {Allison}, {Almela}, {Alvarez Castillo},
  {Alvarez-Mu{\~n}iz}, {Alves Batista}, {Ambrosio}, {Aminaei}, {Anchordoqui},
  {Andringa}, {Aramo}, {Aranda}, {Arqueros}, {Asorey}, {Assis}, {Aublin},
  {Ave}, {Avenier}, {Avila}, {Awal}, {Badescu}, {Barber}, {B{\"a}uml}, {Baus},
  {Beatty}, {Becker}, {Bellido}, {Berat}, {Bertaina}, {Bertou}, {Biermann},
  {Billoir}, {Blaess}, {Blanco}, {Bleve}, {Bl{\"u}mer},
  {Boh{\'a}{\v{c}}ov{\'a}}, {Boncioli}, {Bonifazi}, {Bonino}, {Borodai},
  {Brack}, {Brancus}, {Bridgeman}, {Brogueira}, {Brown}, {Buchholz}, {Bueno},
  {Buitink}, {Buscemi}, {Caballero-Mora}, {Caccianiga}, {Caccianiga},
  {Candusso}, {Caramete}, {Caruso}, {Castellina}, {Cataldi}, {Cazon}, {Cester},
  {Chavez}, {Chiavassa}, {Chinellato}, {Chudoba}, {Cilmo}, {Clay}, {Cocciolo},
  {Colalillo}, {Coleman}, {Collica}, {Coluccia}, {Concei{\c{c}}{\~a}o},
  {Contreras}, {Cooper}, {Cordier}, {Coutu}, {Covault}, {Cronin}, {Curutiu},
  {Dallier}, {Daniel}, {Dasso}, {Daumiller}, {Dawson}, {de Almeida}, {De
  Domenico}, {de Jong}, {de Mello Neto}, {De Mitri}, {de Oliveira}, {de Souza},
  {del Peral}, {Deligny}, {Dembinski}, {Dhital}, {Di Giulio}, {Di Matteo},
  {Diaz}, {D{\'\i}az Castro}, {Diogo}, {Dobrigkeit}, {Docters}, {D'Olivo},
  {Dorofeev}, {Dorosti Hasankiadeh}, {Dova}, {Ebr}, {Engel}, {Erdmann},
  {Erfani}, {Escobar}, {Espadanal}, {Etchegoyen}, {Facal San Luis}, {Falcke},
  {Fang}, {Farrar}, {Fauth}, {Fazzini}, {Ferguson}, {Fernandes}, {Fick},
  {Figueira}, {Filevich}, {Filip{\v{c}}i{\v{c}}}, {Fox}, {Fratu}, {Freire},
  {Fr{\"o}hlich}, {Fuchs}, {Fujii}, {Gaior}, {Garc{\'\i}a}, {Garcia-Gamez},
  {Garcia-Pinto}, {Garilli}, {Gascon Bravo}, {Gate}, {Gemmeke}, {Ghia},
  {Giaccari}, {Giammarchi}, {Giller}, {Glaser}, {Glass}, {G{\'o}mez Berisso},
  {G{\'o}mez Vitale}, {Gon{\c{c}}alves}, {Gonzalez}, {Gonz{\'a}lez}, {Gookin},
  {Gordon}, {Gorgi}, {Gorham}, {Gouffon}, {Grebe}, {Griffith}, {Grillo},
  {Grubb}, {Guarino}, {Guedes}, {Hampel}, {Hansen}, {Harari}, {Harrison},
  {Hartmann}, {Harton}, {Haungs}, {Hebbeker}, {Heck}, {Heimann}, {Herve},
  {Hill}, {Hojvat}, {Hollon}, {Holt}, {Homola}, {H{\"o}randel}, {Horvath},
  {Hrabovsk{\'y}}, {Huber}, {Huege}, {Insolia}, {Isar}, {Jandt}, {Jansen},
  {Jarne}, {Josebachuili}, {K{\"a}{\"a}p{\"a}}, {Kambeitz}, {Kampert},
  {Kasper}, {Katkov}, {K{\'e}gl}, {Keilhauer}, {Keivani}, {Kemp}, {Kieckhafer},
  {Klages}, {Kleifges}, {Kleinfeller}, {Krause}, {Krohm}, {Kr{\"o}mer},
  {Kruppke-Hansen}, {Kuempel}, {Kunka}, {LaHurd}, {Latronico}, {Lauer},
  {Lauscher}, {Lautridou}, {Le Coz}, {Le{\~a}o}, {Lebrun}, {Lebrun}, {Leigui de
  Oliveira}, {Letessier-Selvon}, {Lhenry-Yvon}, {Link}, {L{\'o}pez}, {Louedec},
  {Lozano Bahilo}, {Lu}, {Lucero}, {Ludwig}, {Malacari}, {Maldera},
  {Mallamaci}, {Maller}, {Mandat}, {Mantsch}, {Mariazzi}, {Marin},
  {Mari{\c{s}}}, {Marsella}, {Martello}, {Martin}, {Martinez}, {Mart{\'\i}nez
  Bravo}, {Martraire}, {Mas{\'\i}as Meza}, {Mathes}, {Mathys}, {Matthews},
  {Matthews}, {Matthiae}, {Maurel}, {Maurizio}, {Mayotte}, {Mazur}, {Medina},
  {Medina-Tanco}, {Meissner}, {Melissas}, {Melo}, {Menshikov}, {Messina},
  {Meyhandan}, {Mi{\'c}anovi{\'c}}, {Micheletti}, {Middendorf}, {Minaya},
  {Miramonti}, {Mitrica}, {Molina-Bueno}, {Mollerach}, {Monasor}, {Monnier
  Ragaigne}, {Montanet}, {Morello}, {Mostaf{\'a}}, {Moura}, {Muller},
  {M{\"u}ller}, {M{\"u}ller}, {M{\"u}nchmeyer}, {Mussa}, {Navarra}, {Navas},
  {Necesal}, {Nellen}, {Nelles}, {Neuser}, {Nguyen}, {Niechciol}, {Niemietz},
  {Niggemann}, {Nitz}, {Nosek}, {Novotny}, {No{\v{z}}ka}, {Ochilo},
  {Oikonomou}, {Olinto}, {Oliveira}, {Pacheco}, {Pakk Selmi-Dei}, {Palatka},
  {Pallotta}, {Palmieri}, {Papenbreer}, {Parente}, {Parra}, {Paul}, {Pech},
  {P{\c{e}}kala}, {Pelayo}, {Pepe}, {Perrone}, {Petermann}, {Peters},
  {Petrera}, {Petrov}, {Phuntsok}, {Piegaia}, {Pierog}, {Pieroni}, {Pimenta},
  {Pirronello}, {Platino}, {Plum}, {Porcelli}, {Porowski}, {Prado},
  {Privitera}, {Prouza}, {Purrello}, {Quel}, {Querchfeld}, {Quinn},
  {Rautenberg}, {Ravel}, {Ravignani}, {Revenu}, {Ridky}, {Riggi}, {Risse},
  {Ristori}, {Rizi}, {Rodrigues de Carvalho}, {Rodriguez Fernandez}, {Rodriguez
  Rojo}, {Rodr{\'\i}guez-Fr{\'\i}as}, {Rogozin}, {Ros}, {Rosado}, {Rossler},
  {Roth}, {Roulet}, {Rovero}, {Saffi}, {Saftoiu}, {Salamida}, {Salazar},
  {Saleh}, {Salesa Greus}, {Salina}, {S{\'a}nchez}, {Sanchez-Lucas}, {Santo},
  {Santos}, {Santos}, {Sarazin}, {Sarkar}, {Sarmento}, {Sato}, {Scharf},
  {Scherini}, {Schieler}, {Schiffer}, {Schmidt}, {Scholten}, {Schoorlemmer},
  {Schov{\'a}nek}, {Schr{\"o}der}, {Schulz}, {Schulz}, {Schumacher}, {Sciutto},
  {Segreto}, {Settimo}, {Shadkam}, {Shellard}, {Sidelnik}, {Sigl}, {Sima},
  {{\'S}mia{\l}kowski}, {{\v{S}}m{\'\i}da}, {Snow}, {Sommers}, {Sorokin},
  {Squartini}, {Srivastava}, {Stani{\v{c}}}, {Stapleton}, {Stasielak},
  {Stephan}, {Stutz}, {Suarez}, {Suomij{\"a}rvi}, {Supanitsky}, {Sutherland},
  {Swain}, {Szadkowski}, {Szuba}, {Taborda}, {Tapia}, {Tepe}, {Theodoro},
  {Timmermans}, {Todero Peixoto}, {Toma}, {Tomankova}, {Tom{\'e}}, {Tonachini},
  {Torralba Elipe}, {Torres Machado}, {Travnicek}, {Trovato}, {Ulrich},
  {Unger}, {Urban}, {Vald{\'e}s Galicia}, {Vali{\~n}o}, {Valore}, {van Aar},
  {van Bodegom}, {van den Berg}, {van Velzen}, {van Vliet}, {Varela}, {Vargas
  C{\'a}rdenas}, {Varner}, {V{\'a}zquez}, {V{\'a}zquez}, {Veberi{\v{c}}},
  {Verzi}, {Vicha}, {Videla}, {Villase {\~n}or}, {Vlcek}, {Vorobiov},
  {Wahlberg}, {Wainberg}, {Walz}, {Watson}, {Weber}, {Weidenhaupt}, {Weindl},
  {Werner}, {Widom}, {Wiencke}, {Wilczy{\'n}ska}, {Wilczy{\'n}ski}, {Williams},
  {Winchen}, {Wittkowski}, {Wundheiler}, {Wykes}, {Yamamoto}, {Yapici}, {Yuan},
  {Yushkov}, {Zamorano}, {Zas}, {Zavrtanik}, {Zavrtanik}, {Zepeda}, {Zhou},
  {Zhu}, {Zimbres Silva}, {Ziolkowski}, {Zuccarello}, \& {The Pierre Auger
  Collaboration}}]{PierreAuger:2014yba}
{Aab}, A., {Abreu}, P., {Aglietta}, M., {et~al.} 2015, \apj, 804, 15,
  \dodoi{10.1088/0004-637X/804/1/15}

\bibitem[{{Aab} {et~al.}(2016){Aab}, {Abreu}, {Aglietta}, {Ahn}, {Samarai},
  {Albuquerque}, {Allekotte}, {Allison}, {Almela}, {Alvarez Castillo},
  {Alvarez-Mu{\~n}iz}, {Alves Batista}, {Ambrosio}, {Aminaei}, {Anchordoqui},
  {Andringa}, {Aramo}, {Arqueros}, {Arsene}, {Asorey}, {Assis}, {Aublin},
  {Ave}, {Avenier}, {Avila}, {Awal}, {Badescu}, {Barber}, {B{\"a}uml}, {Baus},
  {Beatty}, {Becker}, {Bellido}, {Berat}, {Bertaina}, {Bertou}, {Biermann},
  {Billoir}, {Blaess}, {Blanco}, {Blanco}, {Blazek}, {Bleve}, {Bl{\"u}mer},
  {Boh{\'a}{\v{c}}ov{\'a}}, {Boncioli}, {Bonifazi}, {Borodai}, {Brack},
  {Brancus}, {Bridgeman}, {Brogueira}, {Brown}, {Buchholz}, {Bueno}, {Buitink},
  {Buscemi}, {Caballero-Mora}, {Caccianiga}, {Caccianiga}, {Candusso},
  {Caramete}, {Caruso}, {Castellina}, {Cataldi}, {Cazon}, {Cester}, {Chavez},
  {Chiavassa}, {Chinellato}, {Chudoba}, {Cilmo}, {Clay}, {Cocciolo},
  {Colalillo}, {Coleman}, {Collica}, {Coluccia}, {Concei{\c{c}}{\~a}o},
  {Contreras}, {Cooper}, {Cordier}, {Coutu}, {Covault}, {Cronin}, {Dallier},
  {Daniel}, {Dasso}, {Daumiller}, {Dawson}, {de Almeida}, {de Jong}, {De
  Mauro}, {de Mello Neto}, {De Mitri}, {de Oliveira}, {de Souza}, {del Peral},
  {Deligny}, {Dembinski}, {Dhital}, {Di Giulio}, {Di Matteo}, {Diaz},
  {D{\'\i}az Castro}, {Diogo}, {Dobrigkeit}, {Docters}, {D'Olivo}, {Dorofeev},
  {Dorosti Hasankiadeh}, {Dova}, {Ebr}, {Engel}, {Erdmann}, {Erfani},
  {Escobar}, {Espadanal}, {Etchegoyen}, {Falcke}, {Fang}, {Farrar}, {Fauth},
  {Fazzini}, {Ferguson}, {Fernandes}, {Fick}, {Figueira}, {Filevich},
  {Filip{\v{c}}i{\v{c}}}, {Fox}, {Fratu}, {Freire}, {Fuchs}, {Fujii},
  {Garc{\'\i}a}, {Garcia-Pinto}, {Gate}, {Gemmeke}, {Gherghel-Lascu}, {Ghia},
  {Giaccari}, {Giammarchi}, {Giller}, {G{\l}as}, {Glaser}, {Glass}, {Golup},
  {G{\'o}mez Berisso}, {G{\'o}mez Vitale}, {Gonz{\'a}lez}, {Gookin}, {Gordon},
  {Gorgi}, {Gorham}, {Gouffon}, {Griffith}, {Grillo}, {Grubb}, {Guarino},
  {Guedes}, {Hampel}, {Hansen}, {Harari}, {Harrison}, {Hartmann}, {Harton},
  {Haungs}, {Hebbeker}, {Heck}, {Heimann}, {Hemery}, {Herve}, {Hill}, {Hojvat},
  {Hollon}, {Holt}, {Homola}, {H{\"o}randel}, {Horvath}, {Hrabovsk{\'y}},
  {Huber}, {Huege}, {Insolia}, {Isar}, {Jandt}, {Jansen}, {Jarne}, {Johnsen},
  {Josebachuili}, {K{\"a}{\"a}p{\"a}}, {Kambeitz}, {Kampert}, {Kasper},
  {Katkov}, {K{\'e}gl}, {Keilhauer}, {Keivani}, {Kemp}, {Kieckhafer}, {Klages},
  {Kleifges}, {Kleinfeller}, {Krause}, {Krohm}, {Kr{\"o}mer}, {Kuempel}, {Kukec
  Mezek}, {Kunka}, {LaHurd}, {Latronico}, {Lauer}, {Lauscher}, {Lautridou}, {Le
  Coz}, {Lebrun}, {Lebrun}, {Leigui de Oliveira}, {Letessier-Selvon},
  {Lhenry-Yvon}, {Link}, {Lopes}, {L{\'o}pez}, {L{\'o}pez Casado}, {Louedec},
  {Lu}, {Lucero}, {Malacari}, {Maldera}, {Mallamaci}, {Maller}, {Mandat},
  {Mantsch}, {Mariazzi}, {Marin}, {Mari{\c{s}}}, {Marsella}, {Martello},
  {Martin}, {Martinez}, {Mart{\'\i}nez Bravo}, {Martraire}, {Mas{\'\i}as Meza},
  {Mathes}, {Mathys}, {Matthews}, {Matthews}, {Matthiae}, {Maurizio},
  {Mayotte}, {Mazur}, {Medina}, {Medina-Tanco}, {Meissner}, {Mello}, {Melo},
  {Menshikov}, {Messina}, {Meyhandan}, {Micheletti}, {Middendorf}, {Minaya},
  {Miramonti}, {Mitrica}, {Molina-Bueno}, {Mollerach}, {Montanet}, {Morello},
  {Mostaf{\'a}}, {Moura}, {Muller}, {M{\"u}ller}, {M{\"u}ller}, {Mussa},
  {Navarra}, {Navas}, {Necesal}, {Nellen}, {Nelles}, {Neuser}, {Nguyen},
  {Niculescu-Oglinzanu}, {Niechciol}, {Niemietz}, {Niggemann}, {Nitz}, {Nosek},
  {Novotny}, {No{\v{z}}ka}, {Ochilo}, {Oikonomou}, {Olinto}, {Pacheco}, {Pakk
  Selmi-Dei}, {Palatka}, {Pallotta}, {Papenbreer}, {Parente}, {Parra}, {Paul},
  {Pech}, {P{\c{e}}kala}, {Pelayo}, {Pepe}, {Perrone}, {Petermann}, {Peters},
  {Petrera}, {Petrov}, {Phuntsok}, {Piegaia}, {Pierog}, {Pieroni}, {Pimenta},
  {Pirronello}, {Platino}, {Plum}, {Porcelli}, {Porowski}, {Prado},
  {Privitera}, {Prouza}, {Purrello}, {Quel}, {Querchfeld}, {Quinn},
  {Rautenberg}, {Ravel}, {Ravignani}, {Reinert}, {Revenu}, {Ridky}, {Riggi},
  {Risse}, {Ristori}, {Rizi}, {Rodrigues de Carvalho}, {Rodriguez Fernandez},
  {Rodriguez Rojo}, {Rodr{\'\i}guez-Fr{\'\i}as}, {Rogozin}, {Rosado}, {Roth},
  {Roulet}, {Rovero}, {Saffi}, {Saftoiu}, {Salamida}, {Salazar}, {Saleh},
  {Salesa Greus}, {Salina}, {S{\'a}nchez}, {Sanchez-Lucas}, {Santos}, {Santos},
  {Sarazin}, {Sarkar}, {Sarmento}, {Sato}, {Scarso}, {Schauer}, {Scherini},
  {Schieler}, {Schmidt}, {Scholten}, {Schoorlemmer}, {Schov{\'a}nek},
  {Schr{\"o}der}, {Schulz}, {Schulz}, {Schumacher}, {Sciutto}, {Segreto},
  {Settimo}, {Shadkam}, {Shellard}, {Sidelnik}, {Sigl}, {Sima},
  {{\'S}mia{\l}kowski}, {{\v{S}}m{\'\i}da}, {Snow}, {Sommers}, {Sorokin},
  {Squartini}, {Srivastava}, {Stanca}, {Stani{\v{c}}}, {Stapleton},
  {Stasielak}, {Stephan}, {Stutz}, {Suarez}, {Suomij{\"a}rvi}, {Supanitsky},
  {Sutherland}, {Swain}, {Szadkowski}, {Taborda}, {Tapia}, {Tepe}, {Theodoro},
  {Timmermans}, {Todero Peixoto}, {Toma}, {Tomankova}, {Tom{\'e}}, {Tonachini},
  {Torralba Elipe}, {Torres Machado}, {Travnicek}, {Trini}, {Ulrich}, {Unger},
  {Urban}, {Vald{\'e}s Galicia}, {Vali{\~n}o}, {Valore}, {van Aar}, {van
  Bodegom}, {van den Berg}, {van Velzen}, {van Vliet}, {Varela}, {Vargas
  C{\'a}rdenas}, {Varner}, {Vasquez}, {V{\'a}zquez}, {V{\'a}zquez},
  {Veberi{\v{c}}}, {Verzi}, {Vicha}, {Videla}, {Villase{\~n}or}, {Vlcek},
  {Vorobiov}, {Wahlberg}, {Wainberg}, {Walz}, {Watson}, {Weber}, {Weidenhaupt},
  {Weindl}, {Werner}, {Widom}, {Wiencke}, {Wilczy{\'n}ski}, {Winchen},
  {Wittkowski}, {Wundheiler}, {Wykes}, {Yang}, {Yapici}, {Yushkov}, {Zas},
  {Zavrtanik}, {Zavrtanik}, {Zepeda}, {Zhu}, {Zimmermann}, {Ziolkowski},
  {Zong}, {Zuccarello}, \& {The Pierre Auger Collaboration}}]{augerPrime2016}
---. 2016, arXiv e-prints, arXiv:1604.03637.
\newblock \doarXiv{1604.03637}

\bibitem[{{Aab} {et~al.}(2017){Aab}, {Abreu}, {Aglietta}, {Samarai},
  {Albuquerque}, {Allekotte}, {Almela}, {Alvarez Castillo},
  {Alvarez-Mu{\~n}iz}, {Anastasi}, {Anchordoqui}, {Andrada}, {Andringa},
  {Aramo}, {Arqueros}, {Arsene}, {Asorey}, {Assis}, {Aublin}, {Avila},
  {Badescu}, {Balaceanu}, {Barbato}, {Barreira Luz}, {Beatty}, {Becker},
  {Bellido}, {Berat}, {Bertaina}, {Bertou}, {Biermann}, {Billoir}, {Biteau},
  {Blaess}, {Blanco}, {Blazek}, {Bleve}, {Boh{\'a}{\v{c}}ov{\'a}}, {Boncioli},
  {Bonifazi}, {Borodai}, {Botti}, {Brack}, {Brancus}, {Bretz}, {Bridgeman},
  {Briechle}, {Buchholz}, {Bueno}, {Buitink}, {Buscemi}, {Caballero-Mora},
  {Caccianiga}, {Cancio}, {Canfora}, {Caramete}, {Caruso}, {Castellina},
  {Cataldi}, {Cazon}, {Chavez}, {Chinellato}, {Chudoba}, {Clay}, {Cobos},
  {Colalillo}, {Coleman}, {Collica}, {Coluccia}, {Concei{\c{c}}{\~a}o},
  {Consolati}, {Contreras}, {Cooper}, {Coutu}, {Covault}, {Cronin}, {D'Amico},
  {Daniel}, {Dasso}, {Daumiller}, {Dawson}, {de Almeida}, {de Jong}, {De
  Mauro}, {de Mello Neto}, {De Mitri}, {de Oliveira}, {de Souza}, {Debatin},
  {Deligny}, {Di Giulio}, {Di Matteo}, {D{\'\i}az Castro}, {Diogo},
  {Dobrigkeit}, {D'Olivo}, {Dorosti}, {dos Anjos}, {Dova}, {Dundovic}, {Ebr},
  {Engel}, {Erdmann}, {Erfani}, {Escobar}, {Espadanal}, {Etchegoyen}, {Falcke},
  {Farrar}, {Fauth}, {Fazzini}, {Fenu}, {Fick}, {Figueira},
  {Filip{\v{c}}i{\v{c}}}, {Fratu}, {Freire}, {Fujii}, {Fuster}, {Gaior},
  {Garc{\'\i}a}, {Garcia-Pinto}, {Gat{\'e}}, {Gemmeke}, {Gherghel-Lascu},
  {Ghia}, {Giaccari}, {Giammarchi}, {Giller}, {G{\l}as}, {Glaser}, {Golup},
  {G{\'o}mez Berisso}, {G{\'o}mez Vitale}, {Gonz{\'a}lez}, {Gorgi}, {Gorham},
  {Grillo}, {Grubb}, {Guarino}, {Guedes}, {Hampel}, {Hansen}, {Harari},
  {Harrison}, {Harton}, {Haungs}, {Hebbeker}, {Heck}, {Heimann}, {Herve},
  {Hill}, {Hojvat}, {Holt}, {Homola}, {H{\"o}randel}, {Horvath},
  {Hrabovsk{\'y}}, {Huege}, {Hulsman}, {Insolia}, {Isar}, {Jandt}, {Jansen},
  {Johnsen}, {Josebachuili}, {Jurysek}, {K{\"a}{\"a}p{\"a}}, {Kambeitz},
  {Kampert}, {Katkov}, {Keilhauer}, {Kemmerich}, {Kemp}, {Kemp}, {Kieckhafer},
  {Klages}, {Kleifges}, {Kleinfeller}, {Krause}, {Krohm}, {Kuempel}, {Kukec
  Mezek}, {Kunka}, {Kuotb Awad}, {LaHurd}, {Lauscher}, {Legumina}, {Leigui de
  Oliveira}, {Letessier-Selvon}, {Lhenry-Yvon}, {Link}, {Lo Presti}, {Lopes},
  {L{\'o}pez}, {L{\'o}pez Casado}, {Luce}, {Lucero}, {Malacari}, {Mallamaci},
  {Mandat}, {Mantsch}, {Mariazzi}, {Mari{\c{s}}}, {Marsella}, {Martello},
  {Martinez}, {Mart{\'\i}nez Bravo}, {Mas{\'\i}as Meza}, {Mathes}, {Mathys},
  {Matthews}, {Matthews}, {Matthiae}, {Mayotte}, {Mazur}, {Medina},
  {Medina-Tanco}, {Melo}, {Menshikov}, {Merenda}, {Michal}, {Micheletti},
  {Middendorf}, {Miramonti}, {Mitrica}, {Mockler}, {Mollerach}, {Montanet},
  {Morello}, {Mostaf{\'a}}, {M{\"u}ller}, {M{\"u}ller}, {Muller}, {M{\"u}ller},
  {Mussa}, {Naranjo}, {Nellen}, {Nguyen}, {Niculescu-Oglinzanu}, {Niechciol},
  {Niemietz}, {Niggemann}, {Nitz}, {Nosek}, {Novotny}, {No{\v{z}}ka},
  {N{\'u}{\~n}ez}, {Ochilo}, {Oikonomou}, {Olinto}, {Palatka}, {Pallotta},
  {Papenbreer}, {Parente}, {Parra}, {Paul}, {Pech}, {Pedreira}, {Pkala},
  {Pelayo}, {Pe{\~n}a-Rodriguez}, {Pereira}, {Perl{\'\i}n}, {Perrone},
  {Peters}, {Petrera}, {Phuntsok}, {Piegaia}, {Pierog}, {Pieroni}, {Pimenta},
  {Pirronello}, {Platino}, {Plum}, {Porowski}, {Prado}, {Privitera}, {Prouza},
  {Quel}, {Querchfeld}, {Quinn}, {Ramos-Pollan}, {Rautenberg}, {Ravignani},
  {Revenu}, {Ridky}, {Riehn}, {Risse}, {Ristori}, {Rizi}, {Rodrigues de
  Carvalho}, {Rodriguez Fernandez}, {Rodriguez Rojo}, {Rogozin}, {Roncoroni},
  {Roth}, {Roulet}, {Rovero}, {Ruehl}, {Saffi}, {Saftoiu}, {Salamida},
  {Salazar}, {Saleh}, {Salesa Greus}, {Salina}, {S{\'a}nchez}, {Sanchez-Lucas},
  {Santos}, {Santos}, {Sarazin}, {Sarmento}, {Sarmiento}, {Sato}, {Schauer},
  {Scherini}, {Schieler}, {Schimp}, {Schmidt}, {Scholten}, {Schov{\'a}nek},
  {Schr{\"o}der}, {Schulz}, {Schumacher}, {Sciutto}, {Segreto}, {Settimo},
  {Shadkam}, {Shellard}, {Sigl}, {Silli}, {Sima}, {{\'S}mia{\l}kowski},
  {{\v{S}}m{\'\i}da}, {Snow}, {Sommers}, {Sonntag}, {Sorokin}, {Squartini},
  {Stanca}, {Stani{\v{c}}}, {Stasielak}, {Stassi}, {Strafella}, {Suarez},
  {Suarez Dur{\'a}n}, {Sudholz}, {Suomij{\"a}rvi}, {Supanitsky},
  {{\v{S}}up{\'\i}k}, {Swain}, {Szadkowski}, {Taboada}, {Taborda}, {Tapia},
  {Theodoro}, {Timmermans}, {Todero Peixoto}, {Tomankova}, {Tom{\'e}},
  {Torralba Elipe}, {Travnicek}, {Trini}, {Ulrich}, {Unger}, {Urban},
  {Vald{\'e}s Galicia}, {Vali{\~n}o}, {Valore}, {van Aar}, {van Bodegom}, {van
  den Berg}, {van Vliet}, {Varela}, {Vargas C{\'a}rdenas}, {Varner},
  {V{\'a}zquez}, {Veberi{\v{c}}}, {Ventura}, {Vergara Quispe}, {Verzi},
  {Vicha}, {Villase{\~n}or}, {Vorobiov}, {Wahlberg}, {Wainberg}, {Walz},
  {Watson}, {Weber}, {Weindl}, {Wiencke}, {Wilczy{\'n}ski}, {Wirtz},
  {Wittkowski}, {Wundheiler}, {Yang}, {Yushkov}, {Zas}, {Zavrtanik},
  {Zavrtanik}, {Zepeda}, {Zimmermann}, {Ziolkowski}, {Zong}, {Zuccarello}, \&
  {The Pierre Auger Collaboration}}]{Aab:2017tyv}
---. 2017, Science, 357, 1266, \dodoi{10.1126/science.aan4338}

\bibitem[{Aab {et~al.}(2018)Aab, Abreu, Aglietta, Albuquerque, Allekotte,
  Almela, Castillo, Alvarez-Mu{\~{n}}iz, Anastasi, Anchordoqui, Andrada,
  Andringa, Aramo, Arsene, Asorey, Assis, Avila, Badescu, Balaceanu, Barbato,
  Luz, Beatty, Becker, Bellido, Berat, Bertaina, Bertou, Biermann, Biteau,
  Blaess, Blanco, Blazek, Bleve, Boh{\'{a}}{\v{c}}ov{\'{a}}, Bonifazi, Borodai,
  Botti, Brack, Brancus, Bretz, Bridgeman, Briechle, Buchholz, Bueno, Buitink,
  Buscemi, Caballero-Mora, Caccianiga, Cancio, Canfora, Caruso, Castellina,
  Catalani, Cataldi, Cazon, Chavez, Chinellato, Chudoba, Clay, Cerutti,
  Colalillo, Coleman, Collica, Coluccia, Concei{\c{c}}{\~{a}}o, Consolati,
  Contreras, Cooper, Coutu, Covault, Cronin, D'Amico, Daniel, Dasso, Daumiller,
  Dawson, de~Almeida, de~Jong, Mauro, de~Mello~Neto, Mitri, de~Oliveira,
  de~Souza, Debatin, Deligny, Castro, Diogo, Dobrigkeit, D'Olivo, Dorosti, dos
  Anjos, Dova, Dundovic, Ebr, Engel, Erdmann, Erfani, Escobar, Espadanal,
  Etchegoyen, Falcke, Farmer, Farrar, Fauth, Fazzini, Fenu, Fick, Figueira,
  Filip{\v{c}}i{\v{c}}, Freire, Fujii, Fuster, Gaïor, Garc{\'{\i}}a,
  Gat{\'{e}}, Gemmeke, Gherghel-Lascu, Ghia, Giaccari, Giammarchi, Giller,
  G{\l}as, Glaser, Golup, Berisso, Vitale, Gonz{\'{a}}lez, Gorgi, Grillo,
  Grubb, Guarino, Guedes, Halliday, Hampel, Hansen, Harari, Harrison, Haungs,
  Hebbeker, Heck, Heimann, Herve, Hill, Hojvat, Holt, Homola, Hörandel,
  Horvath, Hrabovsk{\'{y}}, Huege, Hulsman, Insolia, Isar, Jandt, Johnsen,
  Josebachuili, Jurysek, Kääpä, Kambeitz, Kampert, Keilhauer, Kemmerich,
  Kemp, Kemp, Kieckhafer, Klages, Kleifges, Kleinfeller, Krause, Krohm,
  Kuempel, Mezek, Kunka, Awad, Lago, LaHurd, Lang, Lauscher, Legumina,
  de~Oliveira, Letessier-Selvon, Lhenry-Yvon, Link, Presti, Lopes, L{\'{o}}pez,
  Casado, Lorek, Luce, Lucero, Malacari, Mallamaci, Mandat, Mantsch, Mariazzi,
  Mari{\c{s}}, Marsella, Martello, Martinez, Bravo, Meza, Mathes, Mathys,
  Matthews, Matthiae, Mayotte, Mazur, Medina, Medina-Tanco, Melo, Menshikov,
  Merenda, Michal, Micheletti, Middendorf, Miramonti, Mitrica, Mockler,
  Mollerach, Montanet, Morello, Morlino, Mostaf{\'{a}}, Müller, Müller,
  Muller, Müller, Mussa, Naranjo, Nellen, Nguyen, Niculescu-Oglinzanu,
  Niechciol, Niemietz, Niggemann, Nitz, Nosek, Novotny, No{\v{z}}ka,
  N{\'{u}}{\~{n}}ez, Oikonomou, Olinto, Palatka, Pallotta, Papenbreer, Parente,
  Parra, Paul, Pech, Pedreira, Pȩkala, Pelayo, Pe{\~{n}}a-Rodriguez, Pereira,
  Perlin, Perrone, Peters, Petrera, Phuntsok, Pierog, Pimenta, Pirronello,
  Platino, Plum, Poh, Porowski, Prado, Privitera, Prouza, Quel, Querchfeld,
  Quinn, Ramos-Pollan, Rautenberg, Ravignani, Ridky, Riehn, Risse, Ristori,
  Rizi, de~Carvalho, Fernandez, Rojo, Roncoroni, Roth, Roulet, Rovero, Ruehl,
  Saffi, Saftoiu, Salamida, Salazar, Saleh, Salina, S{\'{a}}nchez,
  Sanchez-Lucas, Santos, Santos, Sarazin, Sarmento, Sarmiento-Cano, Sato,
  Schauer, Scherini, Schieler, Schimp, Schmidt, Scholten, Schov{\'{a}}nek,
  Schröder, Schröder, Schulz, Schumacher, Sciutto, Segreto, Shadkam,
  Shellard, Sigl, Silli, {\v{S}}m{\'{\i}}da, Snow, Sommers, Sonntag, Soriano,
  Squartini, Stanca, Stani{\v{c}}, Stasielak, Stassi, Stolpovskiy, Strafella,
  Streich, Suarez, Dur{\'{a}}n, Sudholz, Suomijärvi, Supanitsky,
  {\v{S}}up{\'{\i}}k, Swain, Szadkowski, Taboada, Taborda, Theodoro,
  Timmermans, Peixoto, Tomankova, Tom{\'{e}}, Elipe, Travnicek, Trini, Ulrich,
  Unger, Urban, Galicia, Vali{\~{n}}o, Valore, van Aar, van Bodegom, van~den
  Berg, van Vliet, Varela, C{\'{a}}rdenas, V{\'{a}}zquez, Veberi{\v{c}},
  Ventura, Quispe, Verzi, Vicha, Villase{\~{n}}or, Vorobiov, Wahlberg,
  Wainberg, Walz, Watson, Weber, Weindl, Wiede{\'{n}}ski, Wiencke,
  Wilczy{\'{n}}ski, Wirtz, Wittkowski, Wundheiler, Yang, Yushkov, Zas,
  Zavrtanik, Zavrtanik, Zepeda, Zimmermann, Ziolkowski, Zong, \&
  and}]{Aab:2018chp}
Aab, A., Abreu, P., Aglietta, M., {et~al.} 2018, The Astrophysical Journal,
  853, L29, \dodoi{10.3847/2041-8213/aaa66d}

\bibitem[{{Aab} {et~al.}(2020){Aab}, {Abreu}, {Aglietta}, {Albury},
  {Allekotte}, {Almela}, {Alvarez Castillo}, {Alvarez-Mu{\~n}iz}, {Alves
  Batista}, {Anastasi}, {Anchordoqui}, {Andrada}, {Andringa}, {Aramo},
  {Ara{\'u}jo Ferreira}, {Asorey}, {Assis}, {Avila}, {Badescu}, {Bakalova},
  {Balaceanu}, {Barbato}, {Barreira Luz}, {Becker}, {Bellido}, {Berat},
  {Bertaina}, {Bertou}, {Biermann}, {Billoir}, {Bister}, {Biteau}, {Blanco},
  {Blazek}, {Bleve}, {Boh{\'a}{\v{c}}ov{\'a}}, {Boncioli}, {Bonifazi}, {Bonneau
  Arbeletche}, {Borodai}, {Botti}, {Brack}, {Bretz}, {Bridgeman}, {Briechle},
  {Buchholz}, {Bueno}, {Buitink}, {Buscemi}, {Caballero-Mora}, {Caccianiga},
  {Calcagni}, {Cancio}, {Canfora}, {Caracas}, {Carceller}, {Caruso},
  {Castellina}, {Catalani}, {Cataldi}, {Cazon}, {Cerda}, {Chinellato}, {Choi},
  {Chudoba}, {Chytka}, {Clay}, {Cobos Cerutti}, {Colalillo}, {Coleman},
  {Coluccia}, {Concei{\c{c}}{\~a}o}, {Condorelli}, {Consolati}, {Contreras},
  {Convenga}, {Covault}, {Dasso}, {Daumiller}, {Dawson}, {Day}, {de Almeida},
  {de Jes{\'u}s}, {de Jong}, {De Mauro}, {de Mello Neto}, {De Mitri}, {de
  Oliveira}, {de Oliveira Franco}, {de Souza}, {De Vito}, {Debatin}, {del
  R{\'\i}o}, {Deligny}, {Dhital}, {Di Matteo}, {D{\'\i}az Castro},
  {Dobrigkeit}, {D'Olivo}, {Dorosti}, {dos Anjos}, {Dova}, {Ebr}, {Engel},
  {Epicoco}, {Erdmann}, {Escobar}, {Etchegoyen}, {Falcke}, {Farmer}, {Farrar},
  {Fauth}, {Fazzini}, {Feldbusch}, {Fenu}, {Fick}, {Figueira},
  {Filip{\v{c}}i{\v{c}}}, {Fodran}, {Freire}, {Fujii}, {Fuster}, {Galea},
  {Galelli}, {Garc{\'\i}a}, {Garcia Vegas}, {Gemmeke}, {Gesualdi},
  {Gherghel-Lascu}, {Ghia}, {Giaccari}, {Giammarchi}, {Giller}, {Glombitza},
  {Gobbi}, {Gollan}, {Golup}, {G{\'o}mez Berisso}, {G{\'o}mez Vitale},
  {Gongora}, {Gonz{\'a}lez}, {Goos}, {G{\'o}ra}, {Gorgi}, {Gottowik}, {Grubb},
  {Guarino}, {Guedes}, {Guido}, {Hahn}, {Halliday}, {Hampel}, {Hansen},
  {Harari}, {Harvey}, {Haungs}, {Hebbeker}, {Heck}, {Hill}, {Hojvat},
  {H{\"o}randel}, {Horvath}, {Hrabovsk{\'y}}, {Huege}, {Hulsman}, {Insolia},
  {Isar}, {Johnsen}, {Jurysek}, {K{\"a}{\"a}p{\"a}}, {Kampert}, {Keilhauer},
  {Kemp}, {Klages}, {Kleifges}, {Kleinfeller}, {K{\"o}pke}, {Kukec Mezek},
  {Lago}, {LaHurd}, {Lang}, {Leigui de Oliveira}, {Lenok}, {Letessier-Selvon},
  {Lhenry-Yvon}, {Lo Presti}, {Lopes}, {L{\'o}pez}, {Lorek}, {Luce}, {Lucero},
  {Machado Payeras}, {Malacari}, {Mancarella}, {Mandat}, {Manning},
  {Manshanden}, {Mantsch}, {Marafico}, {Mariazzi}, {Mari{\c{s}}}, {Marsella},
  {Martello}, {Martinez}, {Mart{\'\i}nez Bravo}, {Mastrodicasa}, {Mathes},
  {Matthews}, {Matthiae}, {Mayotte}, {Mazur}, {Medina-Tanco}, {Melo},
  {Menshikov}, {Merenda}, {Michal}, {Micheletti}, {Miramonti}, {Mockler},
  {Mollerach}, {Montanet}, {Morello}, {Mostaf{\'a}}, {M{\"u}ller}, {Muller},
  {Mulrey}, {Mussa}, {Muzio}, {Namasaka}, {Nellen}, {Niculescu-Oglinzanu},
  {Niechciol}, {Nitz}, {Nosek}, {Novotny}, {No{\v{z}}ka}, {Nucita},
  {N{\'u}{\~n}ez}, {Palatka}, {Pallotta}, {Panetta}, {Papenbreer}, {Parente},
  {Parra}, {Pech}, {Pedreira}, {P{\c{e}}kala}, {Pelayo}, {Pe{\~n}a-Rodriguez},
  {Perez Armand}, {Perlin}, {Perrone}, {Peters}, {Petrera}, {Pierog},
  {Pimenta}, {Pirronello}, {Platino}, {Pont}, {Pothast}, {Privitera}, {Prouza},
  {Puyleart}, {Querchfeld}, {Rautenberg}, {Ravignani}, {Reininghaus}, {Ridky},
  {Riehn}, {Risse}, {Ristori}, {Rizi}, {Rodrigues de Carvalho}, {Rodriguez
  Rojo}, {Roncoroni}, {Roth}, {Roulet}, {Rovero}, {Ruehl}, {Saffi}, {Saftoiu},
  {Salamida}, {Salazar}, {Salina}, {Sanabria Gomez}, {S{\'a}nchez}, {Santos},
  {Santos}, {Sarazin}, {Sarmento}, {Sarmiento-Cano}, {Sato}, {Savina},
  {Sch{\"a}fer}, {Scherini}, {Schieler}, {Schimassek}, {Schimp},
  {Schl{\"u}ter}, {Schmidt}, {Scholten}, {Schov{\'a}nek}, {Schr{\"o}der},
  {Schr{\"o}der}, {Schulz}, {Sciutto}, {Scornavacche}, {Shellard}, {Sigl},
  {Silli}, {Sima}, {{\v{S}}m{\'\i}da}, {Sommers}, {Soriano}, {Souchard},
  {Squartini}, {Stadelmaier}, {Stanca}, {Stani{\v{c}}}, {Stasielak}, {Stassi},
  {Streich}, {Su{\'a}rez-Dur{\'a}n}, {Sudholz}, {Suomij{\"a}rvi}, {Supanitsky},
  {{\v{S}}up{\'\i}k}, {Szadkowski}, {Taboada}, {Tapia}, {Timmermans},
  {Tkachenko}, {Tobiska}, {Todero Peixoto}, {Tom{\'e}}, {Torralba Elipe},
  {Travaini}, {Travnicek}, {Trimarelli}, {Trini}, {Tueros}, {Ulrich}, {Unger},
  {Urban}, {Vaclavek}, {Vacula}, {Vald{\'e}s Galicia}, {Vali{\~n}o}, {Valore},
  {van Vliet}, {Varela}, {Vargas C{\'a}rdenas}, {V{\'a}squez-Ram{\'\i}rez},
  {Veberi{\v{c}}}, {Ventura}, {Vergara Quispe}, {Verzi}, {Vicha},
  {Villase{\~n}or}, {Vink}, {Vorobiov}, {Wahlberg}, {Watson}, {Weber},
  {Weindl}, {Wiencke}, {Wilczy{\'n}ski}, {Winchen}, {Wirtz}, {Wittkowski},
  {Wundheiler}, {Yushkov}, {Zapparrata}, {Zas}, {Zavrtanik}, {Zavrtanik},
  {Zehrer}, {Zepeda}, {Ziolkowski}, \& {Zuccarello}}]{ThePierreAuger:2020SDrec}
{Aab}, A., {Abreu}, P., {Aglietta}, M., {et~al.} 2020, Journal of
  Instrumentation, 15, P10021, \dodoi{10.1088/1748-0221/15/10/P10021}

\bibitem[{Aab {et~al.}(2020)Aab, Abreu, Aglietta, Albury, Allekotte, Almela,
  Castillo, Alvarez-Mu{\~{n}}iz, Batista, Anastasi, Anchordoqui, Andrada,
  Andringa, Aramo, Ferreira, Asorey, Assis, Avila, Badescu, Bakalova,
  Balaceanu, Barbato, Luz, Becker, Bellido, Berat, Bertaina, Bertou, Biermann,
  Bister, Biteau, Blanco, Blazek, Bleve, Boh{\'{a}}{\v{c}}ov{\'{a}}, Boncioli,
  Bonifazi, Arbeletche, Borodai, Botti, Brack, Bretz, Briechle, Buchholz,
  Bueno, Buitink, Buscemi, Caballero-Mora, Caccianiga, Calcagni, Cancio,
  Canfora, Caracas, Carceller, Caruso, Castellina, Catalani, Cataldi, Cazon,
  Cerda, Chinellato, Choi, Chudoba, Chytka, Clay, Cerutti, Colalillo, Coleman,
  Coluccia, Concei{\c{c}}{\~{a}}o, Condorelli, Consolati, Contreras, Convenga,
  Covault, Dasso, Daumiller, Dawson, Day, de~Almeida, de~Jes{\'{u}}s, de~Jong,
  Mauro, de~Mello~Neto, Mitri, de~Oliveira, de~Oliveira~Franco, de~Souza,
  Debatin, del R{\'{\i}}o, Deligny, Dhital, Matteo, Castro, Dobrigkeit,
  D{\textquotesingle}Olivo, Dorosti, dos Anjos, Dova, Ebr, Engel, Epicoco,
  Erdmann, Escobar, Etchegoyen, Falcke, Farmer, Farrar, Fauth, Fazzini,
  Feldbusch, Fenu, Fick, Figueira, Filip{\v{c}}i{\v{c}}, Fodran, Freire, Fujii,
  Fuster, Galea, Galelli, Garc{\'{\i}}a, Vegas, Gemmeke, Gesualdi,
  Gherghel-Lascu, Ghia, Giaccari, Giammarchi, Giller, Glombitza, Gobbi, Golup,
  Berisso, Vitale, Gongora, Gonz{\'{a}}lez, Goos, G{\'{o}}ra, Gorgi, Gottowik,
  Grubb, Guarino, Guedes, Guido, Hahn, Halliday, Hampel, Hansen, Harari,
  Harvey, Haungs, Hebbeker, Heck, Hill, Hojvat, Hörandel, Horvath,
  Hrabovsk{\'{y}}, Huege, Hulsman, Insolia, Isar, Johnsen, Jurysek, Kääpä,
  Kampert, Keilhauer, Kemp, Klages, Kleifges, Kleinfeller, Köpke, Mezek, Lago,
  LaHurd, Lang, de~Oliveira, Lenok, Letessier-Selvon, Lhenry-Yvon, Presti,
  Lopes, L{\'{o}}pez, Lorek, Luce, Lucero, Payeras, Malacari, Mancarella,
  Mandat, Manning, Manshanden, Mantsch, Marafico, Mariazzi, Maris, Marsella,
  Martello, Martinez, Bravo, Mastrodicasa, Mathes, Matthews, Matthiae, Mayotte,
  Mazur, Medina-Tanco, Melo, Menshikov, Merenda, Michal, Micheletti, Miramonti,
  Mockler, Mollerach, Montanet, Morello, Mostaf{\'{a}}, Müller, Muller,
  Mulrey, Mussa, Muzio, Namasaka, Nellen, Niculescu-Oglinzanu, Niechciol, Nitz,
  Nosek, Novotny, No{\v{z}}ka, Nucita, N{\'{u}}{\~{n}}ez, Palatka, Pallotta,
  Panetta, Papenbreer, Parente, Parra, Pech, Pedreira, Pekala, Pelayo,
  Pe{\~{n}}a-Rodriguez, Armand, Perlin, Perrone, Peters, Petrera, Pierog,
  Pimenta, Pirronello, Platino, Pont, Pothast, Privitera, Prouza, Puyleart,
  Querchfeld, Rautenberg, Ravignani, Reininghaus, Ridky, Riehn, Risse, Ristori,
  Rizi, de~Carvalho, Rojo, Roncoroni, Roth, Roulet, Rovero, Ruehl, Saffi,
  Saftoiu, Salamida, Salazar, Salina, Gomez, S{\'{a}}nchez, Santos, Santos,
  Sarazin, Sarmento, Sarmiento-Cano, Sato, Savina, Schäfer, Scherini,
  Schieler, Schimassek, Schimp, Schlüter, Schmidt, Scholten, Schov{\'{a}}nek,
  Schröder, Schröder, Sciutto, Scornavacche, Shellard, Sigl, Silli, Sima,
  Sm{\'{\i}}da, Sommers, Soriano, Souchard, Squartini, Stadelmaier, Stanca,
  Stani{\v{c}}, Stasielak, Stassi, Streich, Su{\'{a}}rez-Dur{\'{a}}n, Sudholz,
  Suomijärvi, Supanitsky, Sup{\'{\i}}k, Szadkowski, Taboada, Tapia,
  Timmermans, Tobiska, Peixoto, Tom{\'{e}}, Elipe, Travaini, Travnicek,
  Trimarelli, Trini, Tueros, Ulrich, Unger, Urban, Vaclavek, Vacula, Galicia,
  Vali{\~{n}}o, Valore, van Vliet, Varela, C{\'{a}}rdenas,
  V{\'{a}}squez-Ram{\'{\i}}rez, Veberi{\v{c}}, Ventura, Quispe, Verzi, Vicha,
  Villase{\~{n}}or, Vink, Vorobiov, Wahlberg, Watson, Weber, Weindl, Wiencke,
  Wilczy{\'{n}}ski, Winchen, Wirtz, Wittkowski, Wundheiler, Yushkov,
  Zapparrata, Zas, Zavrtanik, Zavrtanik, Zehrer, Zepeda, Ziolkowski, \&
  Zuccarello}]{Aab_2020_magn}
Aab, A., Abreu, P., Aglietta, M., {et~al.} 2020, Journal of Cosmology and
  Astroparticle Physics, 2020, 017, \dodoi{10.1088/1475-7516/2020/06/017}

\bibitem[{{Aartsen} {et~al.}(2014{\natexlab{a}}){Aartsen}, {Abbasi},
  {Ackermann}, {Adams}, {Aguilar}, {Ahlers}, {Altmann}, {Arguelles},
  {Auffenberg}, {Bai}, {Baker}, {Barwick}, {Baum}, {Bay}, {Beatty}, {Becker
  Tjus}, {Becker}, {BenZvi}, {Berghaus}, {Berley}, {Bernardini}, {Bernhard},
  {Besson}, {Binder}, {Bindig}, {Bissok}, {Blaufuss}, {Blumenthal}, {Boersma},
  {Bohm}, {Bose}, {B{\"o}ser}, {Botner}, {Brayeur}, {Bretz}, {Brown}, {Bruijn},
  {Casey}, {Casier}, {Chirkin}, {Christov}, {Christy}, {Clark}, {Classen},
  {Clevermann}, {Coenders}, {Cohen}, {Cowen}, {Cruz Silva}, {Danninger},
  {Daughhetee}, {Davis}, {Day}, {De Clercq}, {De Ridder}, {Desiati}, {de
  Vries}, {de With}, {DeYoung}, {D{\'\i}az-V{\'e}lez}, {Dunkman}, {Eagan},
  {Eberhardt}, {Eichmann}, {Eisch}, {Euler}, {Evenson}, {Fadiran}, {Fazely},
  {Fedynitch}, {Feintzeig}, {Feusels}, {Filimonov}, {Finley}, {Fischer-Wasels},
  {Flis}, {Franckowiak}, {Frantzen}, {Fuchs}, {Gaisser}, {Gallagher},
  {Gerhardt}, {Gladstone}, {Gl{\"u}senkamp}, {Goldschmidt}, {Golup},
  {Gonzalez}, {Goodman}, {G{\'o}ra}, {Grandmont}, {Grant}, {Gretskov}, {Groh},
  {Gro{\ss}}, {Ha}, {Haj Ismail}, {Hallen}, {Hallgren}, {Halzen}, {Hanson},
  {Hebecker}, {Heereman}, {Heinen}, {Helbing}, {Hellauer}, {Hickford}, {Hill},
  {Hoffman}, {Hoffmann}, {Homeier}, {Hoshina}, {Huang}, {Huelsnitz}, {Hulth},
  {Hultqvist}, {Hussain}, {Ishihara}, {Jackson}, {Jacobi}, {Jacobsen},
  {Jagielski}, {Japaridze}, {Jero}, {Jlelati}, {Kaminsky}, {Kappes}, {Karg},
  {Karle}, {Kauer}, {Kelley}, {Kiryluk}, {Kl{\"a}s}, {Klein}, {K{\"o}hne},
  {Kohnen}, {Kolanoski}, {K{\"o}pke}, {Kopper}, {Kopper}, {Koskinen},
  {Kowalski}, {Krasberg}, {Kriesten}, {Krings}, {Kroll}, {Kunnen}, {Kurahashi},
  {Kuwabara}, {Labare}, {Landsman}, {Larson}, {Lesiak-Bzdak}, {Leuermann},
  {Leute}, {L{\"u}nemann}, {Mac{\'\i}as}, {Madsen}, {Maggi}, {Maruyama},
  {Mase}, {Matis}, {McNally}, {Meagher}, {Merck}, {Meures}, {Miarecki},
  {Middell}, {Milke}, {Miller}, {Mohrmann}, {Montaruli}, {Morse}, {Nahnhauer},
  {Naumann}, {Niederhausen}, {Nowicki}, {Nygren}, {Obertacke}, {Odrowski},
  {Olivas}, {Omairat}, {O'Murchadha}, {Paul}, {Pepper}, {P{\'e}rez de los
  Heros}, {Pfendner}, {Pieloth}, {Pinat}, {Posselt}, {Price}, {Przybylski},
  {Quinnan}, {R{\"a}del}, {Rameez}, {Rawlins}, {Redl}, {Reimann}, {Resconi},
  {Rhode}, {Ribordy}, {Richman}, {Riedel}, {Robertson}, {Rodrigues}, {Rott},
  {Ruhe}, {Ruzybayev}, {Ryckbosch}, {Saba}, {Sander}, {Santander}, {Sarkar},
  {Schatto}, {Scheriau}, {Schmidt}, {Schmitz}, {Schoenen}, {Sch{\"o}neberg},
  {Sch{\"o}nwald}, {Schukraft}, {Schulte}, {Schulz}, {Seckel}, {Sestayo},
  {Seunarine}, {Shanidze}, {Sheremata}, {Smith}, {Soldin}, {Spiczak},
  {Spiering}, {Stamatikos}, {Stanev}, {Stanisha}, {Stasik}, {Stezelberger},
  {Stokstad}, {St{\"o}{\ss}l}, {Strahler}, {Str{\"o}m}, {Strotjohann},
  {Sullivan}, {Taavola}, {Taboada}, {Tamburro}, {Tepe}, {Ter-Antonyan},
  {Te\{{\v{s}}\}i{\'c}}, {Tilav}, {Toale}, {Tobin}, {Toscano}, {Tselengidou},
  {Unger}, {Usner}, {Vallecorsa}, {van Eijndhoven}, {Van Overloop}, {van
  Santen}, {Vehring}, {Voge}, {Vraeghe}, {Walck}, {Waldenmaier}, {Wallraff},
  {Weaver}, {Wellons}, {Wendt}, {Westerhoff}, {Whelan}, {Whitehorn}, {Wiebe},
  {Wiebusch}, {Williams}, {Wissing}, {Wolf}, {Wood}, {Woschnagg}, {Xu}, {Xu},
  {Yanez}, {Yodh}, {Yoshida}, {Zarzhitsky}, {Ziemann}, {Zierke}, \&
  {Zoll}}]{icEnergyReco2014}
{Aartsen}, M.~G., {Abbasi}, R., {Ackermann}, M., {et~al.} 2014{\natexlab{a}},
  Journal of Instrumentation, 9, P03009, \dodoi{10.1088/1748-0221/9/03/P03009}

\bibitem[{{Aartsen} {et~al.}(2014{\natexlab{b}}){Aartsen}, {Ackermann},
  {Adams}, {Aguilar}, {Ahlers}, {Ahrens}, {Altmann}, {Anderson}, {Arguelles},
  {Arlen}, {Auffenberg}, {Bai}, {Barwick}, {Baum}, {Beatty}, {Becker Tjus},
  {Becker}, {BenZvi}, {Berghaus}, {Berley}, {Bernardini}, {Bernhard}, {Besson},
  {Binder}, {Bindig}, {Bissok}, {Blaufuss}, {Blumenthal}, {Boersma}, {Bohm},
  {Bose}, {B{\"o}ser}, {Botner}, {Brayeur}, {Bretz}, {Brown}, {Casey},
  {Casier}, {Chirkin}, {Christov}, {Christy}, {Clark}, {Classen}, {Clevermann},
  {Coenders}, {Cowen}, {Cruz Silva}, {Danninger}, {Daughhetee}, {Davis}, {Day},
  {de Andr{\'e}}, {De Clercq}, {De Ridder}, {Desiati}, {de Vries}, {de With},
  {DeYoung}, {D{\'\i}az-V{\'e}lez}, {Dunkman}, {Eagan}, {Eberhardt},
  {Eichmann}, {Eisch}, {Euler}, {Evenson}, {Fadiran}, {Fazely}, {Fedynitch},
  {Feintzeig}, {Felde}, {Feusels}, {Filimonov}, {Finley}, {Fischer-Wasels},
  {Flis}, {Franckowiak}, {Frantzen}, {Fuchs}, {Gaisser}, {Gallagher},
  {Gerhardt}, {Gier}, {Gladstone}, {Gl{\"u}senkamp}, {Goldschmidt}, {Golup},
  {Gonzalez}, {Goodman}, {G{\'o}ra}, {Grandmont}, {Grant}, {Gretskov}, {Groh},
  {Gro{\ss}}, {Ha}, {Haack}, {Haj Ismail}, {Hallen}, {Hallgren}, {Halzen},
  {Hanson}, {Hebecker}, {Heereman}, {Heinen}, {Helbing}, {Hellauer}, {Hellwig},
  {Hickford}, {Hill}, {Hoffman}, {Hoffmann}, {Homeier}, {Hoshina}, {Huang},
  {Huelsnitz}, {Hulth}, {Hultqvist}, {Hussain}, {Ishihara}, {Jacobi},
  {Jacobsen}, {Jagielski}, {Japaridze}, {Jero}, {Jlelati}, {Jurkovic},
  {Kaminsky}, {Kappes}, {Karg}, {Karle}, {Kauer}, {Kelley}, {Kheirandish},
  {Kiryluk}, {Kl{\"a}s}, {Klein}, {K{\"o}hne}, {Kohnen}, {Kolanoski}, {Koob},
  {K{\"o}pke}, {Kopper}, {Kopper}, {Koskinen}, {Kowalski}, {Kriesten},
  {Krings}, {Kroll}, {Kunnen}, {Kurahashi}, {Kuwabara}, {Labare}, {Larsen},
  {Larson}, {Lesiak-Bzdak}, {Leuermann}, {Leute}, {L{\"u}nemann},
  {Mac{\'\i}as}, {Madsen}, {Maggi}, {Maruyama}, {Mase}, {Matis}, {McNally},
  {Meagher}, {Meli}, {Meures}, {Miarecki}, {Middell}, {Middlemas}, {Milke},
  {Miller}, {Mohrmann}, {Montaruli}, {Morse}, {Nahnhauer}, {Naumann},
  {Niederhausen}, {Nowicki}, {Nygren}, {Obertacke}, {Odrowski}, {Olivas},
  {Omairat}, {O'Murchadha}, {Palczewski}, {Paul}, {Penek}, {Pepper}, {P{\'e}rez
  de los Heros}, {Pfendner}, {Pieloth}, {Pinat}, {Posselt}, {Price},
  {Przybylski}, {P{\"u}tz}, {Quinnan}, {R{\"a}del}, {Rameez}, {Rawlins},
  {Redl}, {Rees}, {Reimann}, {Resconi}, {Rhode}, {Richman}, {Riedel},
  {Robertson}, {Rodrigues}, {Rongen}, {Rott}, {Ruhe}, {Ruzybayev}, {Ryckbosch},
  {Saba}, {Sander}, {Santander}, {Sarkar}, {Schatto}, {Scheriau}, {Schmidt},
  {Schmitz}, {Schoenen}, {Sch{\"o}neberg}, {Sch{\"o}nwald}, {Schukraft},
  {Schulte}, {Schulz}, {Seckel}, {Sestayo}, {Seunarine}, {Shanidze},
  {Sheremata}, {Smith}, {Soldin}, {Spiczak}, {Spiering}, {Stamatikos},
  {Stanev}, {Stanisha}, {Stasik}, {Stezelberger}, {Stokstad}, {St{\"o}{\ss}l},
  {Strahler}, {Str{\"o}m}, {Strotjohann}, {Sullivan}, {Taavola}, {Taboada},
  {Tamburro}, {Tepe}, {Ter-Antonyan}, {Terliuk}, {Te{\v{s}}i{\'c}}, {Tilav},
  {Toale}, {Tobin}, {Tosi}, {Tselengidou}, {Unger}, {Usner}, {Vallecorsa}, {van
  Eijndhoven}, {Vandenbroucke}, {van Santen}, {Vehring}, {Voge}, {Vraeghe},
  {Walck}, {Wallraff}, {Weaver}, {Wellons}, {Wendt}, {Westerhoff}, {Whelan},
  {Whitehorn}, {Wichary}, {Wiebe}, {Wiebusch}, {Williams}, {Wissing}, {Wolf},
  {Wood}, {Woschnagg}, {Xu}, {Xu}, {Yanez}, {Yodh}, {Yoshida}, {Zarzhitsky},
  {Ziemann}, {Zierke}, {Zoll}, \& {The IceCube
  Collaboration}}]{Aartsen:2014gkd}
{Aartsen}, M.~G., {Ackermann}, M., {Adams}, J., {et~al.} 2014{\natexlab{b}},
  \prl, 113, 101101, \dodoi{10.1103/PhysRevLett.113.101101}

\bibitem[{{Aartsen} {et~al.}(2014{\natexlab{c}}){Aartsen}, {Ackermann},
  {Adams}, {Aguilar}, {Ahlers}, {Ahrens}, {Altmann}, {Anderson}, {Arguelles},
  {Arlen}, {Auffenberg}, {Bai}, {Barwick}, {Baum}, {Beatty}, {Becker Tjus},
  {Becker}, {BenZvi}, {Berghaus}, {Berley}, {Bernardini}, {Bernhard}, {Besson},
  {Binder}, {Bindig}, {Bissok}, {Blaufuss}, {Blumenthal}, {Boersma}, {Bohm},
  {Bos}, {Bose}, {B{\"o}ser}, {Botner}, {Brayeur}, {Bretz}, {Brown}, {Casey},
  {Casier}, {Cheung}, {Chirkin}, {Christov}, {Christy}, {Clark}, {Classen},
  {Clevermann}, {Coenders}, {Cowen}, {Cruz Silva}, {Danninger}, {Daughhetee},
  {Davis}, {Day}, {de Andr{\'e}}, {De Clercq}, {De Ridder}, {Desiati}, {de
  Vries}, {de With}, {DeYoung}, {D{\'\i}az-V{\'e}lez}, {Dunkman}, {Eagan},
  {Eberhardt}, {Eichmann}, {Eisch}, {Euler}, {Evenson}, {Fadiran}, {Fazely},
  {Fedynitch}, {Feintzeig}, {Felde}, {Feusels}, {Filimonov}, {Finley},
  {Fischer-Wasels}, {Flis}, {Franckowiak}, {Frantzen}, {Fuchs}, {Gaisser},
  {Gallagher}, {Gerhardt}, {Gier}, {Gladstone}, {Gl{\"u}senkamp},
  {Goldschmidt}, {Golup}, {Gonzalez}, {Goodman}, {G{\'o}ra}, {Grandmont},
  {Grant}, {Gretskov}, {Groh}, {Gro{\ss}}, {Ha}, {Haack}, {Haj Ismail},
  {Hallen}, {Hallgren}, {Halzen}, {Hanson}, {Hebecker}, {Heereman}, {Heinen},
  {Helbing}, {Hellauer}, {Hellwig}, {Hickford}, {Hill}, {Hoffman}, {Hoffmann},
  {Homeier}, {Hoshina}, {Huang}, {Huelsnitz}, {Hulth}, {Hultqvist}, {Hussain},
  {Ishihara}, {Jacobi}, {Jacobsen}, {Jagielski}, {Japaridze}, {Jero},
  {Jlelati}, {Jurkovic}, {Kaminsky}, {Kappes}, {Karg}, {Karle}, {Kauer},
  {Kelley}, {Kheirandish}, {Kiryluk}, {Kl{\"a}s}, {Klein}, {K{\"o}hne},
  {Kohnen}, {Kolanoski}, {Koob}, {K{\"o}pke}, {Kopper}, {Kopper}, {Koskinen},
  {Kowalski}, {Kriesten}, {Krings}, {Kroll}, {Kroll}, {Kunnen}, {Kurahashi},
  {Kuwabara}, {Labare}, {Larsen}, {Larson}, {Lesiak-Bzdak}, {Leuermann},
  {Leute}, {L{\"u}nemann}, {Mac{\'\i}as}, {Madsen}, {Maggi}, {Maruyama},
  {Mase}, {Matis}, {Maunu}, {McNally}, {Meagher}, {Medici}, {Meli}, {Meures},
  {Miarecki}, {Middell}, {Middlemas}, {Milke}, {Miller}, {Mohrmann},
  {Montaruli}, {Morse}, {Nahnhauer}, {Naumann}, {Niederhausen}, {Nowicki},
  {Nygren}, {Obertacke}, {Odrowski}, {Olivas}, {Omairat}, {O'Murchadha},
  {Palczewski}, {Paul}, {Penek}, {Pepper}, {P{\'e}rez de los Heros},
  {Pfendner}, {Pieloth}, {Pinat}, {Posselt}, {Price}, {Przybylski}, {P{\"u}tz},
  {Quinnan}, {R{\"a}del}, {Rameez}, {Rawlins}, {Redl}, {Rees}, {Reimann},
  {Resconi}, {Rhode}, {Richman}, {Riedel}, {Robertson}, {Rodrigues}, {Rongen},
  {Rott}, {Ruhe}, {Ruzybayev}, {Ryckbosch}, {Saba}, {Sander}, {Sandroos},
  {Santander}, {Sarkar}, {Schatto}, {Scheriau}, {Schmidt}, {Schmitz},
  {Schoenen}, {Sch{\"o}neberg}, {Sch{\"o}nwald}, {Schukraft}, {Schulte},
  {Schulz}, {Seckel}, {Sestayo}, {Seunarine}, {Shanidze}, {Sheremata}, {Smith},
  {Soldin}, {Spiczak}, {Spiering}, {Stamatikos}, {Stanev}, {Stanisha},
  {Stasik}, {Stezelberger}, {Stokstad}, {St{\"o}{\ss}l}, {Strahler},
  {Str{\"o}m}, {Strotjohann}, {Sullivan}, {Taavola}, {Taboada}, {Tamburro},
  {Tepe}, {Ter-Antonyan}, {Terliuk}, {Te{\v{s}}i{\'c}}, {Tilav}, {Toale},
  {Tobin}, {Tosi}, {Tselengidou}, {Unger}, {Usner}, {Vallecorsa}, {van
  Eijndhoven}, {Vandenbroucke}, {van Santen}, {Vehring}, {Voge}, {Vraeghe},
  {Walck}, {Wallraff}, {Weaver}, {Wellons}, {Wendt}, {Westerhoff}, {Whelan},
  {Whitehorn}, {Wichary}, {Wiebe}, {Wiebusch}, {Williams}, {Wissing}, {Wolf},
  {Wood}, {Woschnagg}, {Xu}, {Xu}, {Yanez}, {Yodh}, {Yoshida}, {Zarzhitsky},
  {Ziemann}, {Zierke}, {Zoll}, \& {The IceCube
  Collaboration}}]{Aartsen:2014cva}
---. 2014{\natexlab{c}}, \apj, 796, 109, \dodoi{10.1088/0004-637X/796/2/109}

\bibitem[{{Aartsen} {et~al.}(2016){Aartsen}, {Abraham}, {Ackermann}, {Adams},
  {Aguilar}, {Ahlers}, {Ahrens}, {Altmann}, {Andeen}, {Anderson}, {Ansseau},
  {Anton}, {Archinger}, {Arg{\"u}elles}, {Auffenberg}, {Axani}, {Bai},
  {Barwick}, {Baum}, {Bay}, {Beatty}, {Becker Tjus}, {Becker}, {BenZvi},
  {Berghaus}, {Berley}, {Bernardini}, {Bernhard}, {Besson}, {Binder}, {Bindig},
  {Bissok}, {Blaufuss}, {Blot}, {Bohm}, {B{\"o}rner}, {Bos}, {Bose},
  {B{\"o}ser}, {Botner}, {Braun}, {Brayeur}, {Bretz}, {Burgman}, {Carver},
  {Casier}, {Cheung}, {Chirkin}, {Christov}, {Clark}, {Classen}, {Coenders},
  {Collin}, {Conrad}, {Cowen}, {Cross}, {Day}, {de Andr{\'e}}, {De Clercq},
  {del Pino Rosendo}, {Dembinski}, {De Ridder}, {Desiati}, {de Vries}, {de
  Wasseige}, {de With}, {DeYoung}, {D{\'\i}az-V{\'e}lez}, {di Lorenzo},
  {Dujmovic}, {Dumm}, {Dunkman}, {Eberhardt}, {Ehrhardt}, {Eichmann}, {Eller},
  {Euler}, {Evenson}, {Fahey}, {Fazely}, {Feintzeig}, {Felde}, {Filimonov},
  {Finley}, {Flis}, {F{\"o}sig}, {Franckowiak}, {Friedman}, {Fuchs}, {Gaisser},
  {Gallagher}, {Gerhardt}, {Ghorbani}, {Giang}, {Gladstone}, {Glagla},
  {Gl{\"u}senkamp}, {Goldschmidt}, {Golup}, {Gonzalez}, {Grant}, {Griffith},
  {Haack}, {Haj Ismail}, {Hallgren}, {Halzen}, {Hansen}, {Hansmann},
  {Hansmann}, {Hanson}, {Hebecker}, {Heereman}, {Helbing}, {Hellauer},
  {Hickford}, {Hignight}, {Hill}, {Hoffman}, {Hoffmann}, {Holzapfel},
  {Hoshina}, {Huang}, {Huber}, {Hultqvist}, {In}, {Ishihara}, {Jacobi},
  {Japaridze}, {Jeong}, {Jero}, {Jones}, {Jurkovic}, {Kappes}, {Karg}, {Karle},
  {Katz}, {Kauer}, {Keivani}, {Kelley}, {Kemp}, {Kheirandish}, {Kim},
  {Kintscher}, {Kiryluk}, {Kittler}, {Klein}, {Kohnen}, {Koirala}, {Kolanoski},
  {Konietz}, {K{\"o}pke}, {Kopper}, {Kopper}, {Koskinen}, {Kowalski}, {Krings},
  {Kroll}, {Kr{\"u}ckl}, {Kr{\"u}ger}, {Kunnen}, {Kunwar}, {Kurahashi},
  {Kuwabara}, {Labare}, {Lanfranchi}, {Larson}, {Lauber}, {Lennarz},
  {Lesiak-Bzdak}, {Leuermann}, {Leuner}, {Lu}, {L{\"u}nemann}, {Madsen},
  {Maggi}, {Mahn}, {Mancina}, {Mandelartz}, {Maruyama}, {Mase}, {Maunu},
  {McNally}, {Meagher}, {Medici}, {Meier}, {Meli}, {Menne}, {Merino}, {Meures},
  {Miarecki}, {Mohrmann}, {Montaruli}, {Moulai}, {Nahnhauer}, {Naumann},
  {Neer}, {Niederhausen}, {Nowicki}, {Nygren}, {Obertacke Pollmann}, {Olivas},
  {O'Murchadha}, {Palczewski}, {Pandya}, {Pankova}, {Peiffer}, {Penek},
  {Pepper}, {P{\'e}rez de los Heros}, {Pieloth}, {Pinat}, {Price},
  {Przybylski}, {Quinnan}, {Raab}, {R{\"a}del}, {Rameez}, {Rawlins}, {Reimann},
  {Relethford}, {Relich}, {Resconi}, {Rhode}, {Richman}, {Riedel}, {Robertson},
  {Rongen}, {Rott}, {Ruhe}, {Ryckbosch}, {Rysewyk}, {Sabbatini}, {Sanchez
  Herrera}, {Sandrock}, {Sandroos}, {Sarkar}, {Satalecka}, {Schimp},
  {Schlunder}, {Schmidt}, {Schoenen}, {Sch{\"o}neberg}, {Schumacher}, {Seckel},
  {Seunarine}, {Soldin}, {Song}, {Spiczak}, {Spiering}, {Stahlberg}, {Stanev},
  {Stasik}, {Steuer}, {Stezelberger}, {Stokstad}, {St{\"o}{\ss}l}, {Str{\"o}m},
  {Strotjohann}, {Sullivan}, {Sutherland}, {Taavola}, {Taboada}, {Tatar},
  {Tenholt}, {Ter-Antonyan}, {Terliuk}, {Te{\v{s}}i{\'c}}, {Tilav}, {Toale},
  {Tobin}, {Toscano}, {Tosi}, {Tselengidou}, {Turcati}, {Unger}, {Usner},
  {Vandenbroucke}, {van Eijndhoven}, {Vanheule}, {van Rossem}, {van Santen},
  {Veenkamp}, {Vehring}, {Voge}, {Vraeghe}, {Walck}, {Wallace}, {Wallraff},
  {Wandkowsky}, {Weaver}, {Weiss}, {Wendt}, {Westerhoff}, {Whelan}, {Wickmann},
  {Wiebe}, {Wiebusch}, {Wille}, {Williams}, {Wills}, {Wolf}, {Wood}, {Woolsey},
  {Woschnagg}, {Xu}, {Xu}, {Xu}, {Yanez}, {Yodh}, {Yoshida}, {Zoll}, \& {The
  Icecube Collaboration}}]{Aartsen:2016xlq}
{Aartsen}, M.~G., {Abraham}, K., {Ackermann}, M., {et~al.} 2016, \apj, 833, 3,
  \dodoi{10.3847/0004-637X/833/1/3}

\bibitem[{{Aartsen} {et~al.}(2017{\natexlab{a}}){Aartsen}, {Abraham},
  {Ackermann}, {Adams}, {Aguilar}, {Ahlers}, {Ahrens}, {Altmann}, {Andeen},
  {Anderson}, {Ansseau}, {Anton}, {Archinger}, {Arguelles}, {Arlen},
  {Auffenberg}, {Axani}, {Bai}, {Barwick}, {Baum}, {Bay}, {Beatty}, {Becker
  Tjus}, {Becker}, {BenZvi}, {Berghaus}, {Berley}, {Bernardini}, {Bernhard},
  {Besson}, {Binder}, {Bindig}, {Bissok}, {Blaufuss}, {Blot}, {Boersma},
  {Bohm}, {B{\"o}rner}, {Bos}, {Bose}, {B{\"o}ser}, {Botner}, {Braun},
  {Brayeur}, {Bretz}, {Burgman}, {Casey}, {Casier}, {Cheung}, {Chirkin},
  {Christov}, {Clark}, {Classen}, {Coenders}, {Collin}, {Conrad}, {Cowen},
  {Cruz Silva}, {Daughhetee}, {Davis}, {Day}, {de Andr{\'e}}, {De Clercq}, {del
  Pino Rosendo}, {Dembinski}, {De Ridder}, {Desiati}, {de Vries}, {de
  Wasseige}, {de With}, {DeYoung}, {D{\'\i}az-V{\'e}lez}, {di Lorenzo},
  {Dujmovic}, {Dumm}, {Dunkman}, {Eberhardt}, {Ehrhardt}, {Eichmann}, {Euler},
  {Evenson}, {Fahey}, {Fazely}, {Feintzeig}, {Felde}, {Filimonov}, {Finley},
  {Flis}, {F{\"o}sig}, {Franckowiak}, {Fuchs}, {Gaisser}, {Gaior}, {Gallagher},
  {Gerhardt}, {Ghorbani}, {Giang}, {Gladstone}, {Glagla}, {Gl{\"u}senkamp},
  {Goldschmidt}, {Golup}, {Gonzalez}, {G{\'o}ra}, {Grant}, {Griffith}, {Haack},
  {Haj Ismail}, {Hallgren}, {Halzen}, {Hansen}, {Hansmann}, {Hansmann},
  {Hanson}, {Hebecker}, {Heereman}, {Helbing}, {Hellauer}, {Hickford},
  {Hignight}, {Hill}, {Hoffman}, {Hoffmann}, {Holzapfel}, {Homeier}, {Hoshina},
  {Huang}, {Huber}, {Huelsnitz}, {Hultqvist}, {In}, {Ishihara}, {Jacobi},
  {Japaridze}, {Jeong}, {Jero}, {Jones}, {Jurkovic}, {Kappes}, {Karg}, {Karle},
  {Katz}, {Kauer}, {Keivani}, {Kelley}, {Kemp}, {Kheirandish}, {Kim},
  {Kintscher}, {Kiryluk}, {Kittler}, {Klein}, {Kohnen}, {Koirala}, {Kolanoski},
  {Konietz}, {K{\"o}pke}, {Kopper}, {Kopper}, {Koskinen}, {Kowalski}, {Krings},
  {Kroll}, {Kr{\"u}ckl}, {Kr{\"u}ger}, {Kunnen}, {Kunwar}, {Kurahashi},
  {Kuwabara}, {Labare}, {Lanfranchi}, {Larson}, {Lennarz}, {Lesiak-Bzdak},
  {Leuermann}, {Leuner}, {Lu}, {L{\"u}nemann}, {Madsen}, {Maggi}, {Mahn},
  {Mancina}, {Mandelartz}, {Maruyama}, {Mase}, {Maunu}, {McNally}, {Meagher},
  {Medici}, {Meier}, {Meli}, {Menne}, {Merino}, {Meures}, {Miarecki},
  {Middell}, {Mohrmann}, {Montaruli}, {Moulai}, {Nahnhauer}, {Naumann}, {Neer},
  {Niederhausen}, {Nowicki}, {Nygren}, {Obertacke Pollmann}, {Olivas},
  {Omairat}, {O'Murchadha}, {Palczewski}, {Pandya}, {Pankova}, {Penek},
  {Pepper}, {P{\'e}rez de los Heros}, {Pfendner}, {Pieloth}, {Pinat},
  {Posselt}, {Price}, {Przybylski}, {Quinnan}, {Raab}, {R{\"a}del}, {Rameez},
  {Rawlins}, {Reimann}, {Relich}, {Resconi}, {Rhode}, {Richman}, {Riedel},
  {Robertson}, {Rongen}, {Rott}, {Ruhe}, {Ryckbosch}, {Rysewyk}, {Sabbatini},
  {Sanchez Herrera}, {Sandrock}, {Sandroos}, {Sarkar}, {Satalecka}, {Schimp},
  {Schlunder}, {Schmidt}, {Schoenen}, {Sch{\"o}neberg}, {Sch{\"o}nwald},
  {Schumacher}, {Seckel}, {Seunarine}, {Soldin}, {Song}, {Spiczak}, {Spiering},
  {Stahlberg}, {Stamatikos}, {Stanev}, {Stasik}, {Steuer}, {Stezelberger},
  {Stokstad}, {St{\"o}{\ss}l}, {Str{\"o}m}, {Strotjohann}, {Sullivan},
  {Sutherland}, {Taavola}, {Taboada}, {Tatar}, {Ter-Antonyan}, {Terliuk},
  {Te{\v{s}}i{\'c}}, {Tilav}, {Toale}, {Tobin}, {Toscano}, {Tosi},
  {Tselengidou}, {Turcati}, {Unger}, {Usner}, {Vallecorsa}, {Vandenbroucke},
  {van Eijndhoven}, {Vanheule}, {van Rossem}, {van Santen}, {Veenkamp},
  {Vehring}, {Voge}, {Vraeghe}, {Walck}, {Wallace}, {Wallraff}, {Wandkowsky},
  {Weaver}, {Wendt}, {Westerhoff}, {Whelan}, {Wickmann}, {Wiebe}, {Wiebusch},
  {Wille}, {Williams}, {Wills}, {Wissing}, {Wolf}, {Wood}, {Woolsey},
  {Woschnagg}, {Xu}, {Xu}, {Xu}, {Yanez}, {Yodh}, {Yoshida}, {Zoll}, \& {The
  IceCube Collaboration}}]{Aartsen:2016lir}
---. 2017{\natexlab{a}}, \apj, 835, 45, \dodoi{10.3847/1538-4357/835/1/45}

\bibitem[{{Aartsen} {et~al.}(2017{\natexlab{b}}){Aartsen}, {Ackermann},
  {Adams}, {Aguilar}, {Ahlers}, {Ahrens}, {Altmann}, {Andeen}, {Anderson},
  {Ansseau}, {Anton}, {Archinger}, {Arg{\"u}elles}, {Auer}, {Auffenberg},
  {Axani}, {Baccus}, {Bai}, {Barnet}, {Barwick}, {Baum}, {Bay}, {Beattie},
  {Beatty}, {Becker Tjus}, {Becker}, {Bendfelt}, {BenZvi}, {Berley},
  {Bernardini}, {Bernhard}, {Besson}, {Binder}, {Bindig}, {Bissok}, {Blaufuss},
  {Blot}, {Boersma}, {Bohm}, {B{\"o}rner}, {Bos}, {Bose}, {B{\"o}ser},
  {Botner}, {Bouchta}, {Braun}, {Brayeur}, {Bretz}, {Bron}, {Burgman},
  {Burreson}, {Carver}, {Casier}, {Cheung}, {Chirkin}, {Christov}, {Clark},
  {Classen}, {Coenders}, {Collin}, {Conrad}, {Cowen}, {Cross}, {Day}, {Day},
  {de Andr{\'e}}, {De Clercq}, {del Pino Rosendo}, {Dembinski}, {De Ridder},
  {Descamps}, {Desiati}, {de Vries}, {de Wasseige}, {de With}, {DeYoung},
  {D{\'\i}az-V{\'e}lez}, {di Lorenzo}, {Dujmovic}, {Dumm}, {Dunkman},
  {Eberhardt}, {Edwards}, {Ehrhardt}, {Eichmann}, {Eller}, {Euler}, {Evenson},
  {Fahey}, {Fazely}, {Feintzeig}, {Felde}, {Filimonov}, {Finley}, {Flis},
  {F{\"o}sig}, {Franckowiak}, {Fr{\`e}re}, {Friedman}, {Fuchs}, {Gaisser},
  {Gallagher}, {Gerhardt}, {Ghorbani}, {Giang}, {Gladstone}, {Glauch},
  {Glowacki}, {Gl{\"u}senkamp}, {Goldschmidt}, {Gonzalez}, {Grant}, {Griffith},
  {Gustafsson}, {Haack}, {Hallgren}, {Halzen}, {Hansen}, {Hansmann}, {Hanson},
  {Haugen}, {Hebecker}, {Heereman}, {Helbing}, {Hellauer}, {Heller},
  {Hickford}, {Hignight}, {Hill}, {Hoffman}, {Hoffmann}, {Hoshina}, {Huang},
  {Huber}, {Hulth}, {Hultqvist}, {In}, {Inaba}, {Ishihara}, {Jacobi},
  {Jacobsen}, {Japaridze}, {Jeong}, {Jero}, {Jones}, {Jones}, {Joseph}, {Kang},
  {Kappes}, {Karg}, {Karle}, {Katz}, {Kauer}, {Keivani}, {Kelley}, {Kemp},
  {Kheirandish}, {Kim}, {Kim}, {Kintscher}, {Kiryluk}, {Kitamura}, {Kittler},
  {Klein}, {Kleinfelder}, {Kleist}, {Kohnen}, {Koirala}, {Kolanoski},
  {Konietz}, {K{\"o}pke}, {Kopper}, {Kopper}, {Koskinen}, {Kowalski},
  {Krasberg}, {Krings}, {Kroll}, {Kr{\"u}ckl}, {Kr{\"u}ger}, {Kunnen},
  {Kunwar}, {Kurahashi}, {Kuwabara}, {Labare}, {Laihem}, {Landsman},
  {Lanfranchi}, {Larson}, {Lauber}, {Laundrie}, {Lennarz}, {Leich},
  {Lesiak-Bzdak}, {Leuermann}, {Lu}, {Ludwig}, {L{\"u}nemann}, {Mackenzie},
  {Madsen}, {Maggi}, {Mahn}, {Mancina}, {Mandelartz}, {Maruyama}, {Mase},
  {Matis}, {Maunu}, {McNally}, {McParland}, {Meade}, {Meagher}, {Medici},
  {Meier}, {Meli}, {Menne}, {Merino}, {Meures}, {Miarecki}, {Minor},
  {Montaruli}, {Moulai}, {Murray}, {Nahnhauer}, {Naumann}, {Neer}, {Newcomb},
  {Niederhausen}, {Nowicki}, {Nygren}, {Obertacke Pollmann}, {Olivas},
  {O'Murchadha}, {Palczewski}, {Pandya}, {Pankova}, {Patton}, {Peiffer},
  {Penek}, {Pepper}, {P{\'e}rez de los Heros}, {Pettersen}, {Pieloth}, {Pinat},
  {Price}, {Przybylski}, {Quinnan}, {Raab}, {R{\"a}del}, {Rameez}, {Rawlins},
  {Reimann}, {Relethford}, {Relich}, {Resconi}, {Rhode}, {Richman}, {Riedel},
  {Robertson}, {Rongen}, {Roucelle}, {Rott}, {Ruhe}, {Ryckbosch}, {Rysewyk},
  {Sabbatini}, {Sanchez Herrera}, {Sandrock}, {Sandroos}, {Sandstrom},
  {Sarkar}, {Satalecka}, {Schlunder}, {Schmidt}, {Schoenen}, {Sch{\"o}neberg},
  {Schukraft}, {Schumacher}, {Seckel}, {Seunarine}, {Solarz}, {Soldin}, {Song},
  {Spiczak}, {Spiering}, {Stanev}, {Stasik}, {Stettner}, {Steuer},
  {Stezelberger}, {Stokstad}, {St{\"o}{\ss}l}, {Str{\"o}m}, {Strotjohann},
  {Sulanke}, {Sullivan}, {Sutherland}, {Taavola}, {Taboada}, {Tatar},
  {Tenholt}, {Ter-Antonyan}, {Terliuk}, {Te{\v{s}}i{\'c}}, {Thollander},
  {Tilav}, {Toale}, {Tobin}, {Toscano}, {Tosi}, {Tselengidou}, {Turcati},
  {Unger}, {Usner}, {Vandenbroucke}, {van Eijndhoven}, {Vanheule}, {van
  Rossem}, {van Santen}, {Vehring}, {Voge}, {Vogel}, {Vraeghe}, {Wahl},
  {Walck}, {Wallace}, {Wallraff}, {Wandkowsky}, {Weaver}, {Weiss}, {Wendt},
  {Westerhoff}, {Wharton}, {Whelan}, {Wickmann}, {Wiebe}, {Wiebusch}, {Wille},
  {Williams}, {Wills}, {Wisniewski}, {Wolf}, {Wood}, {Woolsey}, {Woschnagg},
  {Xu}, {Xu}, {Xu}, {Yanez}, {Yodh}, {Yoshida}, \& {Zoll}}]{Aartsen:2016nxy}
{Aartsen}, M.~G., {Ackermann}, M., {Adams}, J., {et~al.} 2017{\natexlab{b}},
  Journal of Instrumentation, 12, P03012,
  \dodoi{10.1088/1748-0221/12/03/P03012}

\bibitem[{{Aartsen} {et~al.}(2017{\natexlab{c}}){Aartsen}, {Ackermann},
  {Adams}, {Aguilar}, {Ahlers}, {Ahrens}, {Altmann}, {Andeen}, {Anderson},
  {Ansseau}, {Anton}, {Archinger}, {Arg{\"u}elles}, {Auffenberg}, {Axani},
  {Bai}, {Barwick}, {Baum}, {Bay}, {Beatty}, {Becker Tjus}, {Becker}, {BenZvi},
  {Berley}, {Bernardini}, {Bernhard}, {Besson}, {Binder}, {Bindig}, {Bissok},
  {Blaufuss}, {Blot}, {Bohm}, {B{\"o}rner}, {Bos}, {Bose}, {B{\"o}ser},
  {Botner}, {Braun}, {Brayeur}, {Bretz}, {Bron}, {Burgman}, {Carver}, {Casier},
  {Cheung}, {Chirkin}, {Christov}, {Clark}, {Classen}, {Coenders}, {Collin},
  {Conrad}, {Cowen}, {Cross}, {Day}, {de Andr{\'e}}, {De Clercq}, {del Pino
  Rosendo}, {Dembinski}, {De Ridder}, {Desiati}, {de Vries}, {de Wasseige}, {de
  With}, {DeYoung}, {D{\'\i}az-V{\'e}lez}, {di Lorenzo}, {Dujmovic}, {Dumm},
  {Dunkman}, {Eberhardt}, {Ehrhardt}, {Eichmann}, {Eller}, {Euler}, {Evenson},
  {Fahey}, {Fazely}, {Feintzeig}, {Felde}, {Filimonov}, {Finley}, {Flis},
  {F{\"o}sig}, {Franckowiak}, {Friedman}, {Fuchs}, {Gaisser}, {Gallagher},
  {Gerhardt}, {Ghorbani}, {Giang}, {Gladstone}, {Glauch}, {Gl{\"u}senkamp},
  {Goldschmidt}, {Gonzalez}, {Grant}, {Griffith}, {Haack}, {Hallgren},
  {Halzen}, {Hansen}, {Hansmann}, {Hanson}, {Hebecker}, {Heereman}, {Helbing},
  {Hellauer}, {Hickford}, {Hignight}, {Hill}, {Hoffman}, {Hoffmann}, {Hoshina},
  {Huang}, {Huber}, {Hultqvist}, {In}, {Ishihara}, {Jacobi}, {Japaridze},
  {Jeong}, {Jero}, {Jones}, {Kang}, {Kappes}, {Karg}, {Karle}, {Katz}, {Kauer},
  {Keivani}, {Kelley}, {Kheirandish}, {Kim}, {Kim}, {Kintscher}, {Kiryluk},
  {Kittler}, {Klein}, {Kohnen}, {Koirala}, {Kolanoski}, {Konietz}, {K{\"o}pke},
  {Kopper}, {Kopper}, {Koskinen}, {Kowalski}, {Krings}, {Kroll}, {Kr{\"u}ckl},
  {Kr{\"u}ger}, {Kunnen}, {Kunwar}, {Kurahashi}, {Kuwabara}, {Labare},
  {Lanfranchi}, {Larson}, {Lauber}, {Lennarz}, {Lesiak-Bzdak}, {Leuermann},
  {Lu}, {L{\"u}nemann}, {Madsen}, {Maggi}, {Mahn}, {Mancina}, {Mandelartz},
  {Maruyama}, {Mase}, {Maunu}, {McNally}, {Meagher}, {Medici}, {Meier}, {Meli},
  {Menne}, {Merino}, {Meures}, {Miarecki}, {Montaruli}, {Moulai}, {Nahnhauer},
  {Naumann}, {Neer}, {Niederhausen}, {Nowicki}, {Nygren}, {Obertacke Pollmann},
  {Olivas}, {O'Murchadha}, {Palczewski}, {Pandya}, {Pankova}, {Peiffer},
  {Penek}, {Pepper}, {P{\'e}rez de los Heros}, {Pieloth}, {Pinat}, {Price},
  {Przybylski}, {Quinnan}, {Raab}, {R{\"a}del}, {Rameez}, {Rawlins}, {Reimann},
  {Relethford}, {Relich}, {Resconi}, {Rhode}, {Richman}, {Riedel}, {Robertson},
  {Rongen}, {Rott}, {Ruhe}, {Ryckbosch}, {Rysewyk}, {Sabbatini}, {Sanchez
  Herrera}, {Sandrock}, {Sandroos}, {Sarkar}, {Satalecka}, {Schlunder},
  {Schmidt}, {Schoenen}, {Sch{\"o}neberg}, {Schumacher}, {Seckel}, {Seunarine},
  {Soldin}, {Song}, {Spiczak}, {Spiering}, {Stanev}, {Stasik}, {Stettner},
  {Steuer}, {Stezelberger}, {Stokstad}, {St{\"o}{\ss}l}, {Str{\"o}m},
  {Strotjohann}, {Sullivan}, {Sutherland}, {Taavola}, {Taboada}, {Tatar},
  {Tenholt}, {Ter-Antonyan}, {Terliuk}, {Te{\v{s}}i{\'c}}, {Tilav}, {Toale},
  {Tobin}, {Toscano}, {Tosi}, {Tselengidou}, {Turcati}, {Unger}, {Usner},
  {Vandenbroucke}, {van Eijndhoven}, {Vanheule}, {van Rossem}, {van Santen},
  {Vehring}, {Voge}, {Vogel}, {Vraeghe}, {Walck}, {Wallace}, {Wallraff},
  {Wandkowsky}, {Weaver}, {Weiss}, {Wendt}, {Westerhoff}, {Whelan}, {Wickmann},
  {Wiebe}, {Wiebusch}, {Wille}, {Williams}, {Wills}, {Wolf}, {Wood}, {Woolsey},
  {Woschnagg}, {Xu}, {Xu}, {Xu}, {Yanez}, {Yodh}, {Yoshida}, \&
  {Zoll}}]{Aartsen:2016lmt}
---. 2017{\natexlab{c}}, Astroparticle Physics, 92, 30,
  \dodoi{10.1016/j.astropartphys.2017.05.002}

\bibitem[{{Aartsen} {et~al.}(2017{\natexlab{d}}){Aartsen}, {Abraham},
  {Ackermann}, {Adams}, {Aguilar}, {Ahlers}, {Ahrens}, {Altmann}, {Andeen},
  {Anderson}, {Ansseau}, {Anton}, {Archinger}, {Arg{\"u}elles}, {Auffenberg},
  {Axani}, {Bai}, {Barwick}, {Baum}, {Bay}, {Beatty}, {Becker Tjus}, {Becker},
  {BenZvi}, {Berley}, {Bernardini}, {Bernhard}, {Besson}, {Binder}, {Bindig},
  {Bissok}, {Blaufuss}, {Blot}, {Bohm}, {B{\"o}rner}, {Bos}, {Bose},
  {B{\"o}ser}, {Botner}, {Braun}, {Brayeur}, {Bretz}, {Bron}, {Burgman},
  {Carver}, {Casier}, {Cheung}, {Chirkin}, {Christov}, {Clark}, {Classen},
  {Coenders}, {Collin}, {Conrad}, {Cowen}, {Cross}, {Day}, {de Andr{\'e}}, {De
  Clercq}, {del Pino Rosendo}, {Dembinski}, {De Ridder}, {Desiati}, {de Vries},
  {de Wasseige}, {de With}, {DeYoung}, {D{\'\i}az-V{\'e}lez}, {di Lorenzo},
  {Dujmovic}, {Dumm}, {Dunkman}, {Eberhardt}, {Ehrhardt}, {Eichmann}, {Eller},
  {Euler}, {Evenson}, {Fahey}, {Fazely}, {Feintzeig}, {Felde}, {Filimonov},
  {Finley}, {Flis}, {F{\"o}sig}, {Franckowiak}, {Friedman}, {Fuchs}, {Gaisser},
  {Gallagher}, {Gerhardt}, {Ghorbani}, {Giang}, {Gladstone}, {Glauch},
  {Gl{\"u}senkamp}, {Goldschmidt}, {Golup}, {Gonzalez}, {Grant}, {Griffith},
  {Haack}, {Haj Ismail}, {Hallgren}, {Halzen}, {Hansen}, {Hansmann}, {Hanson},
  {Hebecker}, {Heereman}, {Helbing}, {Hellauer}, {Hickford}, {Hignight},
  {Hill}, {Hoffman}, {Hoffmann}, {Holzapfel}, {Hoshina}, {Huang}, {Huber},
  {Hultqvist}, {In}, {Ishihara}, {Jacobi}, {Japaridze}, {Jeong}, {Jero},
  {Jones}, {Jurkovic}, {Kappes}, {Karg}, {Karle}, {Katz}, {Kauer}, {Keivani},
  {Kelley}, {Kheirandish}, {Kim}, {Kintscher}, {Kiryluk}, {Kittler}, {Klein},
  {Kohnen}, {Koirala}, {Kolanoski}, {Konietz}, {K{\"o}pke}, {Kopper}, {Kopper},
  {Koskinen}, {Kowalski}, {Krings}, {Kroll}, {Kr{\"u}ckl}, {Kr{\"u}ger},
  {Kunnen}, {Kunwar}, {Kurahashi}, {Kuwabara}, {Labare}, {Lanfranchi},
  {Larson}, {Lauber}, {Lennarz}, {Lesiak-Bzdak}, {Leuermann}, {Lu},
  {L{\"u}nemann}, {Madsen}, {Maggi}, {Mahn}, {Mancina}, {Mandelartz},
  {Maruyama}, {Mase}, {Maunu}, {McNally}, {Meagher}, {Medici}, {Meier}, {Meli},
  {Menne}, {Merino}, {Meures}, {Miarecki}, {Mohrmann}, {Montaruli}, {Moulai},
  {Nahnhauer}, {Naumann}, {Neer}, {Niederhausen}, {Nowicki}, {Nygren},
  {Obertacke Pollmann}, {Olivas}, {O'Murchadha}, {Palczewski}, {Pandya},
  {Pankova}, {Peiffer}, {Penek}, {Pepper}, {P{\'e}rez de los Heros}, {Pieloth},
  {Pinat}, {Price}, {Przybylski}, {Quinnan}, {Raab}, {R{\"a}del}, {Rameez},
  {Rawlins}, {Reimann}, {Relethford}, {Relich}, {Resconi}, {Rhode}, {Richman},
  {Riedel}, {Robertson}, {Rongen}, {Rott}, {Ruhe}, {Ryckbosch}, {Rysewyk},
  {Sabbatini}, {Sanchez Herrera}, {Sandrock}, {Sandroos}, {Sarkar},
  {Satalecka}, {Schlunder}, {Schmidt}, {Schoenen}, {Sch{\"o}neberg},
  {Schumacher}, {Seckel}, {Seunarine}, {Soldin}, {Song}, {Spiczak}, {Spiering},
  {Stanev}, {Stasik}, {Stettner}, {Steuer}, {Stezelberger}, {Stokstad},
  {St{\"o}ssl}, {Str{\"o}m}, {Strotjohann}, {Sullivan}, {Sutherland},
  {Taavola}, {Taboada}, {Tatar}, {Tenholt}, {Ter-Antonyan}, {Terliuk},
  {Te{\v{s}}i{\'c}}, {Tilav}, {Toale}, {Tobin}, {Toscano}, {Tosi},
  {Tselengidou}, {Turcati}, {Unger}, {Usner}, {Vandenbroucke}, {van
  Eijndhoven}, {Vanheule}, {van Rossem}, {van Santen}, {Veenkamp}, {Vehring},
  {Voge}, {Vogel}, {Vraeghe}, {Walck}, {Wallace}, {Wallraff}, {Wandkowsky},
  {Weaver}, {Weiss}, {Wendt}, {Westerhoff}, {Whelan}, {Wickmann}, {Wiebe},
  {Wiebusch}, {Wille}, {Williams}, {Wills}, {Wolf}, {Wood}, {Woolsey},
  {Woschnagg}, {Xu}, {Xu}, {Xu}, {Yanez}, {Yodh}, {Yoshida}, {Zoll}, \& {The
  IceCube Collaboration}}]{Aartsen:2016oji}
{Aartsen}, M.~G., {Abraham}, K., {Ackermann}, M., {et~al.} 2017{\natexlab{d}},
  \apj, 835, 151, \dodoi{10.3847/1538-4357/835/2/151}

\bibitem[{{Aartsen} {et~al.}(2018{\natexlab{a}}){Aartsen}, {Ackermann},
  {Adams}, {Aguilar}, {Ahlers}, {Ahrens}, {Al Samarai}, {Altmann}, {Andeen},
  {Anderson}, {Ansseau}, {Anton}, {Arg{\"u}elles}, {Auffenberg}, {Axani},
  {Bagherpour}, {Bai}, {Barron}, {Barwick}, {Baum}, {Bay}, {Beatty}, {Becker
  Tjus}, {Becker}, {BenZvi}, {Berley}, {Bernardini}, {Besson}, {Binder},
  {Bindig}, {Blaufuss}, {Blot}, {Bohm}, {B{\"o}rner}, {Bos}, {B{\"o}ser},
  {Botner}, {Bourbeau}, {Bourbeau}, {Bradascio}, {Braun}, {Brenzke}, {Bretz},
  {Bron}, {Brostean-Kaiser}, {Burgman}, {Busse}, {Carver}, {Cheung}, {Chirkin},
  {Christov}, {Clark}, {Classen}, {Coenders}, {Collin}, {Conrad}, {Coppin},
  {Correa}, {Cowen}, {Cross}, {Dave}, {Day}, {de Andr{\'e}}, {De Clercq},
  {DeLaunay}, {Dembinski}, {De Ridder}, {Desiati}, {de Vries}, {de Wasseige},
  {de With}, {DeYoung}, {D{\'\i}az-V{\'e}lez}, {di Lorenzo}, {Dujmovic},
  {Dumm}, {Dunkman}, {Dvorak}, {Eberhardt}, {Ehrhardt}, {Eichmann}, {Eller},
  {Evenson}, {Fahey}, {Fazely}, {Felde}, {Filimonov}, {Finley}, {Flis},
  {Franckowiak}, {Friedman}, {Fritz}, {Gaisser}, {Gallagher}, {Gerhardt},
  {Ghorbani}, {Glauch}, {Gl{\"u}senkamp}, {Goldschmidt}, {Gonzalez}, {Grant},
  {Griffith}, {Haack}, {Hallgren}, {Halzen}, {Hanson}, {Hebecker}, {Heereman},
  {Helbing}, {Hellauer}, {Hickford}, {Hignight}, {Hill}, {Hoffman}, {Hoffmann},
  {Hoinka}, {Hokanson-Fasig}, {Hoshina}, {Huang}, {Huber}, {Hultqvist},
  {H{\"u}nnefeld}, {Hussain}, {In}, {Iovine}, {Ishihara}, {Jacobi},
  {Japaridze}, {Jeong}, {Jero}, {Jones}, {Kalaczynski}, {Kang}, {Kappes},
  {Kappesser}, {Karg}, {Karle}, {Katz}, {Kauer}, {Keivani}, {Kelley},
  {Kheirandish}, {Kim}, {Kim}, {Kintscher}, {Kiryluk}, {Kittler}, {Klein},
  {Koirala}, {Kolanoski}, {K{\"o}pke}, {Kopper}, {Kopper}, {Koschinsky},
  {Koskinen}, {Kowalski}, {Krings}, {Kroll}, {Kr{\"u}ckl}, {Kunwar},
  {Kurahashi}, {Kuwabara}, {Kyriacou}, {Labare}, {Lanfranchi}, {Larson},
  {Lauber}, {Leonard}, {Lesiak-Bzdak}, {Leuermann}, {Liu}, {Lozano Mariscal},
  {Lu}, {L{\"u}nemann}, {Luszczak}, {Madsen}, {Maggi}, {Mahn}, {Mancina},
  {Maruyama}, {Mase}, {Maunu}, {Meagher}, {Medici}, {Meier}, {Menne}, {Merino},
  {Meures}, {Miarecki}, {Micallef}, {Moment{\'e}}, {Montaruli}, {Moore},
  {Morse}, {Moulai}, {Nahnhauer}, {Nakarmi}, {Naumann}, {Neer}, {Niederhausen},
  {Nowicki}, {Nygren}, {Obertacke Pollmann}, {Olivas}, {O'Murchadha},
  {O'Sullivan}, {Palczewski}, {Pandya}, {Pankova}, {Peiffer}, {Pepper},
  {P{\'e}rez de los Heros}, {Pieloth}, {Pinat}, {Plum}, {Price}, {Przybylski},
  {Raab}, {R{\"a}del}, {Rameez}, {Rauch}, {Rawlins}, {Rea}, {Reimann},
  {Relethford}, {Relich}, {Resconi}, {Rhode}, {Richman}, {Robertson}, {Rongen},
  {Rott}, {Ruhe}, {Ryckbosch}, {Rysewyk}, {Safa}, {S{\"a}lzer}, {Sanchez
  Herrera}, {Sandrock}, {Sandroos}, {Santander}, {Sarkar}, {Sarkar},
  {Satalecka}, {Schlunder}, {Schmidt}, {Schneider}, {Schoenen},
  {Sch{\"o}neberg}, {Schumacher}, {Sclafani}, {Seckel}, {Seunarine},
  {Soedingrekso}, {Soldin}, {Song}, {Spiczak}, {Spiering}, {Stachurska},
  {Stamatikos}, {Stanev}, {Stasik}, {Stein}, {Stettner}, {Steuer},
  {Stezelberger}, {Stokstad}, {St{\"o}{\ss}l}, {Strotjohann}, {Stuttard},
  {Sullivan}, {Sutherland}, {Taboada}, {Tatar}, {Tenholt}, {Ter-Antonyan},
  {Terliuk}, {Tilav}, {Toale}, {Tobin}, {Toennis}, {Toscano}, {Tosi},
  {Tselengidou}, {Tung}, {Turcati}, {Turley}, {Ty}, {Unger}, {Usner},
  {Vandenbroucke}, {Van Driessche}, {van Eijk}, {van Eijndhoven}, {Vanheule},
  {van Santen}, {Vogel}, {Vraeghe}, {Walck}, {Wallace}, {Wallraff}, {Wandler},
  {Wandkowsky}, {Waza}, {Weaver}, {Weiss}, {Wendt}, {Werthebach}, {Westerhoff},
  {Whelan}, {Whitehorn}, {Wiebe}, {Wiebusch}, {Wille}, {Williams}, {Wills},
  {Wolf}, {Wood}, {Wood}, {Woschnagg}, {Xu}, {Xu}, {Xu}, {Yanez}, {Yodh},
  {Yoshida}, {Yuan}, {Fermi-LAT Collaboration}, {Abdollahi}, {Ajello},
  {Angioni}, {Baldini}, {Ballet}, {Barbiellini}, {Bastieri}, {Bechtol},
  {Bellazzini}, {Berenji}, {Bissaldi}, {Blandford}, {Bonino}, {Bottacini},
  {Bregeon}, {Bruel}, {Buehler}, {Burnett}, {Burns}, {Buson}, {Cameron},
  {Caputo}, {Caraveo}, {Cavazzuti}, {Charles}, {Chen}, {Cheung}, {Chiang},
  {Chiaro}, {Ciprini}, {Cohen-Tanugi}, {Conrad}, {Costantin}, {Cutini},
  {D'Ammando}, {de Palma}, {Digel}, {Di Lalla}, {Di Mauro}, {Di Venere},
  {Dom{\'\i}nguez}, {Favuzzi}, {Franckowiak}, {Fukazawa}, {Funk}, {Fusco},
  {Gargano}, {Gasparrini}, {Giglietto}, {Giomi}, {Giommi}, {Giordano},
  {Giroletti}, {Glanzman}, {Green}, {Grenier}, {Grondin}, {Guiriec}, {Harding},
  {Hayashida}, {Hays}, {Hewitt}, {Horan}, {J{\'o}hannesson}, {Kadler},
  {Kensei}, {Kocevski}, {Krauss}, {Kreter}, {Kuss}, {La Mura}, {Larsson},
  {Latronico}, {Lemoine-Goumard}, {Li}, {Longo}, {Loparco}, {Lovellette},
  {Lubrano}, {Magill}, {Maldera}, {Malyshev}, {Manfreda}, {Mazziotta},
  {McEnery}, {Meyer}, {Michelson}, {Mizuno}, {Monzani}, {Morselli},
  {Moskalenko}, {Negro}, {Nuss}, {Ojha}, {Omodei}, {Orienti}, {Orlando},
  {Palatiello}, {Paliya}, {Perkins}, {Persic}, {Pesce-Rollins}, {Piron},
  {Porter}, {Principe}, {Rain{\`o}}, {Rando}, {Rani}, {Razzano}, {Razzaque},
  {Reimer}, {Reimer}, {Renault-Tinacci}, {Ritz}, {Rochester}, {Saz Parkinson},
  {Sgr{\`o}}, {Siskind}, {Spandre}, {Spinelli}, {Suson}, {Tajima}, {Takahashi},
  {Tanaka}, {Thayer}, {Thompson}, {Tibaldo}, {Torres}, {Torresi}, {Tosti},
  {Troja}, {Valverde}, {Vianello}, {Vogel}, {Wood}, {Wood}, {Zaharijas}, {MAGIC
  Collaboration}, {Ahnen}, {Ansoldi}, {Antonelli}, {Arcaro}, {Baack},
  {Babi{\'c}}, {Banerjee}, {Bangale}, {Barres de Almeida}, {Barrio}, {Becerra
  Gonz{\'a}lez}, {Bednarek}, {Bernardini}, {Berti}, {Bhattacharyya}, {Biland},
  {Blanch}, {Bonnoli}, {Carosi}, {Carosi}, {Ceribella}, {Chatterjee}, {Colak},
  {Colin}, {Colombo}, {Contreras}, {Cortina}, {Covino}, {Cumani}, {Da Vela},
  {Dazzi}, {De Angelis}, {De Lotto}, {Delfino}, {Delgado}, {Di Pierro},
  {Dom{\'\i}nguez}, {Dominis Prester}, {Dorner}, {Doro}, {Einecke},
  {Elsaesser}, {Fallah Ramazani}, {Fern{\'a}ndez-Barral}, {Fidalgo}, {Foffano},
  {Pfrang}, {Fonseca}, {Font}, {Franceschini}, {Fruck}, {Galindo}, {Gallozzi},
  {Garc{\'\i}a L{\'o}pez}, {Garczarczyk}, {Gaug}, {Giammaria}, {Godinovi{\'c}},
  {Gora}, {Guberman}, {Hadasch}, {Hahn}, {Hassan}, {Hayashida}, {Herrera},
  {Hose}, {Hrupec}, {Inoue}, {Ishio}, {Konno}, {Kubo}, {Kushida}, {Lelas},
  {Lindfors}, {Lombardi}, {Longo}, {L{\'o}pez}, {Maggio}, {Majumdar},
  {Makariev}, {Maneva}, {Manganaro}, {Mannheim}, {Maraschi}, {Mariotti},
  {Mart{\'\i}nez}, {Masuda}, {Mazin}, {Minev}, {M}, {Mirzoyan}, {Moralejo},
  {Moreno}, {Moretti}, {Nagayoshi}, {Neustroev}, {Niedzwiecki}, {Nievas
  Rosillo}, {Nigro}, {Nilsson}, {Ninci}, {Nishijima}, {Noda}, {Nogu{\'e}s},
  {Paiano}, {Palacio}, {Paneque}, {Paoletti}, {Paredes}, {Pedaletti},
  {Peresano}, {Persic}, {Prada Moroni}, {Prandini}, {Puljak}, {Rodriguez
  Garcia}, {Reichardt}, {Rhode}, {Rib{\'o}}, {Rico}, {Righi}, {Rugliancich},
  {Saito}, {Satalecka}, {Schweizer}, {Sitarek}, {{\v{S}}nidaric
  {\textasciiacute}}, {Sobczynska}, {Stamerra}, {Strzys}, {Suri{\'c}},
  {Takahashi}, {Tavecchio}, {Temnikov}, {Terzi{\'c}}, {Teshima},
  {Torres-Alb{\`a}}, {Treves}, {Tsujimoto}, {Vanzo}, {Vazquez Acosta}, {Vovk},
  {Ward}, {Will}, {S}, {Zaric {\textasciiacute}}, {AGILE Team}, {Lucarelli},
  {Tavani}, {Piano}, {Donnarumma}, {Pittori}, {Verrecchia}, {Barbiellini},
  {Bulgarelli}, {Caraveo}, {Cattaneo}, {Colafrancesco}, {Costa}, {Di Cocco},
  {Ferrari}, {Gianotti}, {Giuliani}, {Lipari}, {Mereghetti}, {Morselli},
  {Pacciani}, {Paoletti}, {Parmiggiani}, {Pellizzoni}, {Picozza}, {Pilia},
  {Rappoldi}, {Trois}, {Vercellone}, {Vittorini}, {ASAS-SN Team}, {Stanek},
  {Franckowiak}, {Kochanek}, {Beacom}, {Thompson}, {Holoien}, {Dong}, {Prieto},
  {Shappee}, {Holmbo}, {HAWC Collaboration}, {Abeysekara}, {Albert}, {Alfaro},
  {Alvarez}, {Arceo}, {Arteaga-Vel{\'a}zquez}, {Avila Rojas}, {Ayala Solares},
  {Becerril}, {Belmont-Moreno}, {Bernal}, {Caballero-Mora}, {Capistr{\'a}n},
  {Carrami{\~n}ana}, {Casanova}, {Castillo}, {Cotti}, {Cotzomi}, {Couti{\~n}o
  de Le{\'o}n}, {De Le{\'o}n}, {De la Fuente}, {Diaz Hernandez}, {Dichiara},
  {Dingus}, {DuVernois}, {D{\'\i}az-V{\'e}lez}, {Ellsworth}, {Engel},
  {Fiorino}, {Fleischhack}, {Fraija}, {Garc{\'\i}a-Gonz{\'a}lez}, {Garfias},
  {Gonz{\'a}lez Mu{\~n}oz}, {Gonz{\'a}lez}, {Goodman}, {Hampel-Arias},
  {Harding}, {Hernandez}, {Hona}, {Hueyotl-Zahuantitla}, {Hui},
  {H{\"u}ntemeyer}, {Iriarte}, {Jardin-Blicq}, {Joshi}, {Kaufmann}, {Kunde},
  {Lara}, {Lauer}, {Lee}, {Lennarz}, {Le{\'o}n Vargas}, {Linnemann},
  {Longinotti}, {Luis-Raya}, {Luna-Garc{\'\i}a}, {Malone}, {Marinelli},
  {Martinez}, {Martinez-Castellanos}, {Mart{\'\i}nez-Castro},
  {Mart{\'\i}nez-Huerta}, {Matthews}, {Miranda-Romagnoli}, {Moreno},
  {Mostaf{\'a}}, {Nayerhoda}, {Nellen}, {Newbold}, {Nisa}, {Noriega-Papaqui},
  {Pelayo}, {Pretz}, {P{\'e}rez-P{\'e}rez}, {Ren}, {Rho}, {Rivi{\`e}re},
  {Rosa-Gonz{\'a}lez}, {Rosenberg}, {Ruiz-Velasco}, {Ruiz-Velasco}, {Salesa
  Greus}, {Sandoval}, {Schneider}, {Schoorlemmer}, {Sinnis}, {Smith},
  {Springer}, {Surajbali}, {Tibolla}, {Tollefson}, {Torres}, {Villase{\~n}or},
  {Weisgarber}, {Werner}, {Yapici}, {Gaurang}, {Zepeda}, {Zhou}, {{\'A}lvarez},
  {H.~E.~S.~S. Collaboration}, {Abdalla}, {Ang{\"u}ner}, {Armand}, {Backes},
  {Becherini}, {Berge}, {B{\"o}ttcher}, {Boisson}, {Bolmont}, {Bonnefoy},
  {Bordas}, {Brun}, {B{\"u}chele}, {Bulik}, {Caroff}, {Carosi}, {Casanova},
  {Cerruti}, {Chakraborty}, {Chandra}, {Chen}, {Colafrancesco}, {Davids},
  {Deil}, {Devin}, {Djannati-Ata{\"\i}}, {Egberts}, {Emery}, {Eschbach},
  {Fiasson}, {Fontaine}, {Funk}, {F{\"u}{\ss}ling}, {Gallant}, {Gat{\'e}},
  {Giavitto}, {Glawion}, {Glicenstein}, {Gottschall}, {Grondin}, {Haupt},
  {Henri}, {Hinton}, {Hoischen}, {Holch}, {Huber}, {Jamrozy}, {Jankowsky},
  {Jankowsky}, {Jouvin}, {Jung-Richardt}, {Kerszberg}, {Kh{\'e}lifi}, {King},
  {Klepser}, {Kluz {\textasciiacute}niak}, {Komin}, {Kraus}, {Lefaucheur},
  {Lemi{\`e}re}, {Lemoine-Goumard}, {Lenain}, {Leser}, {Lohse},
  {L{\'o}pez-Coto}, {Lorentz}, {Lypova}, {Marandon}, {Guillem
  Mart{\'\i}-Devesa}, {Maurin}, {Mitchell}, {Moderski}, {Mohamed}, {Mohrmann},
  {Moulin}, {Murach}, {de Naurois}, {Niederwanger}, {Niemiec}, {Oakes},
  {O'Brien}, {Ohm}, {Ostrowski}, {Oya}, {Panter}, {Parsons}, {Perennes},
  {Piel}, {Pita}, {Poireau}, {Priyana Noel}, {Prokoph}, {P{\"u}hlhofer},
  {Quirrenbach}, {Raab}, {Rauth}, {Renaud}, {Rieger}, {Rinchiuso}, {Romoli},
  {Rowell}, {Rudak}, {Sasaki}, {Sanchez}, {Schlickeiser}, {Sch{\"u}ssler},
  {Schulz}, {Schwanke}, {Seglar-Arroyo}, {Shafi}, {Simoni}, {Sol}, {Stegmann},
  {Steppa}, {Tavernier}, {Taylor}, {Tiziani}, {Trichard}, {Tsirou}, {van
  Eldik}, {van Rensburg}, {van Soelen}, {Veh}, {Vincent}, {Voisin}, {Wagner},
  {Wagner}, {Wierzcholska}, {Zanin}, {Zdziarski}, {Zech}, {Ziegler}, {Zorn},
  {{\.Z}ywucka}, {INTEGRAL Team}, {Savchenko}, {Ferrigno}, {Bazzano}, {Diehl},
  {Kuulkers}, {Laurent}, {Mereghetti}, {Natalucci}, {Panessa}, {Rodi},
  {Ubertini}, {Kanata}, Teams, {Morokuma}, {Ohta}, {Tanaka}, {Mori},
  {Yamanaka}, {Kawabata}, {Utsumi}, {Nakaoka}, {Kawabata}, {Nagashima},
  {Yoshida}, {Matsuoka}, {Itoh}, {Kapteyn Team}, {Keel}, {Liverpool Telescope
  Team}, {Copperwheat}, {Steele}, {Swift/NuSTAR Team}, {Cenko}, {Cowen},
  {DeLaunay}, {Evans}, {Fox}, {Keivani}, {Kennea}, {Marshall}, {Osborne},
  {Santander}, {Tohuvavohu}, {Turley}, {VERITAS Collaboration}, {Abeysekara},
  {Archer}, {Benbow}, {Bird}, {Brill}, {Brose}, {Buchovecky}, {Buckley},
  {Bugaev}, {Christiansen}, {Connolly}, {Cui}, {Daniel}, {Errando}, {Falcone},
  {Feng}, {Finley}, {Fortson}, {Furniss}, {Gueta}, {H{\"u}tten}, {Hervet},
  {Hughes}, {Humensky}, {Johnson}, {Kaaret}, {Kar}, {Kelley-Hoskins},
  {Kertzman}, {Kieda}, {Krause}, {Krennrich}, {Kumar}, {Lang}, {Lin}, {Maier},
  {McArthur}, {Moriarty}, {Mukherjee}, {Nieto}, {O'Brien}, {Ong}, {Otte},
  {Park}, {Petrashyk}, {Pohl}, {Popkow}, {Pueschel}, {Quinn}, {Ragan},
  {Reynolds}, {Richards}, {Roache}, {Rulten}, {Sadeh}, {Santander}, {Scott},
  {Sembroski}, {Shahinyan}, {Sushch}, {Tr{\'e}panier}, {Tyler}, {Vassiliev},
  {Wakely}, {Weinstein}, {Wells}, {Wilcox}, {Wilhelm}, {Williams}, {Zitzer},
  {VLA/B Team}, {Tetarenko}, {Kimball}, {Miller-Jones}, \&
  {Sivakoff}}]{IceCube:2018dnn}
{Aartsen}, M.~G., {Ackermann}, M., {Adams}, J., {et~al.} 2018{\natexlab{a}},
  Science, 361, eaat1378, \dodoi{10.1126/science.aat1378}

\bibitem[{{Aartsen} {et~al.}(2018{\natexlab{b}}){Aartsen}, {Ackermann},
  {Adams}, {Aguilar}, {Ahlers}, {Ahrens}, {Samarai}, {Altmann}, {Andeen},
  {Anderson}, {Ansseau}, {Anton}, {Arg{\"u}elles}, {Arsioli}, {Auffenberg},
  {Axani}, {Bagherpour}, {Bai}, {Barron}, {Barwick}, {Baum}, {Bay}, {Beatty},
  {Becker Tjus}, {Becker}, {BenZvi}, {Berley}, {Bernardini}, {Besson},
  {Binder}, {Bindig}, {Blaufuss}, {Blot}, {Bohm}, {B{\"o}rner}, {Bos},
  {B{\"o}ser}, {Botner}, {Bourbeau}, {Bourbeau}, {Bradascio}, {Braun},
  {Brenzke}, {Bretz}, {Bron}, {Brostean-Kaiser}, {Burgman}, {Busse}, {Carver},
  {Cheung}, {Chirkin}, {Christov}, {Clark}, {Classen}, {Coenders}, {Collin},
  {Conrad}, {Coppin}, {Correa}, {Cowen}, {Cross}, {Dave}, {Day}, {de
  Andr{\'e}}, {De Clercq}, {DeLaunay}, {Dembinski}, {DeRidder}, {Desiati}, {de
  Vries}, {de Wasseige}, {de With}, {DeYoung}, {D{\'\i}az-V{\'e}lez}, {di
  Lorenzo}, {Dujmovic}, {Dumm}, {Dunkman}, {Dvorak}, {Eberhardt}, {Ehrhardt},
  {Eichmann}, {Eller}, {Evenson}, {Fahey}, {Fazely}, {Felde}, {Filimonov},
  {Finley}, {Flis}, {Franckowiak}, {Friedman}, {Fritz}, {Gaisser}, {Gallagher},
  {Gerhardt}, {Ghorbani}, {Giommi}, {Glauch}, {Gl{\"u}senkamp}, {Goldschmidt},
  {Gonzalez}, {Grant}, {Griffith}, {Haack}, {Hallgren}, {Halzen}, {Hanson},
  {Hebecker}, {Heereman}, {Helbing}, {Hellauer}, {Hickford}, {Hignight},
  {Hill}, {Hoffman}, {Hoffmann}, {Hoinka}, {Hokanson-Fasig}, {Hoshina},
  {Huang}, {Huber}, {Hultqvist}, {H{\"u}nnefeld}, {Hussain}, {In}, {Iovine},
  {Ishihara}, {Jacobi}, {Japaridze}, {Jeong}, {Jero}, {Jones}, {Kalaczynski},
  {Kang}, {Kappes}, {Kappesser}, {Karg}, {Karle}, {Katz}, {Kauer}, {Keivani},
  {Kelley}, {Kheirandish}, {Kim}, {Kim}, {Kintscher}, {Kiryluk}, {Kittler},
  {Klein}, {Koirala}, {Kolanoski}, {K{\"o}pke}, {Kopper}, {Kopper},
  {Koschinsky}, {Koskinen}, {Kowalski}, {Krammer}, {Krings}, {Kroll},
  {Kr{\"u}ckl}, {Kunwar}, {Kurahashi}, {Kuwabara}, {Kyriacou}, {Labare},
  {Lanfranchi}, {Larson}, {Lauber}, {Leonard}, {Lesiak-Bzdak}, {Leuermann},
  {Liu}, {Lozano Mariscal}, {Lu}, {L{\"u}nemann}, {Luszczak}, {Madsen},
  {Maggi}, {Mahn}, {Mancina}, {Maruyama}, {Mase}, {Maunu}, {Meagher}, {Medici},
  {Meier}, {Menne}, {Merino}, {Meures}, {Miarecki}, {Micallef}, {Moment{\'e}},
  {Montaruli}, {Moore}, {Morse}, {Moulai}, {Nahnhauer}, {Nakarmi}, {Naumann},
  {Neer}, {Niederhausen}, {Nowicki}, {Nygren}, {Obertacke Pollmann}, {Olivas},
  {O'Murchadha}, {O'Sullivan}, {Padovani}, {Palczewski}, {Pandya}, {Pankova},
  {Peiffer}, {Pepper}, {P{\'e}rez de los Heros}, {Pieloth}, {Pinat}, {Plum},
  {Price}, {Przybylski}, {Raab}, {R{\"a}del}, {Rameez}, {Rawlins}, {Rea},
  {Reimann}, {Relethford}, {Relich}, {Resconi}, {Rhode}, {Richman},
  {Robertson}, {Rongen}, {Rott}, {Ruhe}, {Ryckbosch}, {Rysewyk}, {Safa},
  {Sahakyan}, {S{\"a}lzer}, {Sanchez Herrera}, {Sandrock}, {Sandroos},
  {Santander}, {Sarkar}, {Sarkar}, {Satalecka}, {Schlunder}, {Schmidt},
  {Schneider}, {Schoenen}, {Sch{\"o}neberg}, {Schumacher}, {Sclafani},
  {Seckel}, {Seunarine}, {Soedingrekso}, {Soldin}, {Song}, {Spiczak},
  {Spiering}, {Stachurska}, {Stamatikos}, {Stanev}, {Stasik}, {Stettner},
  {Steuer}, {Stezelberger}, {Stokstad}, {St{\"o}{\ss}l}, {Strotjohann},
  {Stuttard}, {Sullivan}, {Sutherland}, {Taboada}, {Tatar}, {Tenholt},
  {Ter-Antonyan}, {Terliuk}, {Tilav}, {Toale}, {Tobin}, {Toennis}, {Toscano},
  {Tosi}, {Tselengidou}, {Tung}, {Turcati}, {Turley}, {Ty}, {Unger}, {Usner},
  {Vandenbroucke}, {Van Driessche}, {van Eijk}, {van Eijndhoven}, {Vanheule},
  {van Santen}, {Vogel}, {Vraeghe}, {Walck}, {Wallace}, {Wallraff}, {Wandler},
  {Wandkowsky}, {Waza}, {Weaver}, {Weiss}, {Wendt}, {Werthebach}, {Westerhoff},
  {Whelan}, {Whitehorn}, {Wiebe}, {Wiebusch}, {Wille}, {Williams}, {Wills},
  {Wolf}, {Wood}, {Wood}, {Woschnagg}, {Xu}, {Xu}, {Xu}, {Yanez}, {Yodh},
  {Yoshida}, {Yuan}, \& {The IceCube Collaboration}}]{IceCube:2018cha}
---. 2018{\natexlab{b}}, Science, 361, 147, \dodoi{10.1126/science.aat2890}

\bibitem[{{Aartsen} {et~al.}(2018{\natexlab{c}}){Aartsen}, {Ackermann},
  {Adams}, {Aguilar}, {Ahlers}, {Ahrens}, {Al Samarai}, {Altmann}, {Andeen},
  {Anderson}, {Ansseau}, {Anton}, {Arg{\"u}elles}, {Auffenberg}, {Axani},
  {Backes}, {Bagherpour}, {Bai}, {Barbano}, {Barron}, {Barwick}, {Baum}, {Bay},
  {Beatty}, {Becker Tjus}, {Becker}, {BenZvi}, {Berley}, {Bernardini},
  {Besson}, {Binder}, {Bindig}, {Blaufuss}, {Blot}, {Bohm}, {B{\"o}rner},
  {Bos}, {B{\"o}ser}, {Botner}, {Bourbeau}, {Bourbeau}, {Bradascio}, {Braun},
  {Brenzke}, {Bretz}, {Bron}, {Brostean-Kaiser}, {Burgman}, {Busse}, {Carver},
  {Cheung}, {Chirkin}, {Christov}, {Clark}, {Classen}, {Collin}, {Conrad},
  {Coppin}, {Correa}, {Cowen}, {Cross}, {Dave}, {Day}, {de Andr{\'e}}, {De
  Clercq}, {DeLaunay}, {Dembinski}, {Deoskar}, {De Ridder}, {Desiati}, {de
  Vries}, {de Wasseige}, {de With}, {DeYoung}, {D{\'\i}az-V{\'e}lez}, {di
  Lorenzo}, {Dujmovic}, {Dumm}, {Dunkman}, {Dvorak}, {Eberhardt}, {Ehrhardt},
  {Eichmann}, {Eller}, {Evenson}, {Fahey}, {Fazely}, {Felde}, {Filimonov},
  {Finley}, {Flis}, {Franckowiak}, {Friedman}, {Fritz}, {Gaisser}, {Gallagher},
  {Ganster}, {Gerhardt}, {Ghorbani}, {Giang}, {Glauch}, {Gl{\"u}senkamp},
  {Goldschmidt}, {Gonzalez}, {Grant}, {Griffith}, {Haack}, {Hallgren}, {Halve},
  {Halzen}, {Hanson}, {Hebecker}, {Heereman}, {Helbing}, {Hellauer},
  {Hickford}, {Hignight}, {Hill}, {Hoffman}, {Hoffmann}, {Hoinka},
  {Hokanson-Fasig}, {Hoshina}, {Huang}, {Huber}, {Hultqvist}, {H{\"u}nnefeld},
  {Hussain}, {In}, {Iovine}, {Ishihara}, {Jacobi}, {Japaridze}, {Jeong},
  {Jero}, {Jones}, {Kalaczynski}, {Kang}, {Kappes}, {Kappesser}, {Karg},
  {Karle}, {Katz}, {Kauer}, {Keivani}, {Kelley}, {Kheirandish}, {Kim},
  {Kintscher}, {Kiryluk}, {Kittler}, {Klein}, {Koirala}, {Kolanoski},
  {K{\"o}pke}, {Kopper}, {Kopper}, {Koschinsky}, {Koskinen}, {Kowalski},
  {Krings}, {Kroll}, {Kr{\"u}ckl}, {Kunwar}, {Kurahashi}, {Kyriacou}, {Labare},
  {Lanfranchi}, {Larson}, {Lauber}, {Leonard}, {Leuermann}, {Liu}, {Lohfink},
  {Lozano Mariscal}, {Lu}, {L{\"u}nemann}, {Luszczak}, {Madsen}, {Maggi},
  {Mahn}, {Makino}, {Mancina}, {Mari{\c{s}}}, {Maruyama}, {Mase}, {Maunu},
  {Meagher}, {Medici}, {Meier}, {Menne}, {Merino}, {Meures}, {Miarecki},
  {Micallef}, {Moment{\'e}}, {Montaruli}, {Moore}, {Moulai}, {Nagai},
  {Nahnhauer}, {Nakarmi}, {Naumann}, {Neer}, {Niederhausen}, {Nowicki},
  {Nygren}, {Obertacke Pollmann}, {Olivas}, {O'Murchadha}, {O'Sullivan},
  {Palczewski}, {Pandya}, {Pankova}, {Peiffer}, {Pepper}, {P{\'e}rez de los
  Heros}, {Pieloth}, {Pinat}, {Pizzuto}, {Plum}, {Price}, {Przybylski}, {Raab},
  {R{\"a}del}, {Rameez}, {Rauch}, {Rawlins}, {Rea}, {Reimann}, {Relethford},
  {Renzi}, {Resconi}, {Rhode}, {Richman}, {Robertson}, {Rongen}, {Rott},
  {Ruhe}, {Ryckbosch}, {Rysewyk}, {Safa}, {Sanchez Herrera}, {Sandrock},
  {Sandroos}, {Santander}, {Sarkar}, {Sarkar}, {Satalecka}, {Schaufel},
  {Schlunder}, {Schmidt}, {Schneider}, {Schoenen}, {Sch{\"o}neberg},
  {Schumacher}, {Sclafani}, {Seckel}, {Seunarine}, {Soedingrekso}, {Soldin},
  {Song}, {Spiczak}, {Spiering}, {Stachurska}, {Stamatikos}, {Stanev},
  {Stasik}, {Stein}, {Stettner}, {Steuer}, {Stezelberger}, {Stokstad},
  {St{\"o}{\ss}l}, {Strotjohann}, {Stuttard}, {Sullivan}, {Sutherland},
  {Taboada}, {Tenholt}, {Ter-Antonyan}, {Terliuk}, {Tilav}, {Toale}, {Tobin},
  {T{\"o}nnis}, {Toscano}, {Tosi}, {Tselengidou}, {Tung}, {Turcati}, {Turley},
  {Ty}, {Unger}, {Usner}, {Vandenbroucke}, {Van Driessche}, {van Eijk}, {van
  Eijndhoven}, {Vanheule}, {van Santen}, {Vraeghe}, {Walck}, {Wallace},
  {Wallraff}, {Wandler}, {Wandkowsky}, {Watson}, {Waza}, {Weaver}, {Weiss},
  {Wendt}, {Werthebach}, {Westerhoff}, {Whelan}, {Whitehorn}, {Wiebe},
  {Wiebusch}, {Wille}, {Williams}, {Wills}, {Wolf}, {Wood}, {Wood}, {Woolsey},
  {Woschnagg}, {Wrede}, {Xu}, {Xu}, {Xu}, {Yanez}, {Yodh}, {Yoshida}, {Yuan},
  \& {The IceCube Collaboration}}]{Aartsen:2018vtx}
---. 2018{\natexlab{c}}, \prd, 98, 062003, \dodoi{10.1103/PhysRevD.98.062003}

\bibitem[{{Aartsen} {et~al.}(2019){Aartsen}, {Ackermann}, {Adams}, {Aguilar},
  {Ahlers}, {Ahrens}, {Altmann}, {Andeen}, {Anderson}, {Ansseau}, {Anton},
  {Arg{\"u}elles}, {Auffenberg}, {Axani}, {Backes}, {Bagherpour}, {Bai},
  {Barbano}, {Barron}, {Barwick}, {Baum}, {Bay}, {Beatty}, {Becker Tjus},
  {Becker}, {BenZvi}, {Berley}, {Bernardini}, {Besson}, {Binder}, {Bindig},
  {Blaufuss}, {Blot}, {Bohm}, {B{\"o}rner}, {Bos}, {B{\"o}ser}, {Botner},
  {Bourbeau}, {Bourbeau}, {Bradascio}, {Braun}, {Bretz}, {Bron},
  {Brostean-Kaiser}, {Burgman}, {Busse}, {Carver}, {Chen}, {Cheung}, {Chirkin},
  {Clark}, {Classen}, {Collin}, {Conrad}, {Coppin}, {Correa}, {Cowen}, {Cross},
  {Dave}, {Day}, {de Andr{\'e}}, {De Clercq}, {DeLaunay}, {Dembinski},
  {Deoskar}, {De Ridder}, {Desiati}, {de Vries}, {de Wasseige}, {de With},
  {DeYoung}, {D{\'\i}az-V{\'e}lez}, {Dujmovic}, {Dunkman}, {Dvorak},
  {Eberhardt}, {Ehrhardt}, {Eichmann}, {Eller}, {Evenson}, {Fahey}, {Fazely},
  {Felde}, {Filimonov}, {Finley}, {Franckowiak}, {Friedman}, {Fritz},
  {Gaisser}, {Gallagher}, {Ganster}, {Garrappa}, {Gerhardt}, {Ghorbani},
  {Giang}, {Glauch}, {Gl{\"u}senkamp}, {Goldschmidt}, {Gonzalez}, {Grant},
  {Griffith}, {Haack}, {Hallgren}, {Halve}, {Halzen}, {Hanson}, {Hebecker},
  {Heereman}, {Helbing}, {Hellauer}, {Hickford}, {Hignight}, {Hill}, {Hoffman},
  {Hoffmann}, {Hoinka}, {Hokanson-Fasig}, {Hoshina}, {Huang}, {Huber},
  {Hultqvist}, {H{\"u}nnefeld}, {Hussain}, {In}, {Iovine}, {Ishihara},
  {Jacobi}, {Japaridze}, {Jeong}, {Jero}, {Jones}, {Kalaczynski}, {Kang},
  {Kappes}, {Kappesser}, {Karg}, {Karle}, {Katz}, {Kauer}, {Keivani}, {Kelley},
  {Kheirandish}, {Kim}, {Kintscher}, {Kiryluk}, {Kittler}, {Klein}, {Koirala},
  {Kolanoski}, {K{\"o}pke}, {Kopper}, {Kopper}, {Koskinen}, {Kowalski},
  {Krings}, {Kroll}, {Kr{\"u}ckl}, {Kunwar}, {Kurahashi}, {Kyriacou}, {Labare},
  {Lanfranchi}, {Larson}, {Lauber}, {Leonard}, {Leuermann}, {Liu}, {Lohfink},
  {Mariscal}, {Lu}, {L{\"u}nemann}, {Luszczak}, {Madsen}, {Maggi}, {Mahn},
  {Makino}, {Mancina}, {Mari{\c{s}}}, {Maruyama}, {Mase}, {Maunu}, {Meagher},
  {Medici}, {Meier}, {Menne}, {Merino}, {Meures}, {Miarecki}, {Micallef},
  {Moment{\'e}}, {Montaruli}, {Moore}, {Moulai}, {Nagai}, {Nahnhauer},
  {Nakarmi}, {Naumann}, {Neer}, {Niederhausen}, {Nowicki}, {Nygren}, {Obertacke
  Pollmann}, {Olivas}, {O'Murchadha}, {O'Sullivan}, {Palczewski}, {Pandya},
  {Pankova}, {Peiffer}, {P{\'e}rez de los Heros}, {Pieloth}, {Pinat},
  {Pizzuto}, {Plum}, {Price}, {Przybylski}, {Raab}, {Rameez}, {Rauch},
  {Rawlins}, {Rea}, {Reimann}, {Relethford}, {Renzi}, {Resconi}, {Rhode},
  {Richman}, {Robertson}, {Rongen}, {Rott}, {Ruhe}, {Ryckbosch}, {Rysewyk},
  {Safa}, {Sanchez Herrera}, {Sandrock}, {Sandroos}, {Santander}, {Sarkar},
  {Sarkar}, {Satalecka}, {Schaufel}, {Schlunder}, {Schmidt}, {Schneider},
  {Schneider}, {Sch{\"o}neberg}, {Schumacher}, {Sclafani}, {Seckel},
  {Seunarine}, {Soedingrekso}, {Soldin}, {Song}, {Spiczak}, {Spiering},
  {Stachurska}, {Stamatikos}, {Stanev}, {Stasik}, {Stein}, {Stettner},
  {Steuer}, {Stezelberger}, {Stokstad}, {St{\"o}{\ss}l}, {Strotjohann},
  {Stuttard}, {Sullivan}, {Sutherland}, {Taboada}, {Tenholt}, {Ter-Antonyan},
  {Terliuk}, {Tilav}, {Tobin}, {T{\"o}nnis}, {Toscano}, {Tosi}, {Tselengidou},
  {Tung}, {Turcati}, {Turcotte}, {Turley}, {Ty}, {Unger}, {Unland Elorrieta},
  {Usner}, {Vandenbroucke}, {Van Driessche}, {van Eijk}, {van Eijndhoven},
  {Vanheule}, {van Santen}, {Vraeghe}, {Walck}, {Wallace}, {Wallraff},
  {Wandler}, {Wandkowsky}, {Watson}, {Weaver}, {Weiss}, {Wendt}, {Werthebach},
  {Westerhoff}, {Whelan}, {Whitehorn}, {Wiebe}, {Wiebusch}, {Wille},
  {Williams}, {Wills}, {Wolf}, {Wood}, {Wood}, {Woolsey}, {Woschnagg}, {Wrede},
  {Xu}, {Xu}, {Xu}, {Yanez}, {Yodh}, {Yoshida}, \&
  {Yuan}}]{aartsen8yrDiffusePS2019}
---. 2019, European Physical Journal C, 79, 234,
  \dodoi{10.1140/epjc/s10052-019-6680-0}

\bibitem[{{Aartsen} {et~al.}(2020{\natexlab{a}}){Aartsen}, {Ackermann},
  {Adams}, {Aguilar}, {Ahlers}, {Ahrens}, {Alispach}, {Andeen}, {Anderson},
  {Ansseau}, {Anton}, {Arg{\"u}elles}, {Auffenberg}, {Axani}, {Backes},
  {Bagherpour}, {Bai}, {Balagopal V.}, {Barbano}, {Barwick}, {Bastian}, {Baum},
  {Baur}, {Bay}, {Beatty}, {Becker}, {Becker Tjus}, {BenZvi}, {Berley},
  {Bernardini}, {Besson}, {Binder}, {Bindig}, {Blaufuss}, {Blot}, {Bohm},
  {B{\"o}ser}, {Botner}, {B{\"o}ttcher}, {Bourbeau}, {Bourbeau}, {Bradascio},
  {Braun}, {Bron}, {Brostean-Kaiser}, {Burgman}, {Buscher}, {Busse}, {Carver},
  {Chen}, {Cheung}, {Chirkin}, {Choi}, {Clark}, {Classen}, {Coleman}, {Collin},
  {Conrad}, {Coppin}, {Correa}, {Cowen}, {Cross}, {Dave}, {De Clercq},
  {DeLaunay}, {Dembinski}, {Deoskar}, {De Ridder}, {Desiati}, {de Vries}, {de
  Wasseige}, {de With}, {DeYoung}, {Diaz}, {D{\'\i}az-V{\'e}lez}, {Dujmovic},
  {Dunkman}, {Dvorak}, {Eberhardt}, {Ehrhardt}, {Eller}, {Engel}, {Evenson},
  {Fahey}, {Fazely}, {Felde}, {Filimonov}, {Finley}, {Fox}, {Franckowiak},
  {Friedman}, {Fritz}, {Gaisser}, {Gallagher}, {Ganster}, {Garrappa},
  {Gerhardt}, {Ghorbani}, {Glauch}, {Gl{\"u}senkamp}, {Goldschmidt},
  {Gonzalez}, {Grant}, {Gr{\'e}goire}, {Griffith}, {Griswold}, {G{\"u}nder},
  {G{\"u}nd{\"u}z}, {Haack}, {Hallgren}, {Halliday}, {Halve}, {Halzen},
  {Hanson}, {Haungs}, {Hebecker}, {Heereman}, {Heix}, {Helbing}, {Hellauer},
  {Henningsen}, {Hickford}, {Hignight}, {Hill}, {Hoffman}, {Hoffmann},
  {Hoinka}, {Hokanson-Fasig}, {Hoshina}, {Huang}, {Huber}, {Huber},
  {Hultqvist}, {H{\"u}nnefeld}, {Hussain}, {In}, {Iovine}, {Ishihara},
  {Jansson}, {Japaridze}, {Jeong}, {Jero}, {Jones}, {Jonske}, {Joppe}, {Kang},
  {Kang}, {Kappes}, {Kappesser}, {Karg}, {Karl}, {Karle}, {Katz}, {Kauer},
  {Kelley}, {Kheirandish}, {Kim}, {Kintscher}, {Kiryluk}, {Kittler}, {Klein},
  {Koirala}, {Kolanoski}, {K{\"o}pke}, {Kopper}, {Kopper}, {Koskinen},
  {Kowalski}, {Krings}, {Kr{\"u}ckl}, {Kulacz}, {Kurahashi}, {Kyriacou},
  {Lanfranchi}, {Larson}, {Lauber}, {Lazar}, {Leonard}, {Lesiak-Bzdak},
  {Leszczy{\'n}ska}, {Leuermann}, {Liu}, {Lohfink}, {Lozano Mariscal}, {Lu},
  {Lucarelli}, {L{\"u}nemann}, {Luszczak}, {Lyu}, {Ma}, {Madsen}, {Maggi},
  {Mahn}, {Makino}, {Mallik}, {Mallot}, {Mancina}, {Mari{\c{s}}}, {Maruyama},
  {Mase}, {Maunu}, {McNally}, {Meagher}, {Medici}, {Medina}, {Meier},
  {Meighen-Berger}, {Merino}, {Meures}, {Micallef}, {Mockler}, {Moment{\'e}},
  {Montaruli}, {Moore}, {Morse}, {Moulai}, {Muth}, {Nagai}, {Naumann}, {Neer},
  {Niederhausen}, {Nisa}, {Nowicki}, {Nygren}, {Obertacke Pollmann}, {Oehler},
  {Olivas}, {O'Murchadha}, {O'Sullivan}, {Palczewski}, {Pandya}, {Pankova},
  {Park}, {Peiffer}, {P{\'e}rez de los Heros}, {Philippen}, {Pieloth},
  {Pieper}, {Pinat}, {Pizzuto}, {Plum}, {Porcelli}, {Price}, {Przybylski},
  {Raab}, {Raissi}, {Rameez}, {Rauch}, {Rawlins}, {Rea}, {Rehman}, {Reimann},
  {Relethford}, {Renschler}, {Renzi}, {Resconi}, {Rhode}, {Richman},
  {Robertson}, {Rongen}, {Rott}, {Ruhe}, {Ryckbosch}, {Rysewyk}, {Safa},
  {Sanchez Herrera}, {Sandrock}, {Sandroos}, {Santander}, {Sarkar}, {Sarkar},
  {Satalecka}, {Schaufel}, {Schieler}, {Schlunder}, {Schmidt}, {Schneider},
  {Schneider}, {Schr{\"o}der}, {Schumacher}, {Sclafani}, {Seckel}, {Seunarine},
  {Shefali}, {Silva}, {Snihur}, {Soedingrekso}, {Soldin}, {Song}, {Spiczak},
  {Spiering}, {Stachurska}, {Stamatikos}, {Stanev}, {Stein}, {Stettner},
  {Steuer}, {Stezelberger}, {Stokstad}, {St{\"o}{\ss}l}, {Strotjohann},
  {St{\"u}rwald}, {Stuttard}, {Sullivan}, {Taboada}, {Tenholt}, {Ter-Antonyan},
  {Terliuk}, {Tilav}, {Tollefson}, {Tomankova}, {T{\"o}nnis}, {Toscano},
  {Tosi}, {Trettin}, {Tselengidou}, {Tung}, {Turcati}, {Turcotte}, {Turley},
  {Ty}, {Unger}, {Unland Elorrieta}, {Usner}, {Vandenbroucke}, {Van Driessche},
  {van Eijk}, {van Eijndhoven}, {van Santen}, {Verpoest}, {Vraeghe}, {Walck},
  {Wallace}, {Wallraff}, {Wandkowsky}, {Watson}, {Weaver}, {Weindl}, {Weiss},
  {Weldert}, {Wendt}, {Werthebach}, {Whelan}, {Whitehorn}, {Wiebe}, {Wiebusch},
  {Wille}, {Williams}, {Wills}, {Wolf}, {Wood}, {Wood}, {Woschnagg}, {Wrede},
  {Xu}, {Xu}, {Xu}, {Yanez}, {Yodh}, {Yoshida}, {Yuan}, {Z{\"o}cklein}, \& {The
  IceCube Collaboration}}]{Aartsen:2020aqd}
---. 2020{\natexlab{a}}, \prl, 125, 121104,
  \dodoi{10.1103/PhysRevLett.125.121104}

\bibitem[{{Aartsen} {et~al.}(2020{\natexlab{b}}){Aartsen}, {Ackermann},
  {Adams}, {Aguilar}, {Ahlers}, {Ahrens}, {Alispach}, {Andeen}, {Anderson},
  {Ansseau}, {Anton}, {Arg{\"u}elles}, {Auffenberg}, {Axani}, {Backes},
  {Bagherpour}, {Bai}, {Balagopal}, {Barbano}, {Barwick}, {Bastian}, {Baum},
  {Baur}, {Bay}, {Beatty}, {Becker}, {Becker Tjus}, {BenZvi}, {Berley},
  {Bernardini}, {Besson}, {Binder}, {Bindig}, {Blaufuss}, {Blot}, {Bohm},
  {B{\"o}rner}, {B{\"o}ser}, {Botner}, {B{\"o}ttcher}, {Bourbeau}, {Bourbeau},
  {Bradascio}, {Braun}, {Bron}, {Brostean-Kaiser}, {Burgman}, {Buscher},
  {Busse}, {Carver}, {Chen}, {Cheung}, {Chirkin}, {Choi}, {Clark}, {Classen},
  {Coleman}, {Collin}, {Conrad}, {Coppin}, {Correa}, {Cowen}, {Cross}, {Dave},
  {De Clercq}, {DeLaunay}, {Dembinski}, {Deoskar}, {De Ridder}, {Desiati}, {de
  Vries}, {de Wasseige}, {de With}, {DeYoung}, {Diaz}, {D{\'\i}az-V{\'e}lez},
  {Dujmovic}, {Dunkman}, {Dvorak}, {Eberhardt}, {Ehrhardt}, {Eller}, {Engel},
  {Evenson}, {Fahey}, {Fazely}, {Felde}, {Filimonov}, {Finley}, {Fox},
  {Franckowiak}, {Friedman}, {Fritz}, {Gaisser}, {Gallagher}, {Ganster},
  {Garrappa}, {Gerhardt}, {Ghorbani}, {Glauch}, {Gl{\"u}senkamp},
  {Goldschmidt}, {Gonzalez}, {Grant}, {Griffith}, {Griswold}, {G{\"u}nder},
  {G{\"u}nd{\"u}z}, {Haack}, {Hallgren}, {Halliday}, {Halve}, {Halzen},
  {Hanson}, {Haungs}, {Hebecker}, {Heereman}, {Heix}, {Helbing}, {Hellauer},
  {Henningsen}, {Hickford}, {Hignight}, {Hill}, {Hoffman}, {Hoffmann},
  {Hoinka}, {Hokanson-Fasig}, {Hoshina}, {Huang}, {Huber}, {Huber},
  {Hultqvist}, {H{\"u}nnefeld}, {Hussain}, {In}, {Iovine}, {Ishihara},
  {Japaridze}, {Jeong}, {Jero}, {Jones}, {Jonske}, {Joppe}, {Kang}, {Kang},
  {Kappes}, {Kappesser}, {Karg}, {Karl}, {Karle}, {Katz}, {Kauer}, {Kelley},
  {Kheirandish}, {Kim}, {Kintscher}, {Kiryluk}, {Kittler}, {Klein}, {Koirala},
  {Kolanoski}, {K{\"o}pke}, {Kopper}, {Kopper}, {Koskinen}, {Kowalski},
  {Krings}, {Kr{\"u}ckl}, {Kulacz}, {Kurahashi}, {Kyriacou}, {Labare},
  {Lanfranchi}, {Larson}, {Lauber}, {Lazar}, {Leonard}, {Leszczy{\'n}ska},
  {Leuermann}, {Liu}, {Lohfink}, {Lozano Mariscal}, {Lu}, {Lucarelli},
  {L{\"u}nemann}, {Luszczak}, {Lyu}, {Ma}, {Madsen}, {Maggi}, {Mahn}, {Makino},
  {Mallik}, {Mallot}, {Mancina}, {Mari{\c{s}}}, {Maruyama}, {Mase}, {Matis},
  {Maunu}, {McNally}, {Meagher}, {Medici}, {Medina}, {Meier}, {Meighen-Berger},
  {Menne}, {Merino}, {Meures}, {Micallef}, {Mockler}, {Moment{\'e}},
  {Montaruli}, {Moore}, {Morse}, {Moulai}, {Muth}, {Nagai}, {Naumann}, {Neer},
  {Niederhausen}, {Nisa}, {Nowicki}, {Nygren}, {Obertacke Pollmann}, {Oehler},
  {Olivas}, {O'Murchadha}, {O'Sullivan}, {Palczewski}, {Pandya}, {Pankova},
  {Park}, {Peiffer}, {P{\'e}rez de los Heros}, {Philippen}, {Pieloth}, {Pinat},
  {Pizzuto}, {Plum}, {Porcelli}, {Price}, {Przybylski}, {Raab}, {Raissi},
  {Rameez}, {Rauch}, {Rawlins}, {Rea}, {Reimann}, {Relethford}, {Renschler},
  {Renzi}, {Resconi}, {Rhode}, {Richman}, {Robertson}, {Rongen}, {Rott},
  {Ruhe}, {Ryckbosch}, {Rysewyk}, {Safa}, {Sanchez Herrera}, {Sandrock},
  {Sandroos}, {Santander}, {Sarkar}, {Sarkar}, {Satalecka}, {Schaufel},
  {Schieler}, {Schlunder}, {Schmidt}, {Schneider}, {Schneider}, {Schr{\"o}der},
  {Schumacher}, {Sclafani}, {Seckel}, {Seunarine}, {Shefali}, {Silva},
  {Snihur}, {Soedingrekso}, {Soldin}, {Song}, {Spiczak}, {Spiering},
  {Stachurska}, {Stamatikos}, {Stanev}, {Stein}, {Steinm{\"u}ller}, {Stettner},
  {Steuer}, {Stezelberger}, {Stokstad}, {St{\"o}{\ss}l}, {Strotjohann},
  {St{\"u}rwald}, {Stuttard}, {Sullivan}, {Taboada}, {Tenholt}, {Ter-Antonyan},
  {Terliuk}, {Tilav}, {Tollefson}, {Tomankova}, {T{\"o}nnis}, {Toscano},
  {Tosi}, {Trettin}, {Tselengidou}, {Tung}, {Turcati}, {Turcotte}, {Turley},
  {Ty}, {Unger}, {Unland Elorrieta}, {Usner}, {Vandenbroucke}, {Van Driessche},
  {van Eijk}, {van Eijndhoven}, {Vanheule}, {van Santen}, {Vraeghe}, {Walck},
  {Wallace}, {Wallraff}, {Wandkowsky}, {Watson}, {Weaver}, {Weindl}, {Weiss},
  {Weldert}, {Wendt}, {Werthebach}, {Whelan}, {Whitehorn}, {Wiebe}, {Wiebusch},
  {Wille}, {Williams}, {Wills}, {Wolf}, {Wood}, {Wood}, {Woschnagg}, {Wrede},
  {Xu}, {Xu}, {Xu}, {Yanez}, {Yodh}, {Yoshida}, {Yuan}, \&
  {Z{\"o}cklein}}]{aartsen10yrIntegratedPS2020}
---. 2020{\natexlab{b}}, \prl, 124, 051103,
  \dodoi{10.1103/PhysRevLett.124.051103}

\bibitem[{{Aartsen} {et~al.}(2020{\natexlab{c}}){Aartsen}, {Ackermann},
  {Adams}, {Aguilar}, {Ahlers}, {Ahrens}, {Alispach}, {Andeen}, {Anderson},
  {Ansseau}, {Anton}, {Arg{\"u}elles}, {Auffenberg}, {Axani}, {Backes},
  {Bagherpour}, {Bai}, {Balagopal}, {Barbano}, {Barwick}, {Bastian}, {Baum},
  {Baur}, {Bay}, {Beatty}, {Becker}, {Becker Tjus}, {BenZvi}, {Berley},
  {Bernardini}, {Besson}, {Binder}, {Bindig}, {Blaufuss}, {Blot}, {Bohm},
  {B{\"o}ser}, {Botner}, {B{\"o}ttcher}, {Bourbeau}, {Bourbeau}, {Bradascio},
  {Braun}, {Bron}, {Brostean-Kaiser}, {Burgman}, {Buscher}, {Busse}, {Carver},
  {Chen}, {Cheung}, {Chirkin}, {Choi}, {Clark}, {Classen}, {Coleman}, {Collin},
  {Conrad}, {Coppin}, {Correa}, {Cowen}, {Cross}, {Dave}, {De Clercq},
  {DeLaunay}, {Dembinski}, {Deoskar}, {De Ridder}, {Desiati}, {de Vries}, {de
  Wasseige}, {de With}, {DeYoung}, {Diaz}, {D{\'\i}az-V{\'e}lez}, {Dujmovic},
  {Dunkman}, {Dvorak}, {Eberhardt}, {Ehrhardt}, {Eller}, {Engel}, {Evenson},
  {Fahey}, {Fazely}, {Felde}, {Filimonov}, {Finley}, {Fox}, {Franckowiak},
  {Friedman}, {Fritz}, {Gaisser}, {Gallagher}, {Ganster}, {Garrappa},
  {Gerhardt}, {Ghorbani}, {Glauch}, {Gl{\"u}senkamp}, {Goldschmidt},
  {Gonzalez}, {Grant}, {Griffith}, {Griswold}, {G{\"u}nder}, {G{\"u}nd{\"u}z},
  {Haack}, {Hallgren}, {Halliday}, {Halve}, {Halzen}, {Hanson}, {Haungs},
  {Hebecker}, {Heereman}, {Heix}, {Helbing}, {Hellauer}, {Henningsen},
  {Hickford}, {Hignight}, {Hill}, {Hoffman}, {Hoffmann}, {Hoinka},
  {Hokanson-Fasig}, {Hoshina}, {Huang}, {Huber}, {Huber}, {Hultqvist},
  {H{\"u}nnefeld}, {Hussain}, {In}, {Iovine}, {Ishihara}, {Japaridze}, {Jeong},
  {Jero}, {Jones}, {Jonske}, {Joppe}, {Kang}, {Kang}, {Kappes}, {Kappesser},
  {Karg}, {Karl}, {Karle}, {Katz}, {Kauer}, {Kelley}, {Kheirandish}, {Kim},
  {Kintscher}, {Kiryluk}, {Kittler}, {Klein}, {Koirala}, {Kolanoski},
  {K{\"o}pke}, {Kopper}, {Kopper}, {Koskinen}, {Kowalski}, {Krings},
  {Kr{\"u}ckl}, {Kulacz}, {Kurahashi}, {Kyriacou}, {Lanfranchi}, {Larson},
  {Lauber}, {Lazar}, {Leonard}, {Leszczy{\'n}ska}, {Leuermann}, {Liu},
  {Lohfink}, {Mariscal}, {Lu}, {Lucarelli}, {L{\"u}nemann}, {Luszczak}, {Lyu},
  {Ma}, {Madsen}, {Maggi}, {Mahn}, {Makino}, {Mallik}, {Mallot}, {Mancina},
  {Mari{\c{s}}}, {Maruyama}, {Mase}, {Maunu}, {McNally}, {Meagher}, {Medici},
  {Medina}, {Meier}, {Meighen-Berger}, {Merino}, {Meures}, {Micallef},
  {Mockler}, {Moment{\'e}}, {Montaruli}, {Moore}, {Morse}, {Moulai}, {Muth},
  {Nagai}, {Naumann}, {Neer}, {Niederhausen}, {Nisa}, {Nowicki}, {Nygren},
  {Pollmann}, {Oehler}, {Olivas}, {O'Murchadha}, {O'Sullivan}, {Palczewski},
  {Pandya}, {Pankova}, {Park}, {Peiffer}, {de los Heros}, {Philippen},
  {Pieloth}, {Pinat}, {Pizzuto}, {Plum}, {Porcelli}, {Price}, {Przybylski},
  {Raab}, {Raissi}, {Rameez}, {Rauch}, {Rawlins}, {Rea}, {Reimann},
  {Relethford}, {Renschler}, {Renzi}, {Resconi}, {Rhode}, {Richman},
  {Robertson}, {Rongen}, {Rott}, {Ruhe}, {Ryckbosch}, {Rysewyk}, {Safa},
  {Herrera}, {Sandrock}, {Sandroos}, {Santander}, {Sarkar}, {Sarkar},
  {Satalecka}, {Schaufel}, {Schieler}, {Schlunder}, {Schmidt}, {Schneider},
  {Schneider}, {Schr{\"o}der}, {Schumacher}, {Sclafani}, {Seunarine},
  {Shefali}, {Silva}, {Snihur}, {Soedingrekso}, {Soldin}, {Song}, {Spiczak},
  {Spiering}, {Stachurska}, {Stamatikos}, {Stanev}, {Stein}, {Stettner},
  {Steuer}, {Stezelberger}, {Stokstad}, {St{\"o}{\ss}l}, {Strotjohann},
  {St{\"u}rwald}, {Stuttard}, {Sullivan}, {Taboada}, {Tenholt}, {Ter-Antonyan},
  {Terliuk}, {Tilav}, {Tollefson}, {Tomankova}, {T{\"o}nnis}, {Toscano},
  {Tosi}, {Trettin}, {Tselengidou}, {Tung}, {Turcati}, {Turcotte}, {Turley},
  {Ty}, {Unger}, {Elorrieta}, {Usner}, {Vandenbroucke}, {Van Driessche}, {van
  Eijk}, {van Eijndhoven}, {van Santen}, {Verpoest}, {Vraeghe}, {Walck},
  {Wallace}, {Wallraff}, {Wandkowsky}, {Watson}, {Weaver}, {Weindl}, {Weiss},
  {Weldert}, {Wendt}, {Werthebach}, {Whelan}, {Whitehorn}, {Wiebe}, {Wiebusch},
  {Wille}, {Williams}, {Wills}, {Wolf}, {Wood}, {Wood}, {Woschnagg}, {Wrede},
  {Xu}, {Xu}, {Xu}, {Yanez}, {Yodh}, {Yoshida}, {Yuan}, {Z{\"o}cklein}, \& {The
  IceCube Collaboration}}]{Anita:2020}
---. 2020{\natexlab{c}}, \apj, 892, 53, \dodoi{10.3847/1538-4357/ab791d}

\bibitem[{{Aartsen} {et~al.}(2020{\natexlab{d}}){Aartsen}, {Ackermann},
  {Adams}, {Aguilar}, {Ahlers}, {Ahrens}, {Alispach}, {Andeen}, {Anderson},
  {Ansseau}, {Anton}, {Arg{\"u}elles}, {Auffenberg}, {Axani}, {Bagherpour},
  {Bai}, {A. Balagopal}, {Barbano}, {Bartos}, {Barwick}, {Bastian}, {Baum},
  {Baur}, {Bay}, {Beatty}, {Becker}, {Tjus}, {BenZvi}, {Berley}, {Bernardini},
  {Besson}, {Binder}, {Bindig}, {Blaufuss}, {Blot}, {Bohm}, {B{\"o}ser},
  {Botner}, {B{\"o}ttcher}, {Bourbeau}, {Bourbeau}, {Bradascio}, {Braun},
  {Bron}, {Brostean-Kaiser}, {Burgman}, {Buscher}, {Busse}, {Carver}, {Chen},
  {Cheung}, {Chirkin}, {Choi}, {Clark}, {Clark}, {Classen}, {Coleman},
  {Collin}, {Conrad}, {Coppin}, {Corley}, {Correa}, {Countryman}, {Cowen},
  {Cross}, {Dave}, {Clercq}, {DeLaunay}, {Dembinski}, {Deoskar}, {Ridder},
  {Desiati}, {Vries}, {Wasseige}, {With}, {DeYoung}, {Diaz},
  {D{\'\i}az-V{\'e}lez}, {Dujmovic}, {Dunkman}, {Dvorak}, {Eberhardt},
  {Ehrhardt}, {Eller}, {Engel}, {Evenson}, {Fahey}, {Fazely}, {Felde},
  {Filimonov}, {Finley}, {Fox}, {Franckowiak}, {Friedman}, {Fritz}, {Gaisser},
  {Gallagher}, {Ganster}, {Garrappa}, {Gerhardt}, {Ghorbani}, {Glauch},
  {Gl{\"u}senkamp}, {Goldschmidt}, {Gonzalez}, {Grant}, {Gr{\'e}goire},
  {Griffith}, {Griswold}, {G{\"u}nder}, {G{\"u}nd{\"u}z}, {Haack}, {Hallgren},
  {Halliday}, {Halve}, {Halzen}, {Hanson}, {Haungs}, {Hebecker}, {Heereman},
  {Heix}, {Helbing}, {Hellauer}, {Henningsen}, {Hickford}, {Hignight}, {Hill},
  {Hoffman}, {Hoffmann}, {Hoinka}, {Hokanson-Fasig}, {Hoshina}, {Huang},
  {Huber}, {Huber}, {Hultqvist}, {H{\"u}nnefeld}, {Hussain}, {In}, {Iovine},
  {Ishihara}, {Jansson}, {Japaridze}, {Jeong}, {Jero}, {Jones}, {Jonske},
  {Joppe}, {Kang}, {Kang}, {Kappes}, {Kappesser}, {Karg}, {Karl}, {Karle},
  {Katz}, {Kauer}, {Keivani}, {Kellermann}, {Kelley}, {Kheirandish}, {Kim},
  {Kintscher}, {Kiryluk}, {Kittler}, {Klein}, {Koirala}, {Kolanoski},
  {K{\"o}pke}, {Kopper}, {Kopper}, {Koskinen}, {Kowalski}, {Krings},
  {Kr{\"u}ckl}, {Kulacz}, {Kurahashi}, {Kyriacou}, {Lanfranchi}, {Larson},
  {Lauber}, {Lazar}, {Leonard}, {Leszczy{\'n}ska}, {Liu}, {Lohfink},
  {Mariscal}, {Lu}, {Lucarelli}, {Ludwig}, {L{\"u}nemann}, {Luszczak}, {Lyu},
  {Ma}, {Madsen}, {Maggi}, {Mahn}, {Makino}, {Mallik}, {Mallot}, {Mancina},
  {Mari{\c{s}}}, {Marka}, {Marka}, {Maruyama}, {Mase}, {Maunu}, {McNally},
  {Meagher}, {Medici}, {Medina}, {Meier}, {Meighen-Berger}, {Merino}, {Meures},
  {Micallef}, {Mockler}, {Moment{\'e}}, {Montaruli}, {Moore}, {Morse},
  {Moulai}, {Muth}, {Nagai}, {Naumann}, {Neer}, {Nguyen}, {Niederhausen},
  {Nisa}, {Nowicki}, {Nygren}, {Pollmann}, {Oehler}, {Olivas}, {O'Murchadha},
  {O'Sullivan}, {Palczewski}, {Pandya}, {Pankova}, {Park}, {Peiffer}, {de los
  Heros}, {Philippen}, {Pieloth}, {Pieper}, {Pinat}, {Pizzuto}, {Plum},
  {Porcelli}, {Price}, {Przybylski}, {Raab}, {Raissi}, {Rameez}, {Rauch},
  {Rawlins}, {Rea}, {Rehman}, {Reimann}, {Relethford}, {Renschler}, {Renzi},
  {Resconi}, {Rhode}, {Richman}, {Robertson}, {Rongen}, {Rott}, {Ruhe},
  {Ryckbosch}, {Cantu}, {Safa}, {Herrera}, {Sandrock}, {Sandroos}, {Santander},
  {Sarkar}, {Sarkar}, {Satalecka}, {Schaufel}, {Schieler}, {Schlunder},
  {Schmidt}, {Schneider}, {Schneider}, {Schr{\"o}der}, {Schumacher},
  {Sclafani}, {Seckel}, {Seunarine}, {Shefali}, {Silva}, {Snihur},
  {Soedingrekso}, {Soldin}, {Song}, {Spiczak}, {Spiering}, {Stachurska},
  {Stamatikos}, {Stanev}, {Stein}, {Stettner}, {Steuer}, {Stezelberger},
  {Stokstad}, {St{\"o}{\ss}l}, {Strotjohann}, {St{\"u}rwald}, {Stuttard},
  {Sullivan}, {Taboada}, {Tenholt}, {Ter-Antonyan}, {Terliuk}, {Tilav},
  {Tollefson}, {Tomankova}, {T{\"o}nnis}, {Toscano}, {Tosi}, {Trettin},
  {Tselengidou}, {Tung}, {Turcati}, {Turcotte}, {Turley}, {Ty}, {Unger},
  {Elorrieta}, {Usner}, {Vandenbroucke}, {Driessche}, {Eijk}, {Eijndhoven},
  {Santen}, {Verpoest}, {Veske}, {Vraeghe}, {Walck}, {Wallace}, {Wallraff},
  {Wandkowsky}, {Watson}, {Weaver}, {Weindl}, {Weiss}, {Weldert}, {Wendt},
  {Werthebach}, {Whelan}, {Whitehorn}, {Wiebe}, {Wiebusch}, {Wille},
  {Williams}, {Wills}, {Wolf}, {Wood}, {Wood}, {Woschnagg}, {Wrede}, {Xu},
  {Xu}, {Xu}, {Yanez}, {Yodh}, {Yoshida}, {Yuan}, \&
  {Z{\"o}cklein}}]{GWIcecubeLIGO2020}
---. 2020{\natexlab{d}}, \apjl, 898, L10, \dodoi{10.3847/2041-8213/ab9d24}

\bibitem[{{Abbasi} {et~al.}(2013){Abbasi}, {Abdou}, {Ackermann}, {Adams},
  {Aguilar}, {Ahlers}, {Altmann}, {Andeen}, {Auffenberg}, {Bai}, {Baker},
  {Barwick}, {Baum}, {Bay}, {Beattie}, {Beatty}, {Bechet}, {Becker Tjus},
  {Becker}, {Bell}, {Benabderrahmane}, {BenZvi}, {Berdermann}, {Berghaus},
  {Berley}, {Bernardini}, {Bertrand}, {Besson}, {Bindig}, {Bissok}, {Blaufuss},
  {Blumenthal}, {Boersma}, {Bohm}, {Bose}, {B{\"o}ser}, {Botner}, {Brayeur},
  {Brown}, {Bruijn}, {Brunner}, {Buitink}, {Caballero-Mora}, {Carson}, {Casey},
  {Casier}, {Chirkin}, {Christy}, {Clevermann}, {Cohen}, {Cowen}, {Cruz Silva},
  {Danninger}, {Daughhetee}, {Davis}, {De Clercq}, {Descamps}, {Desiati}, {de
  Vries-Uiterweerd}, {DeYoung}, {D{\'\i}az-V{\'e}lez}, {Dreyer}, {Dumm},
  {Dunkman}, {Eagan}, {Eisch}, {Elliott}, {Ellsworth}, {Engdeg{\r{a}}rd},
  {Euler}, {Evenson}, {Fadiran}, {Fazely}, {Fedynitch}, {Feintzeig}, {Feusels},
  {Filimonov}, {Finley}, {Fischer-Wasels}, {Flis}, {Franckowiak}, {Franke},
  {Frantzen}, {Fuchs}, {Gaisser}, {Gallagher}, {Gerhardt}, {Gladstone},
  {Gl{\"u}senkamp}, {Goldschmidt}, {Goodman}, {G{\'o}ra}, {Grant}, {Gro{\ss}},
  {Grullon}, {Gurtner}, {Ha}, {Haj Ismail}, {Hallgren}, {Halzen}, {Hanson},
  {Heereman}, {Heimann}, {Heinen}, {Helbing}, {Hellauer}, {Hickford}, {Hill},
  {Hoffman}, {Hoffmann}, {Homeier}, {Hoshina}, {Huelsnitz}, {Hulth},
  {Hultqvist}, {Hussain}, {Ishihara}, {Jacobi}, {Jacobsen}, {Japaridze},
  {Jlelati}, {Johansson}, {Kappes}, {Karg}, {Karle}, {Kiryluk}, {Kislat},
  {Kl{\"a}s}, {Klein}, {Klepser}, {K{\"o}hne}, {Kohnen}, {Kolanoski},
  {K{\"o}pke}, {Kopper}, {Kopper}, {Koskinen}, {Kowalski}, {Krasberg}, {Kroll},
  {Kunnen}, {Kurahashi}, {Kuwabara}, {Labare}, {Laihem}, {Landsman}, {Larson},
  {Lauer}, {Lesiak-Bzdak}, {L{\"u}nemann}, {Madsen}, {Maruyama}, {Mase},
  {Matis}, {McDermott}, {McNally}, {Meagher}, {Merck}, {M{\'e}sz{\'a}ros},
  {Meures}, {Miarecki}, {Middell}, {Milke}, {Miller}, {Mohrmann}, {Montaruli},
  {Morse}, {Movit}, {Nahnhauer}, {Naumann}, {Nie{\ss}en}, {Nowicki}, {Nygren},
  {Obertacke}, {Odrowski}, {Olivas}, {Olivo}, {O'Murchadha}, {Panknin}, {Paul},
  {Pepper}, {P{\'e}rez de los Heros}, {Pieloth}, {Pirk}, {Posselt}, {Price},
  {Przybylski}, {R{\"a}del}, {Rawlins}, {Redl}, {Resconi}, {Rhode}, {Ribordy},
  {Richman}, {Riedel}, {Rodrigues}, {Roth}, {Rothmaier}, {Rott}, {Roucelle},
  {Ruhe}, {Rutledge}, {Ruzybayev}, {Ryckbosch}, {Saba}, {Salameh}, {Sander},
  {Santander}, {Sarkar}, {Schatto}, {Scheel}, {Scheriau}, {Schmidt}, {Schmitz},
  {Schoenen}, {Sch{\"o}neberg}, {Sch{\"o}nherr}, {Sch{\"o}nwald}, {Schukraft},
  {Schulte}, {Schulz}, {Seckel}, {Seo}, {Sestayo}, {Seunarine}, {Shulman},
  {Smith}, {Soiron}, {Soldin}, {Spiczak}, {Spiering}, {Stamatikos}, {Stanev},
  {Stasik}, {Stezelberger}, {Stokstad}, {St{\"o}{\ss}l}, {Stoyanov},
  {Strahler}, {Str{\"o}m}, {Sulanke}, {Sullivan}, {Taavola}, {Taboada},
  {Tamburro}, {Ter-Antonyan}, {Tilav}, {Toale}, {Toscano}, {Usner}, {van der
  Drift}, {van Eijndhoven}, {Van Overloop}, {van Santen}, {Vehring}, {Voge},
  {Walck}, {Waldenmaier}, {Wallraff}, {Walter}, {Wasserman}, {Weaver}, {Wendt},
  {Westerhoff}, {Whitehorn}, {Wiebe}, {Wiebusch}, {Williams}, {Wissing},
  {Wolf}, {Wood}, {Woschnagg}, {Xu}, {Xu}, {Xu}, {Yanez}, {Yodh}, {Yoshida},
  {Zarzhitsky}, {Ziemann}, {Zilles}, \& {Zoll}}]{IceTop:2013}
{Abbasi}, R., {Abdou}, Y., {Ackermann}, M., {et~al.} 2013, Nuclear Instruments
  and Methods in Physics Research A, 700, 188,
  \dodoi{10.1016/j.nima.2012.10.067}

\bibitem[{{Abbasi} {et~al.}(2021){Abbasi}, {Ackermann}, {Adams}, {Aguilar},
  {Ahlers}, {Ahrens}, {Alispach}, {Alves}, {Amin}, {Andeen}, {Anderson},
  {Ansseau}, {Anton}, {Arg{\"u}elles}, {Axani}, {Bai}, {Balagopal V.},
  {Barbano}, {Barwick}, {Bastian}, {Basu}, {Baum}, {Baur}, {Bay}, {Beatty},
  {Becker}, {Becker Tjus}, {Bellenghi}, {BenZvi}, {Berley}, {Bernardini},
  {Besson}, {Binder}, {Bindig}, {Blaufuss}, {Blot}, {B{\"o}ser}, {Botner},
  {B{\"o}ttcher}, {Bourbeau}, {Bourbeau}, {Bradascio}, {Braun}, {Bron},
  {Brostean-Kaiser}, {Burgman}, {Busse}, {Campana}, {Chen}, {Chirkin}, {Choi},
  {Clark}, {Clark}, {Classen}, {Coleman}, {Collin}, {Conrad}, {Coppin},
  {Correa}, {Cowen}, {Cross}, {Dave}, {De Clercq}, {DeLaunay}, {Dembinski},
  {Deoskar}, {De Ridder}, {Desai}, {Desiati}, {de Vries}, {de Wasseige}, {de
  With}, {DeYoung}, {Dharani}, {Diaz}, {D{\'\i}az-V{\'e}lez}, {Dujmovic},
  {Dunkman}, {DuVernois}, {Dvorak}, {Ehrhardt}, {Eller}, {Engel}, {Evans},
  {Evenson}, {Fahey}, {Fazely}, {Fiedlschuster}, {Fienberg}, {Filimonov},
  {Finley}, {Fischer}, {Fox}, {Franckowiak}, {Friedman}, {Fritz}, {F{\"u}rst},
  {Gaisser}, {Gallagher}, {Ganster}, {Garrappa}, {Gerhardt}, {Ghadimi},
  {Glauch}, {Gl{\"u}senkamp}, {Goldschmidt}, {Gonzalez}, {Goswami}, {Grant},
  {Gr{\'e}goire}, {Griffith}, {Griswold}, {G{\"u}nd{\"u}z}, {Haack},
  {Hallgren}, {Halliday}, {Halve}, {Halzen}, {Ha Minh}, {Hanson}, {Hardin},
  {Haungs}, {Hauser}, {Hebecker}, {Helbing}, {Henningsen}, {Hickford},
  {Hignight}, {Hill}, {Hill}, {Hoffman}, {Hoffmann}, {Hoinka},
  {Hokanson-Fasig}, {Hoshina}, {Huang}, {Huber}, {Huber}, {Hultqvist},
  {H{\"u}nnefeld}, {Hussain}, {In}, {Iovine}, {Ishihara}, {Jansson},
  {Japaridze}, {Jeong}, {Jones}, {Joppe}, {Kang}, {Kang}, {Kang}, {Kappes},
  {Kappesser}, {Karg}, {Karl}, {Karle}, {Katori}, {Katz}, {Kauer},
  {Kellermann}, {Kelley}, {Kheirandish}, {Kim}, {Kin}, {Kintscher}, {Kiryluk},
  {Klein}, {Koirala}, {Kolanoski}, {K{\"o}pke}, {Kopper}, {Kopper}, {Koskinen},
  {Koundal}, {Kovacevich}, {Kowalski}, {Krings}, {Kr{\"u}ckl}, {Kulacz},
  {Kurahashi}, {Kyriacou}, {Lagunas Gualda}, {Lanfranchi}, {Larson}, {Lauber},
  {Lazar}, {Leonard}, {Leszczy{\'n}ska}, {Li}, {Liu}, {Lohfink}, {Lozano
  Mariscal}, {Lu}, {Lucarelli}, {Ludwig}, {Luszczak}, {Lyu}, {Ma}, {Madsen},
  {Mahn}, {Makino}, {Mallik}, {Mancina}, {Mandalia}, {Mari{\c{s}}}, {Maruyama},
  {Mase}, {McNally}, {Meagher}, {Medina}, {Meier}, {Meighen-Berger}, {Merz},
  {Micallef}, {Mockler}, {Moment{\'e}}, {Montaruli}, {Moore}, {Morse},
  {Moulai}, {Naab}, {Nagai}, {Naumann}, {Necker}, {Neer},
  {Nguy{\'a}{\guillemotright} n}, {Niederhausen}, {Nisa}, {Nowicki}, {Nygren},
  {Obertacke Pollmann}, {Oehler}, {Olivas}, {O'Sullivan}, {Pandya}, {Pankova},
  {Park}, {Parker}, {Paudel}, {Peiffer}, {P{\'e}rez de los Heros}, {Philippen},
  {Pieloth}, {Pieper}, {Pizzuto}, {Plum}, {Popovych}, {Porcelli}, {Prado
  Rodriguez}, {Price}, {Przybylski}, {Raab}, {Raissi}, {Rameez}, {Rawlins},
  {Rea}, {Rehman}, {Reimann}, {Renschler}, {Renzi}, {Resconi}, {Reusch},
  {Rhode}, {Richman}, {Riedel}, {Robertson}, {Roellinghoff}, {Rongen}, {Rott},
  {Ruhe}, {Ryckbosch}, {Rysewyk Cantu}, {Safa}, {Sanchez Herrera}, {Sandrock},
  {Sandroos}, {Santander}, {Sarkar}, {Sarkar}, {Satalecka}, {Scharf},
  {Schaufel}, {Schieler}, {Schlunder}, {Schmidt}, {Schneider}, {Schneider},
  {Schr{\"o}der}, {Schumacher}, {Sclafani}, {Seckel}, {Seunarine}, {Shefali},
  {Silva}, {Smithers}, {Snihur}, {Soedingrekso}, {Soldin}, {Spiczak},
  {Spiering}, {Stachurska}, {Stamatikos}, {Stanev}, {Stein}, {Stettner},
  {Steuer}, {Stezelberger}, {Stokstad}, {Strotjohann}, {Stuttard}, {Sullivan},
  {Taboada}, {Tenholt}, {Ter-Antonyan}, {Tilav}, {Tischbein}, {Tollefson},
  {Tomankova}, {T{\"o}nnis}, {Toscano}, {Tosi}, {Trettin}, {Tselengidou},
  {Tung}, {Turcati}, {Turcotte}, {Turley}, {Twagirayezu}, {Ty}, {Unger},
  {Unland Elorrieta}, {Vandenbroucke}, {van Eijk}, {van Eijndhoven},
  {Vannerom}, {van Santen}, {Verpoest}, {Vraeghe}, {Walck}, {Wallace},
  {Wandkowsky}, {Watson}, {Weaver}, {Weindl}, {Weiss}, {Weldert}, {Wendt},
  {Werthebach}, {Weyrauch}, {Whelan}, {Whitehorn}, {Wiebe}, {Wiebusch},
  {Williams}, {Wolf}, {Wood}, {Woschnagg}, {Wrede}, {Wulff}, {Xu}, {Xu},
  {Yanez}, {Yoshida}, {Yuan}, {Zhang}, \& {The IceCube
  Collaboration}}]{Abbasi:2020jmh}
{Abbasi}, R., {Ackermann}, M., {Adams}, J., {et~al.} 2021, \prd, 104, 022002,
  \dodoi{10.1103/PhysRevD.104.022002}

\bibitem[{{Abbasi} {et~al.}(2014){Abbasi}, {Abe}, {Abu-Zayyad}, {Allen},
  {Anderson}, {Azuma}, {Barcikowski}, {Belz}, {Bergman}, {Blake}, {Cady},
  {Chae}, {Cheon}, {Chiba}, {Chikawa}, {Cho}, {Fujii}, {Fukushima}, {Goto},
  {Hanlon}, {Hayashi}, {Hayashida}, {Hibino}, {Honda}, {Ikeda}, {Inoue},
  {Ishii}, {Ishimori}, {Ito}, {Ivanov}, {Jui}, {Kadota}, {Kakimoto},
  {Kalashev}, {Kasahara}, {Kawai}, {Kawakami}, {Kawana}, {Kawata}, {Kido},
  {Kim}, {Kim}, {Kim}, {Kitamura}, {Kitamura}, {Kuzmin}, {Kwon}, {Lan}, {Lim},
  {Lundquist}, {Machida}, {Martens}, {Matsuda}, {Matsuyama}, {Matthews},
  {Minamino}, {Mukai}, {Myers}, {Nagasawa}, {Nagataki}, {Nakamura}, {Nonaka},
  {Nozato}, {Ogio}, {Ogura}, {Ohnishi}, {Ohoka}, {Oki}, {Okuda}, {Ono},
  {Oshima}, {Ozawa}, {Park}, {Pshirkov}, {Rodriguez}, {Rubtsov}, {Ryu},
  {Sagawa}, {Sakurai}, {Sampson}, {Scott}, {Shah}, {Shibata}, {Shibata},
  {Shimodaira}, {Shin}, {Smith}, {Sokolsky}, {Springer}, {Stokes}, {Stratton},
  {Stroman}, {Suzawa}, {Takamura}, {Takeda}, {Takeishi}, {Taketa}, {Takita},
  {Tameda}, {Tanaka}, {Tanaka}, {Tanaka}, {Thomas}, {Thomson}, {Tinyakov},
  {Tkachev}, {Tokuno}, {Tomida}, {Troitsky}, {Tsunesada}, {Tsutsumi},
  {Uchihori}, {Udo}, {Urban}, {Vasiloff}, {Wong}, {Yamane}, {Yamaoka},
  {Yamazaki}, {Yang}, {Yashiro}, {Yoneda}, {Yoshida}, {Yoshii}, {Zollinger}, \&
  {Zundel}}]{Abbasi:2014lda}
{Abbasi}, R.~U., {Abe}, M., {Abu-Zayyad}, T., {et~al.} 2014, \apjl, 790, L21,
  \dodoi{10.1088/2041-8205/790/2/L21}

\bibitem[{{Abbasi} {et~al.}(2018){Abbasi}, {Abe}, {Abu-Zayyad}, {Allen},
  {Azuma}, {Barcikowski}, {Belz}, {Bergman}, {Blake}, {Cady}, {Cheon}, {Chiba},
  {Chikawa}, {di Matteo}, {Fujii}, {Fujita}, {Fukushima}, {Furlich}, {Goto},
  {Hanlon}, {Hayashi}, {Hayashi}, {Hayashida}, {Hibino}, {Honda}, {Ikeda},
  {Inoue}, {Ishii}, {Ishimori}, {Ito}, {Ivanov}, {Jeong}, {Jeong}, {Jui},
  {Kadota}, {Kakimoto}, {Kalashev}, {Kasahara}, {Kawai}, {Kawakami}, {Kawana},
  {Kawata}, {Kido}, {Kim}, {Kim}, {Kim}, {Kishigami}, {Kitamura}, {Kitamura},
  {Kuzmin}, {Kuznetsov}, {Kwon}, {Lee}, {Lubsandorzhiev}, {Lundquist},
  {Machida}, {Martens}, {Matsuyama}, {Matthews}, {Mayta}, {Minamino}, {Mukai},
  {Myers}, {Nagasawa}, {Nagataki}, {Nakamura}, {Nakamura}, {Nonaka}, {Oda},
  {Ogio}, {Ogura}, {Ohnishi}, {Ohoka}, {Okuda}, {Omura}, {Ono}, {Onogi},
  {Oshima}, {Ozawa}, {Park}, {Pshirkov}, {Remington}, {Rodriguez}, {Rubtsov},
  {Ryu}, {Sagawa}, {Sahara}, {Saito}, {Saito}, {Sakaki}, {Sakurai}, {Scott},
  {Seki}, {Sekino}, {Shah}, {Shibata}, {Shibata}, {Shimodaira}, {Shin}, {Shin},
  {Smith}, {Sokolsky}, {Stokes}, {Stratton}, {Stroman}, {Suzawa}, {Takagi},
  {Takahashi}, {Takamura}, {Takeda}, {Takeishi}, {Taketa}, {Takita}, {Tameda},
  {Tanaka}, {Tanaka}, {Tanaka}, {Thomas}, {Thomson}, {Tinyakov}, {Tkachev},
  {Tokuno}, {Tomida}, {Troitsky}, {Tsunesada}, {Tsutsumi}, {Uchihori}, {Udo},
  {Urban}, {Wong}, {Yamamoto}, {Yamane}, {Yamaoka}, {Yamazaki}, {Yang},
  {Yashiro}, {Yoneda}, {Yoshida}, {Yoshii}, {Zhezher}, \&
  {Zundel}}]{abbasi:2018ApJ867L27A}
---. 2018, \apjl, 867, L27, \dodoi{10.3847/2041-8213/aaebf9}

\bibitem[{{Abreu} {et~al.}(2011){Abreu}, {Aglietta}, {Ahn}, {Albuquerque},
  {Allard}, {Allekotte}, {Allen}, {Allison}, {Alvarez Castillo},
  {Alvarez-Mu{\~n}iz}, {Ambrosio}, {Aminaei}, {Anchordoqui}, {Andringa},
  {Anti{\v{c}}i{\'c}}, {Anzalone}, {Aramo}, {Arganda}, {Arqueros}, {Asorey},
  {Assis}, {Aublin}, {Ave}, {Avenier}, {Avila}, {B{\"a}cker}, {Balzer},
  {Barber}, {Barbosa}, {Bardenet}, {Barroso}, {Baughman}, {B{\"a}uml},
  {Beatty}, {Becker}, {Becker}, {Bell{\'e}toile}, {Bellido}, {BenZvi}, {Berat},
  {Bertou}, {Biermann}, {Billoir}, {Blanco}, {Blanco}, {Bleve}, {Bl{\"u}mer},
  {Boh{\'a}{\v{c}}ov{\'a}}, {Boncioli}, {Bonifazi}, {Bonino}, {Borodai},
  {Brack}, {Brogueira}, {Brown}, {Bruijn}, {Buchholz}, {Bueno}, {Burton},
  {Caballero-Mora}, {Caramete}, {Caruso}, {Castellina}, {Catalano}, {Cataldi},
  {Cazon}, {Cester}, {Chauvin}, {Cheng}, {Chiavassa}, {Chinellato}, {Chou},
  {Chudoba}, {Clay}, {Coluccia}, {Concei{\c{c}}{\~a}o}, {Contreras}, {Cook},
  {Cooper}, {Coppens}, {Cordier}, {Cotti}, {Coutu}, {Covault}, {Creusot},
  {Criss}, {Cronin}, {Curutiu}, {Dagoret-Campagne}, {Dallier}, {Dasso},
  {Daumiller}, {Dawson}, {de Almeida}, {De Domenico}, {De Donato}, {de Jong},
  {De La Vega}, {de Mello Junior}, {de Mello Neto}, {De Mitri}, {de Souza}, {de
  Vries}, {Decerprit}, {del Peral}, {Deligny}, {Dembinski}, {Dhital}, {Di
  Giulio}, {Diaz}, {D{\'\i}az Castro}, {Diep}, {Dobrigkeit}, {Docters},
  {D'Olivo}, {Dong}, {Dorofeev}, {dos Anjos}, {Dova}, {D'Urso}, {Dutan}, {Ebr},
  {Engel}, {Erdmann}, {Escobar}, {Etchegoyen}, {Facal San Luis}, {Fajardo
  Tapia}, {Falcke}, {Farrar}, {Fauth}, {Fazzini}, {Ferguson}, {Ferrero},
  {Fick}, {Filevich}, {Filip{\v{c}}i{\v{c}}}, {Fliescher}, {Fracchiolla},
  {Fraenkel}, {Fr{\"o}hlich}, {Fuchs}, {Gaior}, {Gamarra}, {Gambetta},
  {Garc{\'\i}a}, {Garc{\'\i}a G{\'a}mez}, {Garcia-Pinto}, {Gascon}, {Gemmeke},
  {Gesterling}, {Ghia}, {Giaccari}, {Giller}, {Glass}, {Gold}, {Golup}, {Gomez
  Albarracin}, {G{\'o}mez Berisso}, {Gon{\c{c}}alves}, {Gonzalez}, {Gonzalez},
  {Gookin}, {G{\'o}ra}, {Gorgi}, {Gouffon}, {Gozzini}, {Grashorn}, {Grebe},
  {Griffith}, {Grigat}, {Grillo}, {Guardincerri}, {Guarino}, {Guedes},
  {Guzman}, {Hague}, {Hansen}, {Harari}, {Harmsma}, {Harton}, {Haungs},
  {Hebbeker}, {Heck}, {Herve}, {Hojvat}, {Hollon}, {Holmes}, {Homola},
  {H{\"o}randel}, {Horneffer}, {Hrabovsk{\'y}}, {Huege}, {Insolia}, {Ionita},
  {Italiano}, {Jarne}, {Jiraskova}, {Josebachuili}, {Kadija}, {Kampert},
  {Karhan}, {Kasper}, {K{\'e}gl}, {Keilhauer}, {Keivani}, {Kelley}, {Kemp},
  {Kieckhafer}, {Klages}, {Kleifges}, {Kleinfeller}, {Knapp}, {Koang},
  {Kotera}, {Krohm}, {Kr{\"o}mer}, {Kruppke-Hansen}, {Kuehn}, {Kuempel},
  {Kulbartz}, {Kunka}, {La Rosa}, {Lachaud}, {Lautridou}, {Le{\~a}o}, {Lebrun},
  {Lebrun}, {Leigui de Oliveira}, {Lemiere}, {Letessier-Selvon}, {Lhenry-Yvon},
  {Link}, {L{\'o}pez}, {Lopez Ag{\"u}era}, {Louedec}, {Lozano Bahilo},
  {Lucero}, {Ludwig}, {Lyberis}, {Maccarone}, {Macolino}, {Maldera}, {Mandat},
  {Mantsch}, {Mariazzi}, {Marin}, {Marin}, {Maris}, {Marquez Falcon},
  {Marsella}, {Martello}, {Martin}, {Martinez}, {Mart{\'\i}nez Bravo},
  {Mathes}, {Matthews}, {Matthews}, {Matthiae}, {Maurizio}, {Mazur},
  {Medina-Tanco}, {Melissas}, {Melo}, {Menichetti}, {Menshikov}, {Mertsch},
  {Meurer}, {Mi{\'c}anovi{\'c}}, {Micheletti}, {Miller}, {Miramonti},
  {Mollerach}, {Monasor}, {Monnier Ragaigne}, {Montanet}, {Morales}, {Morello},
  {Moreno}, {Moreno}, {Morris}, {Mostaf{\'a}}, {Moura}, {Mueller}, {Muller},
  {M{\"u}ller}, {M{\"u}nchmeyer}, {Mussa}, {Navarra}, {Navarro}, {Navas},
  {Necesal}, {Nellen}, {Nelles}, {Neuser}, {Nhung}, {Niemietz},
  {Nierstenhoefer}, {Nitz}, {Nosek}, {No{\v{z}}ka}, {Nyklicek},
  {Oehlschl{\"a}ger}, {Olinto}, {Olmos-Gilbaja}, {Ortiz}, {Pacheco}, {Pakk
  Selmi-Dei}, {Palatka}, {Pallotta}, {Palmieri}, {Parente}, {Parizot}, {Parra},
  {Parsons}, {Pastor}, {Paul}, {Pech}, {P{\c{e}}kala}, {Pelayo}, {Pepe},
  {Perrone}, {Pesce}, {Petermann}, {Petrera}, {Petrinca}, {Petrolini},
  {Petrov}, {Petrovic}, {Pfendner}, {Phan}, {Piegaia}, {Pierog}, {Pieroni},
  {Pimenta}, {Pirronello}, {Platino}, {Ponce}, {Pontz}, {Privitera}, {Prouza},
  {Quel}, {Querchfeld}, {Rautenberg}, {Ravel}, {Ravignani}, {Revenu}, {Ridky},
  {Riggi}, {Risse}, {Ristori}, {Rivera}, {Rizi}, {Roberts}, {Robledo},
  {Rodrigues de Carvalho}, {Rodriguez}, {Rodriguez Martino}, {Rodriguez Rojo},
  {Rodriguez-Cabo}, {Rodr{\'\i}guez-Fr{\'\i}as}, {Ros}, {Rosado}, {Rossler},
  {Roth}, {Rouill{\'e}-d'Orfeuil}, {Roulet}, {Rovero}, {R{\"u}hle}, {Salamida},
  {Salazar}, {Salina}, {S{\'a}nchez}, {Santander}, {Santo}, {Santos}, {Santos},
  {Sarazin}, {Sarkar}, {Sarkar}, {Sato}, {Scharf}, {Scherini}, {Schieler},
  {Schiffer}, {Schmidt}, {Schmidt}, {Scholten}, {Schoorlemmer}, {Schovancova},
  {Schov{\'a}nek}, {Schr{\"o}der}, {Schulte}, {Schuster}, {Sciutto}, {Scuderi},
  {Segreto}, {Settimo}, {Shadkam}, {Shellard}, {Sidelnik}, {Sigl}, {Silva
  Lopez}, {{\'S}mia{\l} kowski}, {{\v{S}}m{\'\i}da}, {Snow}, {Sommers},
  {Sorokin}, {Spinka}, {Squartini}, {Stanic}, {Stapleton}, {Stasielak},
  {Stephan}, {Strazzeri}, {Stutz}, {Suarez}, {Suomij{\"a}rvi}, {Supanitsky},
  {{\v{S}}u{\v{s}}a}, {Sutherland}, {Swain}, {Szadkowski}, {Szuba},
  {Tamashiro}, {Tapia}, {Tartare}, {Ta{\c{s}}c{\u{a}}u}, {Tavera Ruiz},
  {Tcaciuc}, {Tegolo}, {Thao}, {Thomas}, {Tiffenberg}, {Tiwari}, {Tkaczyk},
  {Todero Peixoto}, {Tom{\'e}}, {Tonachini}, {Travnicek}, {Tridapalli},
  {Tristram}, {Trovato}, {Tueros}, {Ulrich}, {Unger}, {Urban}, {Vald{\'e}s
  Galicia}, {Vali{\~n}o}, {Valore}, {van den Berg}, {Varela}, {Vargas
  C{\'a}rdenas}, {V{\'a}zquez}, {V{\'a}zquez}, {Veberi{\v{c}}}, {Verzi},
  {Vicha}, {Videla}, {Villase{\~n}or}, {Wahlberg}, {Wahrlich}, {Wainberg},
  {Walz}, {Warner}, {Watson}, {Weber}, {Weidenhaupt}, {Weindl}, {Westerhoff},
  {Whelan}, {Wieczorek}, {Wiencke}, {Wilczy{\'n}ska}, {Wilczy{\'n}ski}, {Will},
  {Williams}, {Winchen}, {Winders}, {Winnick}, {Wommer}, {Wundheiler},
  {Yamamoto}, {Yapici}, {Younk}, {Yuan}, {Yushkov}, {Zamorano}, {Zas},
  {Zavrtanik}, {Zavrtnik}, {Zaw}, {Zepeda}, {Zimbres-Silva}, {Ziolkowski}, \&
  {The Pierre Auger Collaboration}}]{Abreu:2011pj}
{Abreu}, P., {Aglietta}, M., {Ahn}, E.~J., {et~al.} 2011, arXiv e-prints,
  arXiv:1107.4809.
\newblock \doarXiv{1107.4809}

\bibitem[{Abreu {et~al.}(2021)Abreu, Aglietta, Albury, Allekotte, Almela,
  Alvarez-Muniz, Alves~Batista, Anastasi, Anchordoqui, Andrada, Andringa,
  Aramo, Araújo~Ferreira, Arteaga~Velazquez, Asorey, Assis, Avila, Badescu,
  Bakalova, Balaceanu, Barbato, Barreira~Luz, Becker, Bellido, Berat, Bertaina,
  Bertou, Biermann, Binet, Bismark, Bister, Biteau, Blazek, Bleve, Bohacova,
  boncioli, Bonifazi, Bonneau~Arbeletche, Borodai, Botti, Brack, Bretz,
  Brichetto~Orchera, Briechle, Buchholz, Bueno, Buitink, Buscemi, Büsken,
  Caballero-Mora, Caccianiga, Canfora, Caracas, Carceller, Caruso, Castellina,
  Catalani, Cataldi, Cazon, Cerda, Chinellato, Chudoba, Chytka, Clay,
  Cobos~Cerutti, Colalillo, Coleman, Coluccia, Conceição, Condorelli,
  Consolati, Contreras, Convenga, Correia~dos Santos, Covault, Dasso,
  Daumiller, Dawson, Day, de~Almeida, de~Jesús, de~Jong, De~Mauro,
  de~Mello~Neto, De~Mitri, de~Oliveira, de~Oliveira~Franco, de~Palma, de~Souza,
  De~Vito, del Río, Deligny, Deval, di~Matteo, Dobrigkeit, D'Olivo,
  Domingues~Mendes, dos Anjos, dos Santos, Dova, Ebr, Engel, Epicoco, Erdmann,
  Escobar, Etchegoyen, Falcke, Farmer, Farrar, Fauth, Fazzini, Feldbusch, Fenu,
  Fick, Figueira, Filipcic, Fitoussi, Fodran, Freire, Fujii, Fuster, Galea,
  Galelli, García, García~Vegas, Gemmeke, Gesualdi, Gherghel-Lascu, Ghia,
  GIACCARI, Giammarchi, Glombitza, Gobbi, Gollan, Golup, Gómez~Berisso,
  Gómez~Vitale, Gongora, González, Gonzalez, Goos, Gora, Gorgi, Gottowik,
  Grubb, Guarino, Guedes, Guido, Hahn, Hamal, Hampel, Hansen, Harari, Harvey,
  Haungs, Hebbeker, Heck, Hill, Hojvat, Hörandel, Horvath, Hrabovsky, Huege,
  Insolia, Isar, Janecek, Johnsen, Juryšek, Kääpä, Kampert, Karastathis,
  Keilhauer, Kemp, Khakurdikar, Kizakke~Covilakam, Klages, Kleifges,
  Kleinfeller, Köpke, Kunka, Lago, Lang, Langner, Leigui~de Oliveira, Lenok,
  Letessier-Selvon, Lhenry-Yvon, Lo~Presti, LOPES, López, Lu, Luce, Lundquist,
  Machado~Payeras, Mancarella, Mandat, Manning, Manshanden, Mantsch, Marafico,
  Mariazzi, Maris, Marsella, Martello, Martinelli, Martínez~Bravo,
  Mastrodicasa, Mathes, Matthews, Matthiae, Mayotte, Mazur, Medina-Tanco, Melo,
  Menshikov, Merenda, Michal, Micheletti, Miramonti, Mollerach, Montanet,
  Morello, Mostafa, Müller, Muller, Mulrey, Mussa, Muzio, Namasaka,
  Nasr-Esfahani, Nellen, Niculescu-Oglinzanu, Niechciol, Nitz, Nosek, Novotný,
  Nozka, Nucita, Nunez, Palatka, Pallotta, Papenbreer, Parente, Parra,
  Pawlowsky, Pech, Pedreira, Pękala, Pelayo, Peña-Rodríguez,
  Pereira~Martins, Perez~Armand, Pérez~Bertolli, Perlin, Perrone, Petrera,
  Pierog, Pimenta, Pirronello, Platino, Pont, Pothast, Privitera, Prouza,
  Puyleart, Querchfeld, Rautenberg, Ravignani, Reininghaus, Ridky, Riehn,
  Risse, Rizi, Rodrigues~de Carvalho, Rodriguez~Rojo, Roncoroni, Rossoni, Roth,
  Roulet, Rovero, Ruehl, Saftoiu, Salamida, Salazar, Salina, Sanabria~Gomez,
  Sánchez, Santos, Santos, Sarazin, Sarmento, Sarmiento-Cano, Sato, Savina,
  Schäfer, Scherini, Schieler, Schimassek, Schimp, Schlüter, Schmidt,
  Scholten, Schovanek, Schröder, Schröder, Schulte, Sciutto, Scornavacche,
  Segreto, Sehgal, Shellard, Sigl, Silli, Sima, Smida, Sommers, Soriano,
  Souchard, Squartini, Stadelmaier, Stanca, Stanič, Stasielak, Stassi,
  Streich, Suárez-Durán, Sudholz, Suomijarvi, Supanitsky, Szadkowski, Tapia,
  Taricco, Timmermans, Tkachenko, Tobiska, Todero~Peixoto, Tomé, Torrès,
  Travaini, Travnicek, Trimarelli, Tueros, Ulrich, Unger, Vaclavek, Vacula,
  Valdés~Galicia, Valore, Varela, Vásquez-Ramírez, Veberic, Ventura,
  Vergara~Quispe, Verzi, Vicha, Vink, Vorobiov, Wahlberg, Watanabe, Watson,
  Weber, Weindl, Wiencke, Wilczyński, Wirtz, Wittkowski, Wundheiler, Yushkov,
  Zapparrata, Zas, Zavrtanik, Zavrtanik, \& Zehrer}]{Abreu:2021eL}
Abreu, P., Aglietta, M., Albury, J.~M., {et~al.} 2021, in Proceedings of 37th
  International Cosmic Ray Conference {\textemdash} PoS(ICRC2021), Vol. 395,
  307, \dodoi{10.22323/1.395.0307}

\bibitem[{{Abu-Zayyad} {et~al.}(2012){Abu-Zayyad}, {Aida}, {Allen}, {Anderson},
  {Azuma}, {Barcikowski}, {Belz}, {Bergman}, {Blake}, {Cady}, {Cheon}, {Chiba},
  {Chikawa}, {Cho}, {Cho}, {Fujii}, {Fujii}, {Fukuda}, {Fukushima}, {Gorbunov},
  {Hanlon}, {Hayashi}, {Hayashi}, {Hayashida}, {Hibino}, {Hiyama}, {Honda},
  {Iguchi}, {Ikeda}, {Ikuta}, {Inoue}, {Ishii}, {Ishimori}, {Ivanov},
  {Iwamoto}, {Jui}, {Kadota}, {Kakimoto}, {Kalashev}, {Kanbe}, {Kasahara},
  {Kawai}, {Kawakami}, {Kawana}, {Kido}, {Kim}, {Kim}, {Kim}, {Kim},
  {Kitamoto}, {Kobayashi}, {Kobayashi}, {Kondo}, {Kuramoto}, {Kuzmin}, {Kwon},
  {Lim}, {Machida}, {Martens}, {Martineau}, {Matsuda}, {Matsuura}, {Matsuyama},
  {Matthews}, {Myers}, {Minamino}, {Miyata}, {Miyauchi}, {Murano}, {Nakamura},
  {Nam}, {Nonaka}, {Ogio}, {Ohnishi}, {Ohoka}, {Oki}, {Oku}, {Okuda}, {Oshima},
  {Ozawa}, {Park}, {Pshirkov}, {Rodriguez}, {Roh}, {Rubtsov}, {Ryu}, {Sagawa},
  {Sakurai}, {Sampson}, {Scott}, {Shah}, {Shibata}, {Shibata}, {Shimodaira},
  {Shin}, {Shin}, {Shirahama}, {Smith}, {Sokolsky}, {Sonley}, {Springer},
  {Stokes}, {Stratton}, {Stroman}, {Suzuki}, {Takahashi}, {Takeda}, {Taketa},
  {Takita}, {Tameda}, {Tanaka}, {Tanaka}, {Tanaka}, {Thomas}, {Thomson},
  {Tinyakov}, {Tkachev}, {Tokuno}, {Tomida}, {Troitsky}, {Tsunesada},
  {Tsutsumi}, {Tsuyuguchi}, {Uchihori}, {Udo}, {Ukai}, {Vasiloff}, {Wada},
  {Wong}, {Wood}, {Yamakawa}, {Yamaoka}, {Yamazaki}, {Yang}, {Yoshida},
  {Yoshii}, {Zollinger}, \& {Zundel}}]{AbuZayyad:2012kk}
{Abu-Zayyad}, T., {Aida}, R., {Allen}, M., {et~al.} 2012, Nuclear Instruments
  and Methods in Physics Research A, 689, 87,
  \dodoi{10.1016/j.nima.2012.05.079}

\bibitem[{{Ageron} {et~al.}(2011){Ageron}, {Aguilar}, {Al Samarai}, {Albert},
  {Ameli}, {Andr{\'e}}, {Anghinolfi}, {Anton}, {Anvar}, {Ardid}, {Arnaud},
  {Aslanides}, {Assis Jesus}, {Astraatmadja}, {Aubert}, {Auer}, {Barbarito},
  {Baret}, {Basa}, {Bazzotti}, {Becherini}, {Beltramelli}, {Bersani}, {Bertin},
  {Beurthey}, {Biagi}, {Bigongiari}, {Billault}, {Blaes}, {Bogazzi}, {de
  Botton}, {Bou-Cabo}, {Boudahef}, {Bouwhuis}, {Brown}, {Brunner}, {Busto},
  {Caillat}, {Calzas}, {Camarena}, {Capone}, {Caponetto}, {C{\^a}rloganu},
  {Carminati}, {Carmona}, {Carr}, {Carton}, {Cassano}, {Castorina}, {Cecchini},
  {Ceres}, {Chaleil}, {Charvis}, {Chauchot}, {Chiarusi}, {Circella},
  {Comp{\`e}re}, {Coniglione}, {Coppolani}, {Cosquer}, {Costantini}, {Cottini},
  {Coyle}, {Cuneo}, {Curtil}, {D'Amato}, {Damy}, {van Dantzig}, {de Bonis},
  {Decock}, {Decowski}, {Dekeyser}, {Delagnes}, {Desages-Ardellier},
  {Deschamps}, {Destelle}, {di Maria}, {Dinkespiler}, {Distefano}, {Dominique},
  {Donzaud}, {Dornic}, {Dorosti}, {Drogou}, {Drouhin}, {Druillole}, {Durand},
  {Durand}, {Eberl}, {Emanuele}, {Engelen}, {Ernenwein}, {Escoffier},
  {Falchini}, {Favard}, {Fehr}, {Feinstein}, {Ferri}, {Ferry}, {Fiorello},
  {Flaminio}, {Folger}, {Fritsch}, {Fuda}, {Galat{\'a}}, {Galeotti}, {Gay},
  {Gensolen}, {Giacomelli}, {Gojak}, {G{\'o}mez-Gonz{\'a}lez}, {Goret}, {Graf},
  {Guillard}, {Halladjian}, {Hallewell}, {van Haren}, {Hartmann}, {Heijboer},
  {Heine}, {Hello}, {Henry}, {Hern{\'a}ndez-Rey}, {Herold}, {H{\"o}{\ss}l},
  {Hogenbirk}, {Hsu}, {Hubbard}, {Jaquet}, {Jaspers}, {de Jong}, {Jourde},
  {Kadler}, {Kalantar-Nayestanaki}, {Kalekin}, {Kappes}, {Karg}, {Karkar},
  {Karolak}, {Katz}, {Keller}, {Kestener}, {Kok}, {Kok}, {Kooijman}, {Kopper},
  {Kouchner}, {Kretschmer}, {Kruijer}, {Kuch}, {Kulikovskiy}, {Lachartre},
  {Lafoux}, {Lagier}, {Lahmann}, {Lahonde-Hamdoun}, {Lamare}, {Lambard},
  {Languillat}, {Larosa}, {Lavalle}, {Le Guen}, {Le Provost}, {Levansuu},
  {Lef{\`e}vre}, {Legou}, {Lelaizant}, {L{\'e}v{\'e}que}, {Lim}, {Lo Presti},
  {Loehner}, {Loucatos}, {Louis}, {Lucarelli}, {Lyashuk}, {Magnier}, {Mangano},
  {Marcel}, {Marcelin}, {Margiotta}, {Martinez-Mora}, {Masullo}, {Maz{\'e}as},
  {Mazure}, {Meli}, {Melissas}, {Migneco}, {Mongelli}, {Montaruli}, {Morganti},
  {Moscoso}, {Motz}, {Musumeci}, {Naumann}, {Naumann-Godo}, {Neff}, {Niess},
  {Nooren}, {Oberski}, {Olivetto}, {Palanque-Delabrouille}, {Palioselitis},
  {Papaleo}, {P{\u{a}}v{\u{a}}la{\c{s}}}, {Payet}, {Payre}, {Peek}, {Petrovic},
  {Piattelli}, {Picot-Clemente}, {Picq}, {Piret}, {Poinsignon}, {Popa},
  {Pradier}, {Presani}, {Prono}, {Racca}, {Raia}, {van Randwijk}, {Real},
  {Reed}, {R{\'e}thor{\'e}}, {Rewiersma}, {Riccobene}, {Richardt}, {Richter},
  {Ricol}, {Rigaud}, {Roca}, {Roensch}, {Rolin}, {Rostovtsev}, {Rottura},
  {Roux}, {Rujoiu}, {Ruppi}, {Russo}, {Salesa}, {Salomon}, {Sapienza},
  {Schmitt}, {Sch{\"o}ck}, {Schuller}, {Sch{\"u}ssler}, {Sciliberto},
  {Shanidze}, {Shirokov}, {Simeone}, {Sottoriva}, {Spies}, {Spona}, {Spurio},
  {Steijger}, {Stolarczyk}, {Streeb}, {Sulak}, {Taiuti}, {Tamburini}, {Tao},
  {Tasca}, {Terreni}, {Tezier}, {Toscano}, {Urbano}, {Valdy}, {Vallage}, {van
  Elewyck}, {Vannoni}, {Vecchi}, {Venekamp}, {Verlaat}, {Vernin}, {Virique},
  {de Vries}, {van Wijk}, {Wijnker}, {Wobbe}, {de Wolf}, {Yakovenko}, {Yepes},
  {Zaborov}, {Zaccone}, {Zornoza}, \& {Z{\'u}{\~n}iga}}]{Collaboration:2011nsa}
{Ageron}, M., {Aguilar}, J.~A., {Al Samarai}, I., {et~al.} 2011, Nuclear
  Instruments and Methods in Physics Research A, 656, 11,
  \dodoi{10.1016/j.nima.2011.06.103}

\bibitem[{{Aguilar} {et~al.}(2007){Aguilar}, {Albert}, {Ameli}, {Anghinolfi},
  {Anton}, {Anvar}, {Aslanides}, {Aubert}, {Barbarito}, {Basa}, {Battaglieri},
  {Becherini}, {Bellotti}, {Beltramelli}, {Bertin}, {Bigi}, {Billault},
  {Blaes}, {de Botton}, {Bouwhuis}, {Bradbury}, {Bruijn}, {Brunner}, {Burgio},
  {Busto}, {Cafagna}, {Caillat}, {Calzas}, {Capone}, {Caponetto}, {Carmona},
  {Carr}, {Cartwright}, {Castel}, {Castorina}, {Cavasinni}, {Cecchini},
  {Ceres}, {Charvis}, {Chauchot}, {Chiarusi}, {Circella}, {Colnard},
  {Comp{\`e}re}, {Coniglione}, {Cottini}, {Coyle}, {Cuneo}, {Cussatlegras},
  {Damy}, {van Dantzig}, {de Marzo}, {Dekeyser}, {Delagnes}, {Denans},
  {Deschamps}, {Dessages-Ardellier}, {Destelle}, {Dinkespieler}, {Distefano},
  {Donzaud}, {Drogou}, {Druillole}, {Durand}, {Ernenwein}, {Escoffier},
  {Falchini}, {Favard}, {Feinstein}, {Ferry}, {Festy}, {Fiorello}, {Flaminio},
  {Galeotti}, {Gallone}, {Giacomelli}, {Girard}, {Gojak}, {Goret}, {Graf},
  {Hallewell}, {Harakeh}, {Hartmann}, {Heijboer}, {Heine}, {Hello},
  {Hern{\'a}ndez-Rey}, {H{\"o}{\ss}l}, {Hoffman}, {Hogenbirk}, {Hubbard},
  {Jaquet}, {Jaspers}, {de Jong}, {Jouvenot}, {Kalantar-Nayestanaki}, {Kappes},
  {Karg}, {Karkar}, {Katz}, {Keller}, {Kok}, {Kooijman}, {Kopper}, {Korolkova},
  {Kouchner}, {Kretschmer}, {Kruijer}, {Kuch}, {Kudryavstev}, {Lachartre},
  {Lafoux}, {Lagier}, {Lahmann}, {Lamanna}, {Lamare}, {Languillat},
  {Laschinsky}, {Le Guen}, {Le Provost}, {Le van Suu}, {Legou}, {Lim}, {Lo
  Nigro}, {Lo Presti}, {Loehner}, {Loucatos}, {Louis}, {Lucarelli}, {Lyashuk},
  {Marcelin}, {Margiotta}, {Masullo}, {Maz{\'e}as}, {Mazure}, {McMillan},
  {Megna}, {Melissas}, {Migneco}, {Milovanovic}, {Mongelli}, {Montaruli},
  {Morganti}, {Moscoso}, {Musumeci}, {Naumann}, {Naumann-Godo}, {Niess},
  {Olivetto}, {Ostasch}, {Palanque-Delabrouille}, {Payre}, {Peek}, {Petta},
  {Piattelli}, {Pineau}, {Poinsignon}, {Popa}, {Pradier}, {Racca}, {Randazzo},
  {van Randwijk}, {Real}, {van Rens}, {R{\'e}thor{\'e}}, {Rewiersma},
  {Riccobene}, {Rigaud}, {Ripani}, {Roca}, {Roda}, {Rolin}, {Romita}, {Rose},
  {Rostovtsev}, {Roux}, {Ruppi}, {Russo}, {Salesa}, {Salomon}, {Sapienza},
  {Schmitt}, {Schuller}, {Shanidze}, {Sokalski}, {Spona}, {Spurio}, {van der
  Steenhoven}, {Stolarczyk}, {Streeb}, {Stubert}, {Sulak}, {Taiuti},
  {Tamburini}, {Tao}, {Terreni}, {Thompson}, {Valdy}, {Valente}, {Vallage},
  {Venekamp}, {Verlaat}, {Vernin}, {de Vita}, {de Vries}, {van Wijk}, {de Witt
  Huberts}, {Wobbe}, {de Wolf}, {Yao}, {Zaborov}, {Zaccone}, {Zornoza}, \&
  {Z{\'u}{\~n}iga}}]{Aguilar:2006pd}
{Aguilar}, J.~A., {Albert}, A., {Ameli}, F., {et~al.} 2007, Nuclear Instruments
  and Methods in Physics Research A, 570, 107,
  \dodoi{10.1016/j.nima.2006.09.098}

\bibitem[{{Ahlers} \& {Halzen}(2018)}]{Ahlers:2018fkn}
{Ahlers}, M., \& {Halzen}, F. 2018, Progress in Particle and Nuclear Physics,
  102, 73, \dodoi{10.1016/j.ppnp.2018.05.001}

\bibitem[{{Albert} {et~al.}(2017){Albert}, {Andr{\'e}}, {Anghinolfi}, {Anton},
  {Ardid}, {Aubert}, {Avgitas}, {Baret}, {Barrios-Mart{\'\i}}, {Basa},
  {Belhorma}, {Bertin}, {Biagi}, {Bormuth}, {Bourret}, {Bouwhuis},
  {Br{\^a}nza{\c{s}}}, {Bruijn}, {Brunner}, {Busto}, {Capone}, {Caramete},
  {Carr}, {Celli}, {Cherkaoui El Moursli}, {Chiarusi}, {Circella}, {Coelho},
  {Coleiro}, {Coniglione}, {Costantini}, {Coyle}, {Creusot}, {D{\'\i}az},
  {Deschamps}, {de Bonis}, {Distefano}, {di Palma}, {Domi}, {Donzaud},
  {Dornic}, {Drouhin}, {Eberl}, {El Bojaddaini}, {El Khayati}, {Els{\"a}sser},
  {Enzenh{\"o}fer}, {Ettahiri}, {Fassi}, {Felis}, {Fusco}, {Galat{\`a}}, {Gay},
  {Giordano}, {Glotin}, {Gr{\'e}goire}, {Gracia Ruiz}, {Graf}, {Hallmann}, {van
  Haren}, {Heijboer}, {Hello}, {Hern{\'a}ndez-Rey}, {H{\"o}{\ss}l},
  {Hofest{\"a}dt}, {Hugon}, {Illuminati}, {James}, {de Jong}, {Jongen},
  {Kadler}, {Kalekin}, {Katz}, {Kie{\ss}ling}, {Kouchner}, {Kreter},
  {Kreykenbohm}, {Kulikovskiy}, {Lachaud}, {Lahmann}, {Lef{\`e}vre}, {Leonora},
  {Lotze}, {Loucatos}, {Marcelin}, {Margiotta}, {Marinelli},
  {Mart{\'\i}nez-Mora}, {Mele}, {Melis}, {Michael}, {Migliozzi}, {Moussa},
  {Navas}, {Nezri}, {Organokov}, {P{\v{a}}v{\v{a}}la{\c{s}}}, {Pellegrino},
  {Perrina}, {Piattelli}, {Popa}, {Pradier}, {Quinn}, {Racca}, {Riccobene},
  {S{\'a}nchez-Losa}, {Salda{\~n}a}, {Salvadori}, {Samtleben}, {Sanguineti},
  {Sapienza}, {Sch{\"u}ssler}, {Sieger}, {Spurio}, {Stolarczyk}, {Taiuti},
  {Tayalati}, {Trovato}, {Turpin}, {T{\"o}nnis}, {Vallage}, {van Elewyck},
  {Versari}, {Vivolo}, {Vizzoca}, {Wilms}, {Zornoza}, {Z{\'u}{\~n}iga}, \&
  {ANTARES Collaboration}}]{Albert:2017ohr}
{Albert}, A., {Andr{\'e}}, M., {Anghinolfi}, M., {et~al.} 2017, \prd, 96,
  082001, \dodoi{10.1103/PhysRevD.96.082001}

\bibitem[{{Albert} {et~al.}(2018){Albert}, {Andr{\'e}}, {Anghinolfi}, {Anton},
  {Ardid}, {Aubert}, {Aublin}, {Avgitas}, {Baret}, {Barrios-Mart{\'\i}},
  {Basa}, {Belhorma}, {Bertin}, {Biagi}, {Bormuth}, {Boumaaza}, {Bourret},
  {Bouwhuis}, {Br{\^a}nza{\c{s}}}, {Bruijn}, {Brunner}, {Busto}, {Capone},
  {Caramete}, {Carr}, {Celli}, {Chabab}, {Cherkaoui El Moursli}, {Chiarusi},
  {Circella}, {Coelho}, {Coleiro}, {Colomer}, {Coniglione}, {Costantini},
  {Coyle}, {Creusot}, {D{\'\i}az}, {Deschamps}, {Distefano}, {Di Palma},
  {Domi}, {Don{\`a}}, {Donzaud}, {Dornic}, {Drouhin}, {Eberl}, {El Bojaddaini},
  {El Khayati}, {Els{\"a}sser}, {Enzenh{\"o}fer}, {Ettahiri}, {Fassi}, {Felis},
  {Fermani}, {Ferrara}, {Fusco}, {Gay}, {Glotin}, {Gr{\'e}goire}, {Ruiz},
  {Graf}, {Hallmann}, {van Haren}, {Heijboer}, {Hello}, {Hern{\'a}ndez-Rey},
  {H{\"o}{\ss}l}, {Hofest{\"a}dt}, {Illuminati}, {de Jong}, {Jongen}, {Kadler},
  {Kalekin}, {Katz}, {Khan-Chowdhury}, {Kouchner}, {Kreter}, {Kreykenbohm},
  {Kulikovskiy}, {Lachaud}, {Lahmann}, {Lef{\`e}vre}, {Leonora}, {Lotze},
  {Loucatos}, {Marcelin}, {Margiotta}, {Marinelli}, {Mart{\'\i}nez-Mora},
  {Mele}, {Melis}, {Migliozzi}, {Moussa}, {Navas}, {Nezri}, {Nu{\~n}ez},
  {Organokov}, {P{\u{a}}v{\u{a}}la{\c{s}}}, {Pellegrino}, {Piattelli}, {Popa},
  {Pradier}, {Quinn}, {Racca}, {Randazzo}, {Riccobene}, {S{\'a}nchez-Losa},
  {Salda{\~n}a}, {Salvadori}, {Samtleben}, {Sanguineti}, {Sapienza},
  {Sch{\"u}ssler}, {Spurio}, {Stolarczyk}, {Taiuti}, {Tayalati}, {Trovato},
  {Vallage}, {Van Elewyck}, {Versari}, {Vivolo}, {Wilms}, {Zaborov}, {Zornoza},
  {Z{\'u}{\~n}iga}, \& {ANTARES Collaboration}}]{Albert:2018kjg}
---. 2018, \apjl, 863, L30, \dodoi{10.3847/2041-8213/aad8c0}

\bibitem[{Albert {et~al.}(2020)Albert, André, Anghinolfi, Anton, Ardid,
  Aubert, Aublin, Baret, Basa, Belhorma, Bertin, Biagi, Bissinger, Boumaaza,
  Bourret, Bouta, Bouwhuis, Brânzaş, Bruijn, Brunner, Busto, Capone,
  Caramete, Carr, Celli, Chabab, Chau, Moursli, Chiarusi, Circella, Coleiro,
  Colomer, Coniglione, Costantini, Coyle, Creusot, Díaz, de~Wasseige,
  Deschamps, Distefano, Di~Palma, Domi, Donzaud, Dornic, Drouhin, Eberl,
  Bojaddaini, Khayati, Elsässer, Enzenhöfer, Ettahiri, Fassi, Fermani,
  Ferrara, Filippini, Fusco, Gay, Glotin, Gozzini, Ruiz, Graf, Guidi, Hallmann,
  van Haren, Heijboer, Hello, Hernández-Rey, Hößl, Hofestädt, Illuminati,
  James, de~Jong, de~Jong, Jongen, Kadler, Kalekin, Katz, Khan-Chowdhury,
  Kouchner, Kreter, Kreykenbohm, Kulikovskiy, Lahmann, Breton, Lefèvre,
  Leonora, Levi, Lincetto, Lopez-Coto, Loucatos, Maggi, Manczak, Marcelin,
  Margiotta, Marinelli, Martínez-Mora, Mele, Melis, Migliozzi, Moser, Moussa,
  Muller, Nauta, Navas, Nezri, Nielsen, Nuñez-Castiñeyra, O'Fearraigh,
  Organokov, Păvălaş, Pellegrino, Perrin-Terrin, Piattelli, Poirè, Popa,
  Pradier, Quinn, Randazzo, Riccobene, Sánchez-Losa, Salah-Eddine, Samtleben,
  Sanguineti, Sapienza, Schüssler, Spurio, Stolarczyk, Strandberg, Taiuti,
  Tayalati, Thakore, Tingay, Trovato, Vallage, Van~Elewyck, Versari, Viola,
  Vivolo, Wilms, Zaborov, Zegarelli, Zornoza, Zúñiga, Collaboration, Aartsen,
  Ackermann, Adams, Aguilar, Ahlers, Ahrens, Alispach, Andeen, Anderson,
  Ansseau, Anton, Argüelles, Auffenberg, Axani, Backes, Bagherpour, Bai, V.,
  Barbano, Barwick, Bastian, Baum, Baur, Bay, Beatty, Becker, Tjus, BenZvi,
  Berley, Bernardini, Besson, Binder, Bindig, Blaufuss, Blot, Bohm, Böser,
  Botner, Böttcher, Bourbeau, Bourbeau, Bradascio, Braun, Bron,
  Brostean-Kaiser, Burgman, Buscher, Busse, Carver, Chen, Cheung, Chirkin,
  Choi, Clark, Classen, Coleman, Collin, Conrad, Coppin, Correa, Cowen, Cross,
  Dave, De~Clercq, DeLaunay, Dembinski, Deoskar, De~Ridder, Desiati, de~Vries,
  de~Wasseige, de~With, DeYoung, Diaz, Díaz-Vélez, Dujmovic, Dunkman, Dvorak,
  Eberhardt, Ehrhardt, Eller, Engel, Evenson, Fahey, Fazely, Felde, Filimonov,
  Finley, Fox, Franckowiak, Friedman, Fritz, Gaisser, Gallagher, Ganster,
  Garrappa, Gerhardt, Ghorbani, Glauch, Glüsenkamp, Goldschmidt, Gonzalez,
  Grant, Grégoire, Griffith, Griswold, Günder, Gündüz, Haack, Hallgren,
  Halliday, Halve, Halzen, Hanson, Haungs, Hebecker, Heereman, Heix, Helbing,
  Hellauer, Henningsen, Hickford, Hignight, Hill, Hoffman, Hoffmann, Hoinka,
  Hokanson-Fasig, Hoshina, Huang, Huber, Huber, Hultqvist, Hünnefeld, Hussain,
  In, Iovine, Ishihara, Jansson, Japaridze, Jeong, Jero, Jones, Jonske, Joppe,
  Kang, Kang, Kappes, Kappesser, Karg, Karl, Karle, Katz, Kauer, Kelley,
  Kheirandish, Kim, Kintscher, Kiryluk, Kittler, Klein, Koirala, Kolanoski,
  Köpke, Kopper, Kopper, Koskinen, Kowalski, Krings, Krückl, Kulacz,
  Kurahashi, Kyriacou, Lanfranchi, Larson, Lauber, Lazar, Leonard,
  Leszczyńska, Leuermann, Liu, Lohfink, Mariscal, Lu, Lucarelli, Lünemann,
  Luszczak, Lyu, Ma, Madsen, Maggi, Mahn, Makino, Mallik, Mallot, Mancina,
  Mariş, Maruyama, Mase, Maunu, McNally, Meagher, Medici, Medina, Meier,
  Meighen-Berger, Merino, Meures, Micallef, Mockler, Momenté, Montaruli,
  Moore, Morse, Moulai, Muth, Nagai, Naumann, Neer, Niederhausen, Nisa,
  Nowicki, Nygren, Pollmann, Oehler, Olivas, O'Murchadha, O'Sullivan,
  Palczewski, Pandya, Pankova, Park, Peiffer, de~los Heros, Philippen, Pieloth,
  Pieper, Pinat, Pizzuto, Plum, Porcelli, Price, Przybylski, Raab, Raissi,
  Rameez, Rauch, Rawlins, Rea, Reimann, Relethford, Renschler, Renzi, Resconi,
  Rhode, Richman, Robertson, Rongen, Rott, Ruhe, Ryckbosch, Rysewyk, Safa,
  Herrera, Sandrock, Sandroos, Santander, Sarkar, Sarkar, Satalecka, Schaufel,
  Schieler, Schlunder, Schmidt, Schneider, Schneider, Schröder, Schumacher,
  Sclafani, Seckel, Seunarine, Shefali, Silva, Snihur, Soedingrekso, Soldin,
  Song, Spiczak, Spiering, Stachurska, Stamatikos, Stanev, Stein, Stettner,
  Steuer, Stezelberger, Stokstad, Stößl, Strotjohann, Stürwald, Stuttard,
  Sullivan, Taboada, Tenholt, Ter-Antonyan, Terliuk, Tilav, Tollefson,
  Tomankova, Tönnis, Toscano, Tosi, Trettin, Tselengidou, Tung, Turcati,
  Turcotte, Turley, Ty, Unger, Elorrieta, Usner, Vandenbroucke, Van~Driessche,
  van Eijk, van Eijndhoven, van Santen, Verpoest, Vraeghe, Walck, Wallace,
  Wallraff, Wandkowsky, Watson, Weaver, Weindl, Weiss, Weldert, Wendt,
  Werthebach, Whelan, Whitehorn, Wiebe, Wiebusch, Wille, Williams, Wills, Wolf,
  Wood, Wood, Woschnagg, Wrede, Xu, Xu, Xu, Yanez, Yodh, Yoshida, Yuan, \&
  Zöcklein}]{illuminatiANTARESIceCube2020}
Albert, A., André, M., Anghinolfi, M., {et~al.} 2020, The Astrophysical
  Journal, 892, 92, \dodoi{10.3847/1538-4357/ab7afb}

\bibitem[{{Alves Batista} {et~al.}(2019){Alves Batista}, {Biteau},
  {Bustamante}, {Dolag}, {Engel}, {Fang}, {Kampert}, {Kostunin}, {Mostafa},
  {Murase}, {Oikonomou}, {Olinto}, {Panasyuk}, {Sigl}, {Taylor}, \&
  {Unger}}]{Alves_Batista_2019}
{Alves Batista}, R., {Biteau}, J., {Bustamante}, M., {et~al.} 2019, Frontiers
  in Astronomy and Space Sciences, 6, 23, \dodoi{10.3389/fspas.2019.00023}

\bibitem[{{Anchordoqui} {et~al.}(2008){Anchordoqui}, {Hooper}, {Sarkar}, \&
  {Taylor}}]{Anchordoqui:2007tn}
{Anchordoqui}, L.~A., {Hooper}, D., {Sarkar}, S., \& {Taylor}, A.~M. 2008,
  Astroparticle Physics, 29, 1, \dodoi{10.1016/j.astropartphys.2007.10.006}

\bibitem[{{ANTARES Collaboration} {et~al.}(2002){ANTARES Collaboration},
  {Amram}, {Anghinolfi}, {Anvar}, {Ardellier-Desages}, {Aslanides}, {Aubert},
  {Azoulay}, {Bailey}, {Basa}, {Battaglieri}, {Bellotti}, {Benhammou},
  {Bernard}, {Berthier}, {Bertin}, {Billault}, {Blaes}, {Bland}, {Blondeau},
  {de Botton}, {Boulesteix}, {Brooks}, {Brunner}, {Cafagna}, {Calzas},
  {Capone}, {Caponetto}, {C{\^a}rloganu}, {Carmona}, {Carr}, {Carton},
  {Cartwright}, {Cassol}, {Cecchini}, {Ciacio}, {Circella}, {Comp{\`e}re},
  {Cooper}, {Coyle}, {Croquette}, {Cuneo}, {Danilov}, {van Dantzig}, {De
  Marzo}, {DeVita}, {Deck}, {Destelle}, {Dispau}, {Drougou}, {Druillole},
  {Engelen}, {Feinstein}, {Festy}, {Fopma}, {Gallone}, {Giacomelli}, {Goret},
  {Gosset}, {Gournay}, {Heijboer}, {Hern{\'a}ndez-Rey}, {Herrouin}, {Hubbard},
  {Jaquet}, {de Jong}, {Karolak}, {Kooijman}, {Kouchner}, {Kudryavtsev},
  {Lachartre}, {Lafoux}, {Lamare}, {Languillat}, {Laubier}, {Laugier}, {Le
  Guen}, {Le Provost}, {Le Van Suu}, {Lemoine}, {Lo Nigro}, {Lo Presti},
  {Loucatos}, {Louis}, {Lyashuk}, {Magnier}, {Marcelin}, {Margiotta}, {Massol},
  {Masullo}, {Maz{\'e}as}, {Mazeau}, {Mazure}, {McMillan}, {Michel}, {Migneco},
  {Millot}, {Mols}, {Montanet}, {Montaruli}, {Morel}, {Moscoso}, {Musumeci},
  {Navas}, {Nezri}, {Nooren}, {Oberski}, {Olivetto}, {Oppelt-Pohl},
  {Palanque-Delabrouille}, {Papaleo}, {Payre}, {Perrin}, {Petruccetti},
  {Petta}, {Piattelli}, {Poinsignon}, {Potheau}, {Queinec}, {Racca}, {Raia},
  {Randazzo}, {Rethore}, {Riccobene}, {Ricol}, {Ripani}, {Roca-Blay}, {Rolin},
  {Rostovstev}, {Russo}, {Sacquin}, {Salusti}, {Schuller}, {Schuster},
  {Soirat}, {Souvorova}, {Spooner}, {Spurio}, {Stolarczyk}, {Stubert},
  {Taiuti}, {Tao}, {Tayalati}, {Thompson}, {Tilav}, {Triay}, {Valente},
  {Varlamov}, {Vaudaine}, {Vernin}, {de Witt Huberts}, {de Wolf}, {Zakharov},
  {Zavatarelli}, {de D. Zornoza}, \& {Z{\'u}{\~n}iga}}]{Amram:2001mi}
{ANTARES Collaboration}, {Amram}, P., {Anghinolfi}, M., {et~al.} 2002, Nuclear
  Instruments and Methods in Physics Research A, 484, 369,
  \dodoi{10.1016/S0168-9002(01)02026-5}

\bibitem[{Aublin {et~al.}(2019)Aublin, Coleiro, Kouchner, {for the ANTARES
  Collaboration}, Al~Samarai, Barbano, Montaruli, Schumacher, Wiebusch, {for
  the IceCube Collaboration}, Caccianiga, Ghia, Giaccari, Golup, {for the
  Pierre Auger Collaboration}, Sagawa, Tinyakov, \& {for the Telescope Array
  Collaboration}}]{Aublin:2019irc}
Aublin, J., Coleiro, A., Kouchner, A., {et~al.} 2019, EPJ Web Conf., 210,
  03003, \dodoi{10.1051/epjconf/201921003003}

\bibitem[{{Bahcall} \& {Waxman}(2001)}]{Bahcall:1999yr}
{Bahcall}, J., \& {Waxman}, E. 2001, \prd, 64, 023002,
  \dodoi{10.1103/PhysRevD.64.023002}

\bibitem[{Barbano(2019)}]{barbanoICRC2019}
Barbano, A.~M. 2019, in Proceedings of 36th International Cosmic Ray Conference
  {\textemdash} PoS(ICRC2019), Vol. 358, 842, \dodoi{10.22323/1.358.0842}

\bibitem[{{Biteau} {et~al.}(2019){Biteau}, {Bister}, {Caccianiga}, {Deligny},
  {di Matteo}, {Fujii}, {Harari}, {Kawata}, {Ivanov}, {Lundquist}, {Menezes de
  Almeida}, {Mockler}, {Nonaka}, {Sagawa}, {Tinyakov}, {Tkachev}, \&
  {Troitsky}}]{biteauCoveringCelestialSphere2019}
{Biteau}, J., {Bister}, T., {Caccianiga}, L., {et~al.} 2019, in European
  Physical Journal Web of Conferences, Vol. 210, European Physical Journal Web
  of Conferences, 01005, \dodoi{10.1051/epjconf/201921001005}

\bibitem[{{Bonifazi} \& {The Pierre Auger
  Collaboration}(2009)}]{Bonifazi:2009ma}
{Bonifazi}, C., \& {The Pierre Auger Collaboration}. 2009, Nuclear Physics B
  Proceedings Supplements, 190, 20, \dodoi{10.1016/j.nuclphysbps.2009.03.063}

\bibitem[{Davoudifar(2011)}]{Davoudifar:2011nxv}
Davoudifar, P. 2011, in {32nd International Cosmic Ray Conference}, Vol.~2,
  230, \dodoi{10.7529/ICRC2011/V02/1183}

\bibitem[{Dawson(2019)}]{Dawson:2020bkp}
Dawson, B. 2019, in Proceedings of 36th International Cosmic Ray Conference
  {\textemdash} PoS(ICRC2019), Vol. 358, 231, \dodoi{10.22323/1.358.0231}

\bibitem[{Durrer \& Neronov(2013)}]{Durrer:2013pga}
Durrer, R., \& Neronov, A. 2013, Astron. Astrophys. Rev., 21, 62,
  \dodoi{10.1007/s00159-013-0062-7}

\bibitem[{{G{\'o}rski} {et~al.}(2005){G{\'o}rski}, {Hivon}, {Banday},
  {Wandelt}, {Hansen}, {Reinecke}, \& {Bartelmann}}]{gorskiHEALPix2005}
{G{\'o}rski}, K.~M., {Hivon}, E., {Banday}, A.~J., {et~al.} 2005, \apj, 622,
  759, \dodoi{10.1086/427976}

\bibitem[{Haack \& Wiebusch(2017)}]{haack2018measurement}
Haack, C., \& Wiebusch, C. 2017, in Proceedings of 35th International Cosmic
  Ray Conference {\textemdash} PoS(ICRC2017), Vol. 301, 1005,
  \dodoi{10.22323/1.301.1005}

\bibitem[{Illuminati {et~al.}(2019)Illuminati, Aublin, \&
  Navas}]{Aublin:2019zzn}
Illuminati, G., Aublin, J., \& Navas, S. 2019, in Proceedings of 36th
  International Cosmic Ray Conference {\textemdash} PoS(ICRC2019), Vol. 358,
  920, \dodoi{10.22323/1.358.0920}

\bibitem[{Inoue {et~al.}(2020)Inoue, Khangulyan, \& Doi}]{Inoue_2020}
Inoue, Y., Khangulyan, D., \& Doi, A. 2020, The Astrophysical Journal, 891,
  L33, \dodoi{10.3847/2041-8213/ab7661}

\bibitem[{Jansson \& Farrar(2012)}]{JFgmf2012}
Jansson, R., \& Farrar, G.~R. 2012, The Astrophysical Journal, 757, 14,
  \dodoi{10.1088/0004-637X/757/1/14}

\bibitem[{Kawai {et~al.}(2008)}]{Kawai:2008zza}
Kawai, H., {et~al.} 2008, Nuclear Physics B - Proceedings Supplements, 175-176,
  221, \dodoi{10.1016/j.nuclphysbps.2007.11.002}

\bibitem[{Kopper(2017)}]{Kopper:2017zzm}
Kopper, C. 2017, in Proceedings of 35th International Cosmic Ray Conference
  {\textemdash} PoS(ICRC2017), Vol. 301, 981, \dodoi{10.22323/1.301.0981}

\bibitem[{Kronberg(1994)}]{Kronberg:1993vk}
Kronberg, P.~P. 1994, Rept. Prog. Phys., 57, 325,
  \dodoi{10.1088/0034-4885/57/4/001}

\bibitem[{Meszaros(2018)}]{meszaros2018astrophysical}
Meszaros, P. 2018, PhT, 71, 36

\bibitem[{Murase(2015)}]{Murase:2014tsa}
Murase, K. 2015, AIP Conference Proceedings, 1666, 040006,
  \dodoi{10.1063/1.4915555}

\bibitem[{Murase {et~al.}(2013)Murase, Ahlers, \& Lacki}]{Murase:2013rfa}
Murase, K., Ahlers, M., \& Lacki, B.~C. 2013, Physical Review D, 88, 121301,
  \dodoi{10.1103/PhysRevD.88.121301}

\bibitem[{Murase {et~al.}(2014)Murase, Inoue, \& Dermer}]{murase_agn_2014}
Murase, K., Inoue, Y., \& Dermer, C.~D. 2014, Phys. Rev. D, 90, 023007,
  \dodoi{10.1103/PhysRevD.90.023007}

\bibitem[{Murase \& Waxman(2016)}]{Murase:2016gly}
Murase, K., \& Waxman, E. 2016, Physical Review D, 94, 103006,
  \dodoi{10.1103/PhysRevD.94.103006}

\bibitem[{Nagano \& Watson(2000)}]{Nagano:2000ve}
Nagano, M., \& Watson, A.~A. 2000, Reviews of Modern Physics, 72, 689,
  \dodoi{10.1103/RevModPhys.72.689}

\bibitem[{{Paiano} {et~al.}(2018){Paiano}, {Falomo}, {Treves}, \&
  {Scarpa}}]{paiano:2018txs}
{Paiano}, S., {Falomo}, R., {Treves}, A., \& {Scarpa}, R. 2018, \apjl, 854,
  L32, \dodoi{10.3847/2041-8213/aaad5e}

\bibitem[{Palladino {et~al.}(2020)Palladino, van Vliet, Winter, \&
  Franckowiak}]{palladino:2020}
Palladino, A., van Vliet, A., Winter, W., \& Franckowiak, A. 2020, Monthly
  Notices of the Royal Astronomical Society, 494, 4255,
  \dodoi{10.1093/mnras/staa1003}

\bibitem[{{Particle Data Group} {et~al.}(2020){Particle Data Group}, {Zyla},
  {Barnett}, {Beringer}, {Dahl}, {Dwyer}, {Groom}, {Lin}, {Lugovsky},
  {Pianori}, {Robinson}, {Wohl}, {Yao}, {Agashe}, {Aielli}, {Allanach},
  {Amsler}, {Antonelli}, {Aschenauer}, {Asner}, {Baer}, {Banerjee}, {Baudis},
  {Bauer}, {Beatty}, {Belousov}, {Bethke}, {Bettini}, {Biebel}, {Black},
  {Blucher}, {Buchmuller}, {Burkert}, {Bychkov}, {Cahn}, {Carena}, {Ceccucci},
  {Cerri}, {Chakraborty}, {Chivukula}, {Cowan}, {D'Ambrosio}, {Damour}, {de
  Florian}, {de Gouv{\^e}a}, {DeGrand}, {de Jong}, {Dissertori}, {Dobrescu},
  {D'Onofrio}, {Doser}, {Drees}, {Dreiner}, {Eerola}, {Egede}, {Eidelman},
  {Ellis}, {Erler}, {Ezhela}, {Fetscher}, {Fields}, {Foster}, {Freitas},
  {Gallagher}, {Garren}, {Gerber}, {Gerbier}, {Gershon}, {Gershtein},
  {Gherghetta}, {Godizov}, {Gonzalez-Garcia}, {Goodman}, {Grab}, {Gritsan},
  {Grojean}, {Gr{\"u}newald}, {Gurtu}, {Gutsche}, {Haber}, {Hanhart},
  {Hashimoto}, {Hayato}, {Hebecker}, {Heinemeyer}, {Heltsley},
  {Hern{\'a}ndez-Rey}, {Hikasa}, {Hisano}, {H{\"o}cker}, {Holder}, {Holtkamp},
  {Huston}, {Hyodo}, {Johnson}, {Kado}, {Karliner}, {Katz}, {Kenzie}, {Khoze},
  {Klein}, {Klempt}, {Kowalewski}, {Krauss}, {Kreps}, {Krusche}, {Kwon},
  {Lahav}, {Laiho}, {Lellouch}, {Lesgourgues}, {Liddle}, {Ligeti}, {Lippmann},
  {Liss}, {Littenberg}, {Lourengo}, {Lugovsky}, {Lusiani}, {Makida}, {Maltoni},
  {Mannel}, {Manohar}, {Marciano}, {Masoni}, {Matthews}, {Mei{\ss}ner},
  {Mikhasenko}, {Miller}, {Milstead}, {Mitchell}, {M{\"o}nig}, {Molaro},
  {Moortgat}, {Moskovic}, {Nakamura}, {Narain}, {Nason}, {Navas}, {Neubert},
  {Nevski}, {Nir}, {Olive}, {Patrignani}, {Peacock}, {Petcov}, {Petrov},
  {Pich}, {Piepke}, {Pomarol}, {Profumo}, {Quadt}, {Rabbertz}, {Rademacker},
  {Raffelt}, {Ramani}, {Ramsey-Musolf}, {Ratcliff}, {Richardson}, {Ringwald},
  {Roesler}, {Rolli}, {Romaniouk}, {Rosenberg}, {Rosner}, {Rybka}, {Ryskin},
  {Ryutin}, {Sakai}, {Salam}, {Sarkar}, {Sauli}, {Schneider}, {Scholberg},
  {Schwartz}, {Schwiening}, {Scott}, {Sharma}, {Sharpe}, {Shutt}, {Silari},
  {Sj{\"o}strand}, {Skands}, {Skwarnicki}, {Smoot}, {Soffer}, {Sozzi},
  {Spanier}, {Spiering}, {Stahl}, {Stone}, {Sumino}, {Sumiyoshi}, {Syphers},
  {Takahashi}, {Tanabashi}, {Tanaka}, {Ta{\v{s}}evsk{\'y}}, {Terashi},
  {Terning}, {Thoma}, {Thorne}, {Tiator}, {Titov}, {Tkachenko}, {Tovey},
  {Trabelsi}, {Urquijo}, {Valencia}, {Van de Water}, {Varelas}, {Venanzoni},
  {Verde}, {Vincter}, {Vogel}, {Vogelsang}, {Vogt}, {Vorobyev}, {Wakely},
  {Walkowiak}, {Walter}, {Wands}, {Wascko}, {Weinberg}, {Weinberg}, {White},
  {Wiencke}, {Willocq}, {Woody}, {Workman}, {Yokoyama}, {Yoshida},
  {Zanderighi}, {Zeller}, {Zenin}, {Zhu}, {Zhu}, {Zimmermann}, {Anderson},
  {Basaglia}, {Lugovsky}, {Schaffner}, \& {Zheng}}]{Zyla:2020zbs}
{Particle Data Group}, {Zyla}, P.~A., {Barnett}, R.~M., {et~al.} 2020, Progress
  of Theoretical and Experimental Physics, 2020, 083C01,
  \dodoi{10.1093/ptep/ptaa104}

\bibitem[{Pshirkov {et~al.}(2011)Pshirkov, Tinyakov, Kronberg, \&
  Newton-McGee}]{PTgmf2011}
Pshirkov, M.~S., Tinyakov, P.~G., Kronberg, P.~P., \& Newton-McGee, K.~J. 2011,
  The Astrophysical Journal, 738, 192, \dodoi{10.1088/0004-637X/738/2/192}

\bibitem[{Pshirkov {et~al.}(2016)Pshirkov, Tinyakov, \&
  Urban}]{Pshirkov:2015tua}
Pshirkov, M.~S., Tinyakov, P.~G., \& Urban, F.~R. 2016, Phys. Rev. Lett., 116,
  191302, \dodoi{10.1103/PhysRevLett.116.191302}

\bibitem[{Rodrigues {et~al.}(2021)Rodrigues, Heinze, Palladino, van Vliet, \&
  Winter}]{rodrigues_agn_2021}
Rodrigues, X., Heinze, J., Palladino, A., van Vliet, A., \& Winter, W. 2021,
  Phys. Rev. Lett., 126, 191101, \dodoi{10.1103/PhysRevLett.126.191101}

\bibitem[{{Schumacher}(2019)}]{Schumacher:2019qdx}
{Schumacher}, L. 2019, in European Physical Journal Web of Conferences, Vol.
  207, European Physical Journal Web of Conferences, 02010,
  \dodoi{10.1051/epjconf/201920702010}

\bibitem[{{Sommers}(2001)}]{SOMMERS2001271}
{Sommers}, P. 2001, Astroparticle Physics, 14, 271,
  \dodoi{10.1016/S0927-6505(00)00130-4}

\bibitem[{Stettner(2019)}]{Stettner:2019tok}
Stettner, J. 2019, in Proceedings of 36th International Cosmic Ray Conference
  {\textemdash} PoS(ICRC2019), Vol. 358, 1017, \dodoi{10.22323/1.358.1017}

\bibitem[{{The IceCube Collaboration}(2013)}]{Aartsen:2013jdh}
{The IceCube Collaboration}. 2013, Science, 342, 1242856,
  \dodoi{10.1126/science.1242856}

\bibitem[{{The IceCube Collaboration} {et~al.}(2016){The IceCube
  Collaboration}, {The Pierre Auger Collaboration}, \& {The Telescope Array
  Collaboration}}]{Aartsen:2015dml}
{The IceCube Collaboration}, {The Pierre Auger Collaboration}, \& {The
  Telescope Array Collaboration}. 2016, \jcap, 2016, 037,
  \dodoi{10.1088/1475-7516/2016/01/037}

\bibitem[{{The Pierre Auger Collaboration}(2015)}]{ThePierreAuger:2015rma}
{The Pierre Auger Collaboration}. 2015, Nuclear Instruments and Methods in
  Physics Research A, 798, 172, \dodoi{10.1016/j.nima.2015.06.058}

\bibitem[{Tkachev {et~al.}(2021)Tkachev, Fujii, Ivanov, Jui, Kawata, Kim,
  Kuznetsov, Nonaka, Ogio, Rubtsov, Sagawa, Thomson, Tinyakov, \&
  Troitsky}]{Tkachev:20211D}
Tkachev, I., Fujii, T., Ivanov, D., {et~al.} 2021, in Proceedings of 37th
  International Cosmic Ray Conference {\textemdash} PoS(ICRC2021), Vol. 395,
  392, \dodoi{10.22323/1.395.0392}

\bibitem[{{Tokuno} {et~al.}(2012){Tokuno}, {Tameda}, {Takeda}, {Kadota},
  {Ikeda}, {Chikawa}, {Fujii}, {Fukushima}, {Honda}, {Inoue}, {Kakimoto},
  {Kawana}, {Kido}, {Matthews}, {Nonaka}, {Ogio}, {Okuda}, {Ozawa}, {Sagawa},
  {Sakurai}, {Shibata}, {Taketa}, {Thomas}, {Tomida}, {Tsunesada}, {Udo},
  {Abu-zayyad}, {Aida}, {Allen}, {Anderson}, {Azuma}, {Barcikowski}, {Belz},
  {Bergman}, {Blake}, {Cady}, {Cheon}, {Chiba}, {Cho}, {Cho}, {Fujii},
  {Fukuda}, {Gorbunov}, {Hanlon}, {Hayashi}, {Hayashi}, {Hayashida}, {Hibino},
  {Hiyama}, {Iguchi}, {Ikuta}, {Ishii}, {Ishimori}, {Ivanov}, {Iwamoto}, {Jui},
  {Kalashev}, {Kanbe}, {Kasahara}, {Kawai}, {Kawakami}, {Kim}, {Kim}, {Kim},
  {Kim}, {Kitamoto}, {Kobayashi}, {Kobayashi}, {Kondo}, {Kuramoto}, {Kuzmin},
  {Kwon}, {Lim}, {Machida}, {Martens}, {Martineau}, {Matsuda}, {Matsuura},
  {Matsuyama}, {Myers}, {Minamino}, {Miyata}, {Miyauchi}, {Murano}, {Nakamura},
  {Nam}, {Ohnishi}, {Ohoka}, {Oki}, {Oku}, {Oshima}, {Park}, {Pshirkov},
  {Rodriguez}, {Roh}, {Rubtsov}, {Ryu}, {Sampson}, {Scott}, {Shah}, {Shibata},
  {Shimodaira}, {Shin}, {Shin}, {Shirahama}, {Smith}, {Sokolsky}, {Sonley},
  {Springer}, {Stokes}, {Stratton}, {Stroman}, {Suzuki}, {Takahashi}, {Takita},
  {Tanaka}, {Tanaka}, {Tanaka}, {Thomson}, {Tinyakov}, {Tkachev}, {Troitsky},
  {Tsutsumi}, {Tsuyuguchi}, {Uchihori}, {Ukai}, {Vasiloff}, {Wada}, {Wong},
  {Wood}, {Yamakawa}, {Yamaoka}, {Yamazaki}, {Yang}, {Yoshida}, {Yoshii},
  {Zollinger}, \& {Zundel}}]{Tokuno:2012mi}
{Tokuno}, H., {Tameda}, Y., {Takeda}, M., {et~al.} 2012, Nuclear Instruments
  and Methods in Physics Research A, 676, 54,
  \dodoi{10.1016/j.nima.2012.02.044}

\bibitem[{Troitsky {et~al.}(2017)Troitsky, Fukushima, Ikeda, Ivanov, Kawata,
  Kido, Lundquist, Matthews, Nonaka, Okuda, Rubtsov, Sagawa, Sakurai, Takeda,
  Takeishi, Taketa, Thomson, Tinyakov, Tkachev, \& Tokuno}]{Troitsky:20171D}
Troitsky, S., Fukushima, M., Ikeda, D., {et~al.} 2017, in Proceedings of 35th
  International Cosmic Ray Conference {\textemdash} PoS(ICRC2017), Vol. 301,
  548, \dodoi{10.22323/1.301.0548}

\bibitem[{{Wandkowsky}(2018)}]{wandkowsky2018latest}
{Wandkowsky}, N. 2018, in XXVIII International Conference on Neutrino Physics
  and Astrophysics, 501, \dodoi{10.5281/zenodo.1301088}

\bibitem[{Waxman \& Bahcall(1999)}]{Waxman:1998yy}
Waxman, E., \& Bahcall, J.~N. 1999, Physical Review D, 59, 023002,
  \dodoi{10.1103/PhysRevD.59.023002}

\end{thebibliography}

\end{document}